\documentclass[useAMS,usenatbib]{mn2e}
\input psfig.tex
\usepackage{latexsym}
\usepackage{amssymb}
\usepackage{amsmath}
\usepackage{graphics}
\usepackage{graphicx}
\usepackage{natbib}
\usepackage{subfigure}

\title[The Role of Dust in Models of Population Synthesis]{The Role of Dust in Models of Population Synthesis}
\author[L. P. Cassar\`{a}, L. Piovan, A. Weiss, M. Salaris and C. Chiosi]{L. P. Cassar\`{a}$^{1,2}$\thanks{E-mail:
letizia@lambrate.inaf.it (LPC); lorenzo.piovan@gmail.com (LP); weiss@mpa-garching.mpg.de (AW); M.Salaris@ljmu.ac.uk (MS); 
cesare.chiosi@unipd.it (CC)}, L. Piovan$^{1,3}$\footnotemark[1], A. Weiss$^{3}$\footnotemark[1],
M. Salaris$^{4}$\footnotemark[1] and C. Chiosi$^{1}$\footnotemark[1] \\
 $^{1}$Department of Physics and Astronomy, University of Padova,
       Via Marzolo 8-I, 35131, Padova, Italy\\
 $^{2}$INAF-IASF Milano, Via E. Bassini 15, 20133 Milano, Italy\\
 $^{3}$Max-Planck-Institut f\"ur Astrophysik, Karl-Schwarzschild-Str. 1, Garching bei M\"unchen, Germany\\
 $^{4}$Astrophysics Research Institute, Liverpool John Moores University, IC2, Liverpool Science Park, United Kingdom}

\begin{document}

\date{Accepted 2013 September 19.  Received 2013 September 18; in original form 2013 June 15}

\pagerange{\pageref{firstpage}--\pageref{lastpage}} \pubyear{2013}

\maketitle

\label{firstpage}

\begin{abstract}

We have employed
state-of-the-art evolutionary models of low and intermediate-mass AGB stars, and included
the effect of circumstellar dust shells on the spectral
energy distribution (SED) of AGB stars, to revise the Padua
library of isochrones \citep{Bertelli1994} that covers an extended range of
ages and initial chemical compositions.
The major revision involves the thermally pulsing AGB phase, that is now taken from fully
evolutionary calculations by \citet{Weiss2009}.
Two libraries of about 600 AGB
dust-enshrouded SEDs each, have also been calculated, one for oxygen-rich M-stars and one
for carbon-rich C-stars. Each library accounts for
different values of input parameters like the optical depth
$\tau$, dust composition, and temperature of the inner boundary of the
dust shell. These libraries of dusty AGB spectra have been implemented
into a large composite library of theoretical stellar spectra,
to cover all regions of the Hertzsprung-Russell Diagram (HRD) crossed by the
isochrones.

With the aid of the above isochrones and libraries of stellar SEDs, we
have calculated the spectro-photometric properties (SEDs, magnitudes, and
colours) of single-generation stellar populations (SSPs) for six metallicities, more 
than fifty ages (from $\sim$3 Myr to 15 Gyr),
and nine choices of the Initial Mass Function.
The new isochrones and SSPs have been compared to the colour-magnitude diagrams
(CMDs) of field populations in the LMC and SMC, with particular
emphasis on AGB stars, and the integrated colours of star clusters in
the same galaxies, using data from the SAGE (\textit{Surveying the Agents of Galaxy Evolution}) catalogues.
We have also examined the integrated colours of a small sample of star clusters located in the outskirts of M31.
The agreement between theory and observations is generally good.
In particular, the new SSPs reproduce the red tails of the AGB star distribution in the
CMDs of field stars in the Magellanic Clouds. Some discrepancies still exist and need to be investigated further.
\end{abstract}


\section{Introduction} \label{intro}

The frontier for high-z objects has been continuously and quickly
extended by the HST WFC3 camera from z$\sim$4-5 \citep{Madau1996,Steidel1999}, and z$\sim$6
 \citep{Stanway2003,Dickinson2004} to z$\sim$10 \citep{Zheng2012,Bouwens2012,Oesch2012}.\\
\indent According to the current view, first galaxies formed at
 z$\sim$10-20 \citep{RowanRobinson2012}, and this high redshift universe is
 obscured by copious amounts of
dust \citep[see][]{Shapley2001,Carilli2001,Robson2004,Wang2008a,Wang2008b,Michalowski2010a,Michalowski2010b},
whose origin and composition are a matter of debate \citep{Gall2011a,Gall2011b,Dwek2009,Draine2009,Dwek2011}.
Understanding the properties of this interstellar dust, and modelling its
coupling with stellar populations are critical to determine the properties of the high-z universe, and
obtain precious
clues on the fundamental question of when and how galaxies formed and evolved.
A major effort is thus being made in the theoretical spectro-photometric,
dynamical, and chemical modelling of dusty galaxies \citep[see for
instance][]{Narayanan2010,Jonsson2010,Grassi2010,Pipino2011,Popescu2011}.
\indent Stellar radiation is absorbed by dust, and reemitted at longer wavelengths, resulting in a change of its
spectral energy distribution (SED) \citep{Silva1998,Piovan2006b,Popescu2011}.
Dust also strongly affects the production of molecular hydrogen and the local amount
of UV radiation in galaxies, thus playing a major role in the star formation process \citep{Yamasawa2011}.\\
\indent The inclusion of dust in the theoretical models of galaxy spectra
leads to a growing
complexity and typically to a much larger set of parameters. We can
identify two main circumstances in which dust interacts with the
stellar light. First, massive stars
are embedded in their parental molecular clouds (MCs), during the early evolution; the
duration of this phase is short,
but the effect of dust on the stellar spectra is not negligible,
and a significant fraction of light is shifted to the IR region.
Second, during the asymptotic giant branch (AGB) phase low and intermediate mass stars
may form an outer dust-rich shell of material, that obscures and
reprocesses the radiation emitted from the photosphere. \\
\indent Stars and dust are tightly interwoven not only
\emph{locally} (stars-MCs, stars-circumstellar dust shell), but also
\emph{globally} (stars, gas and dust mixed in the galactic
environment). In general, dust is partly associated  with the
diffuse interstellar medium (ISM), partly with star forming  molecular regions,
and partly with the circumstellar envelopes of AGB stars. In all
cases, the effect is the absorption of the stellar light at
UV-optical wavelength, with consequent re-emission in the
NIR-MIR-FIR (near, middle and far infrared, respectively). It is clear from these 
considerations that dust affects the observed SEDs of high-z objects,
hampering their interpretation in terms of
fundamental physical parameters like stellar ages, metallicities, initial
mass function (IMF), and the determination of the galaxy star formation histories (SFHs).\\
\indent This paper is the first of a series devoted to study
the spectro-photometric evolution of star clusters and
galaxies, taking into account the key role played by
dust in determining the spectro-photometric properties of single-generation stellar populations (SSPs). 
The final goal is to derive new state-of-the-art
isochrones and integrated properties of SSPs, and to model the spectro-photometric
properties of galaxies, considering the local and global effects of dust formation, destruction and evolution.\\
\indent We have set up an extended library of isochrones and
SSPs of different chemical composition, age and IMFs, that take into account the effect of circumstellar dust around AGB stars. 
Although we will show that the IMF has a marginal effect on the SED, hence magnitudes and colours of
SSPs, it plays an important role in determining properties
of galaxies, that can be interpreted as the sum of many SSPs of different age, weighted by the SFH. In fact, the IMF
affects both the chemical enrichment of the galactic ISM by the stellar ejecta, and the galaxy stellar mass.\\
\indent The outline of the paper is as follows. Section
\ref{iso_state_art} provides a brief review of the state-of-the art regarding theoretical isochrones and SSPs, the building blocks of the
  evolutionary population synthesis (EPS) models. In Section \ref{star_models}
	we describe the new models for AGB stars by \citet{Weiss2009} and how
	they have been included in the Padua Library of stellar models and isochrones by \citet{Bertelli1994}.
        Section \ref{new_isocs} presents our new isochrones,
	whereas in Section \ref{dust_free_SSP} we describe the companion SSPs without the inclusion of dust.
 Section \ref{AGBShell} analyzes the effects of dust shells around AGB stars on the radiation
 emitted by the central object. In particular, we model the dust-rich envelope of
AGB stars at varying optical depth, as a function of the efficiency of mass-loss and the dust to gas ratio.
We finally calculate two libraries of stellar spectra for oxygen-rich M-type stars and
carbon-rich C-stars, respectively. The results are described in Section \ref{dustyspectra}.
The SSPs including the effect of dust, are presented in
Section \ref{SSPs_dustyAGB}.
In Section \ref{StarClusters}, we validate our isochrones and SSPs
on Small and
Large Magellanic Cloud (SMC and LMC) field stars, and clusters in the SMC, LMC and M31.
Finally, Section \ref{disc_conc}, summarizes the main results of this study.

\section{Isochrones with AGB stars } \label{iso_state_art}

Stellar evolutionary tracks, isochrones and SSPs can be used to
study photometric and spectroscopic observations of resolved and
unresolved stellar populations, from the simple age-dating of star
clusters, to the derivation of star formation histories of resolved
galaxies. To mention just a few recent applications, we recall here  \citet{Pessev2006,Pessev2008}, \citet{Ma2012} and references therein.\\
\indent They are also necessary to study the
spectro-photometric evolution of galaxies, using either EPS classical
models \citep{Arimoto1987,Bressan1994,Silva1998,Buzzoni2002,Bruzual2003,Buzzoni2005,Piovan2006b},
or models based on chemo-dynamical simulations, like the ones presented in \citet{Tantalo2010}.
For a recent review of the EPS theory, see, e.g., \citet{Conroy2013}.\\
\indent Many groups have published large grids of stellar isochrones,
covering a wide range of stellar parameters (age, mass, metal content, metal mixture, helium abundance)
that can be used in stellar population synthesis models of galaxies. To give just a few examples, we
refer the reader to
the Geneva database of stellar evolution tracks and isochrones \citep{Lejeune2001}, the various
 releases of stellar tracks and isochrones from Padua \citep{Bertelli1994,Girardi2002,Marigo2008,Bertelli2008},
the BaSTI database \citep{Pietrinferni2004,Pietrinferni2006,Cordier2007},
the Dartmouth database \citep{Dotter2008}, the Yunnan-I \citep{Zhang2002},  Yunnan-II \citep{Zhang2004,Zhang2005}
and most recently the  Yunnan-III models \citep{Zhang2012}. A more detailed overview is given
by \citet{Zhang2012} and will not be repeated here.\\
\indent One of the major uncertainties is the inclusion of the AGB evolutionary phase.
In brief, AGB stars play an important role for
populations with an age larger than about one hundred million years.
Even though the AGB phase is short lived, these stars are very
bright, they may reach very low effective temperatures, and can
get enshrouded in a shell of self-produced dust that
reprocesses the radiation emitted by the central object.
Thanks to their luminosity, they
contribute significantly to the total light emitted by a SSP. Also, because of their low
surface temperatures, they dominate the NIR spectra and colours.
All stars in the mass range from about 0.8 M$_{\odot}$ to $\sim$
6 M$_{\odot}$, are known to become AGB stars towards the end of their
evolution, before moving to the Planetary Nebula (PN) and carbon-oxygen
White Dwarf (CO-WD) phases, after having lost their envelope.
The AGB phase is characterized by the so-called thermal
pulsing instability of the He-burning shell (TP-AGB phase) that causes recurrent
expansions/contractions of the envelope and other surface phenomena that make the AGB
phase particularly difficult to follow.
There are currently two classes of models for the TP-AGB phase.
The first one includes the
semi-analytical or synthetic TP-AGB models; these calculations model the evolution of
the layers above the inert CO-core, by
adopting suitable inner boundary conditions,
and account for mass-loss from the photosphere and envelope burning (EB; also called Hot Bottom Burning HBB).
By employing analytical relations obtained from fully evolutionary calculations regarding, i.e.,
the CO-core mass-luminosity relation, the evolution through the thermal pulses is followed,
taking into account the growth of
the CO-core, the change of the surface abundances, its effect on the surface
opacities, the decrease of the total mass, and the increase of the mean luminosity
\citep[see][and references therein]{Marigo1996,Wagenhuber1998,Marigo1998,Marigo2002,
Izzard2004,Cordier2007,Marigo2007,Buell2012}.
The second type of models includes time-consuming, full evolutionary AGB calculations
\citep{Karakas2002,Straniero2003,Kitsikis2007,Weiss2009,Karakas2011}. Additionally,
models can be grouped according to the opacity
adopted for the outer layers, e.g. opacities with fixed carbon to
oxygen abundance ratio (denoted here as [C/O], with [C/O]$< 1$ typical of the envelopes of M-stars), and
opacities dependent on [C/O], that can increase above unity as the
abundance of carbon increases
during the third dredge-up.\\

\indent \textbf{\textsf{The old past: short AGB tracks}}. We consider the isochrones of \citet{Bertelli1994}
to illustrate the past
situation with classical models of AGB stars,
i.e. synthetic models with envelope  opacities for [C/O]-ratios typical of M-stars.
The points to note are
(i) the limited redward extension of the AGB in the HR diagram (HRD),
due to  the low opacity in the C-O-rich envelopes of these stars
\citep[see][and below]{Marigo2002}; (ii) isochrones (and SSPs in turn)
of metallicity significantly higher than solar (e.g. $Z$=0.05 and
$Z$=0.1) miss the AGB phase and directly evolve from core
He-burning to the White Dwarf WD stage. Stars of this type are  good
candidates to explain the UV-excess of elliptical galaxies
and its correlation with metallicity  \citep{Bressan1994}. In brief:
low mass stars (and stars at the lower end of the intermediate-mass range)
with  metallicities $\sim 2.5\, Z_{\odot}$ undergo the
He-burning at the red side of their HB (red-HB) but miss the TP-AGB.
Soon after the early-AGB (E-AGB) phase is completed, they move to the WD
stage. When the metallicity is higher, ($3\,Z_{\odot}$), low-mass
He-burning stars ($0.55-0.6\, M_{\odot}$) spend a
significant fraction of their evolution at rather high
T$_{\rm eff}$, and soon after He-exhaustion in the core, they
evolve directly to the WD stage. They are called
Hot-HB and AGB-manqu\'{e} objects, and  play a crucial role in the UV-upturn of massive
elliptical galaxies \citep{Greggio1990,Castellani1991,Bressan1994}. This behaviour results from a 
combination of both the lower hydrogen content in the envelope, and the enhanced CNO efficiency in the
H-burning shell, that both concur to burn the hydrogen-rich envelope much faster than in stars 
of the same mass but lower metallicity and helium content.\\

\indent \textbf{\textsf{The recent past: extended AGB tracks}}. The
insufficient extension of the classical models for AGB stars has been
cured by the new models calculated over the past decade, thanks
in particular to the adoption of opacities for the model envelopes, that  increase
  significantly when passing from oxygen- to carbon-dominated abundances
\citep{Marigo2002,Marigo2007,Marigo2008,Weiss2009}.
The Padua and BaSTI stellar model libraries have included the  TP-AGB
phase according to the prescriptions by \citet{Marigo2008} and
\citet{Cordier2007}, using \textit{synthetic AGB-models}
\citep{Iben1978,Renzini1981,Groenewegen1993,Marigo1996}, as
  described above.
Synthetic models are in turn calibrated against the full stellar
models
and observational data.\\

\indent \textbf{\textsf{This study}}. Despite the more
extended AGB phase brought by the improved opacities \citep[][]{Marigo2002}, the refined
prescriptions for synthetic models adopted by \citet{Marigo2008},
and new sets of stellar models and isochrones presented by
\citet{Bertelli2007,Bertelli2008,Bertelli2009} and \citet{Nasi2008},
there are some properties of the classical \textit{Padua Library} \citep[][- http://pleiadi.pd.astro.it/]{Bertelli1994}
that went lost in the subsequent releases. First of all,
the large range of metallicities and initial masses (including massive stars), and
the Hot-HB and AGB-manqu{\'e} evolutionary channels, plus others not relevant to this discussion.
As the \citet{Bertelli1994} isochrones have been widely
used to study spectro-photometric properties of a large variety of
astrophysical objects, from star clusters to galaxies of different
morphological types \citep[see][and the many papers referring to
  it]{Bertelli1994} both in the nearby and high redshift Universe,
instead of generating new isochrones and SSPs based entirely on the new
stellar models by \citet{Weiss2009} -- that cover a much smaller
range of initial masses and chemical compositions (see below) -- we
consider the \citet{Bertelli1994} isochrones until
the E-AGB phase, and add the TP-AGB models of
\citet{Weiss2009}. Important similarities between these two model sets
ensure that the match can be performed safely.
The new AGB models by \citet{Weiss2009} allows us to
discriminate between carbon-rich and oxygen-rich stages of the AGB evolution
of stars of different mass and initial chemical composition. This improves upon the previous SSP models with dust by
\citet{Piovan2003}, that could not follow the evolution
of the C and O surface abundances of the AGB stars, and
the oxygen- to carbon-rich envelopes was roughly estimated from the old synthetic AGB
models by \citet{Marigo2001}.
Updating SSP and SED calculation in presence of dust was not possible
for long time, because the synthetic AGB models with variable opacities in outer envelopes
and a more realistic description of  oxygen- and carbon-rich regimes by \citet[][and references]{Marigo2008} were not
public. \\

\indent The main characteristics of our adopted model libraries can be summarized as follows:\\
(i) The stellar models of the \citet{Bertelli1994} library are those
of \citet{Alongi1993}, \citet{Bressan1993}, \citet{Fagotto1994a,Fagotto1994b,Fagotto1994c}, \citet{Girardi1996}
and were calculated with the Padua stellar evolution code. All
evolutionary phases, from the zero age main sequence to the start of the TP-AGB stage or
central C-ignition are included, as appropriate for the mass
of the model.
We have considered the metallicities $Z$=0.0001, 0.0004, 0.004, 0.008, 0.02, and 0.05.
The case with $Z$=0.1 is not included (see below).
For all models, the primordial He-content is $Y$=0.23 and
the He enrichment law is $\Delta Y/\Delta Z$=2.5.\\
(ii) The stellar models of \citet{Weiss2009} have initial masses in
the range from 1 to 6 $M\odot$ and metallicities from $Z$=0.004 to
$Z$=0.05; they cannot be used to calculate both very young and very old isochrones
and  neither deal with very low and/or very high
metallicities. The models of \citet{Weiss2009} were calculated with the
GARSTEC code \citep[see][for a description of the code]{Weiss2008}.\\
(iii) The very high metallicity $Z$=0.1 cannot be included because  \citet{Weiss2009} new
AGB models are not available with this composition.
Although the very high metallicity stars may appear as Hot-HB and even AGB manqu\`{e} objects,
(models predict that at $Z$=0.1 this should occur for ages above $\sim$8.5 Gyr),
still a large number of objects is expected to evolve through the standard AGB phase and
develop a thick dust-rich envelope.
Neglecting the presence of stars of very high $Z$ -- albeit in small percentages --
could affect comparisons of models with the MIR-FIR emission of stellar populations in the nuclear regions of  giant
elliptical galaxies \citep{Bressan1994}.
We have a similar problem also at the very low metallicity $Z$=0.0001. The lowest
metallicity in the AGB models by \citet{Weiss2009} is $Z$=0.0004. The problem is here less severe, and can
be easily cured by extrapolating the properties of the $Z$=0.0004 AGB models, down to $Z$=0.0001.\\
(iv) Finally, both groups of models make use of the
\citet{Grevesse1993} solar metal mixture.

\begin{figure}
\begin{center}
\subfigure
{\includegraphics[width=0.40\textwidth]{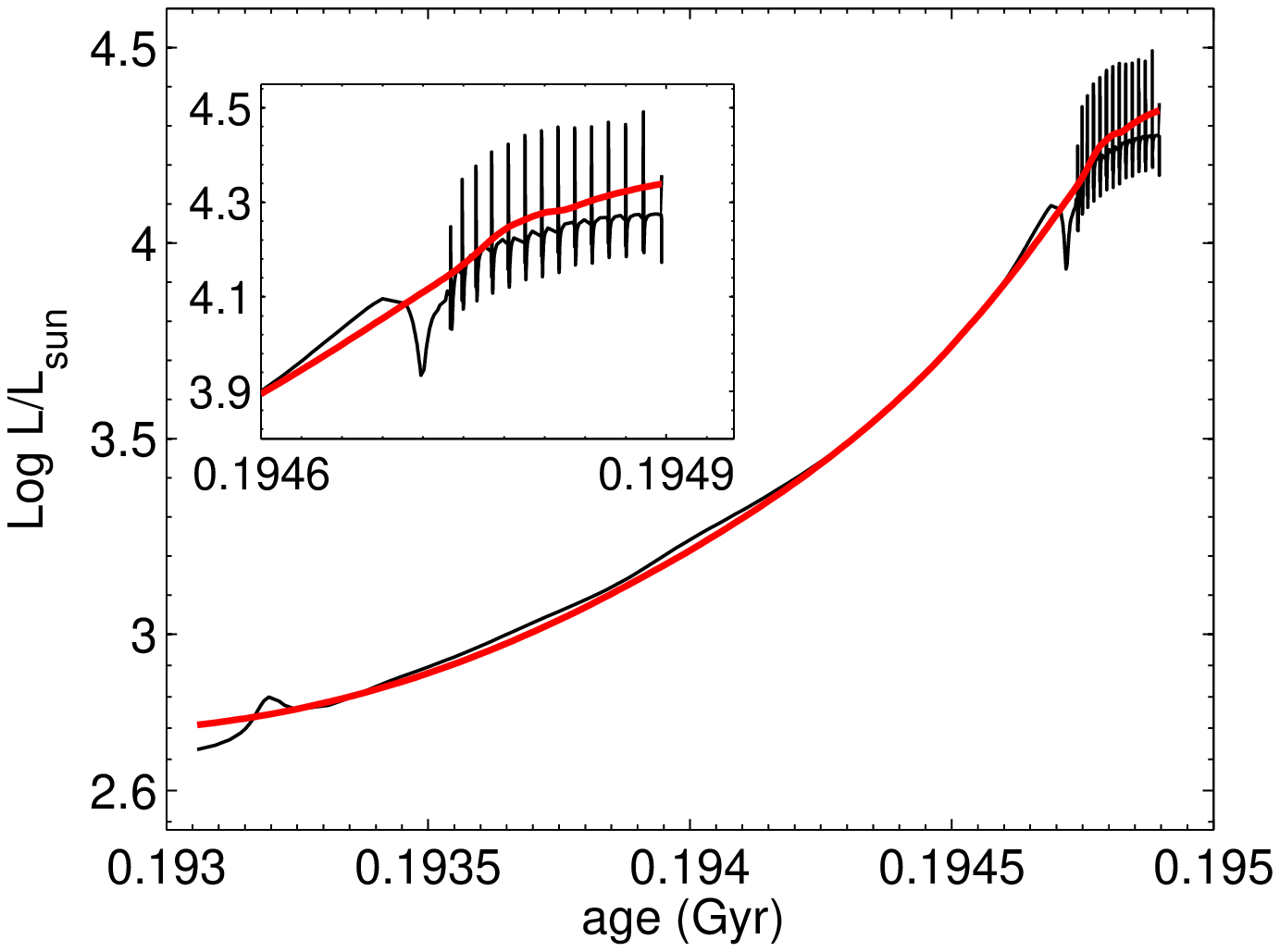}}
\subfigure
{\includegraphics[width=0.40\textwidth]{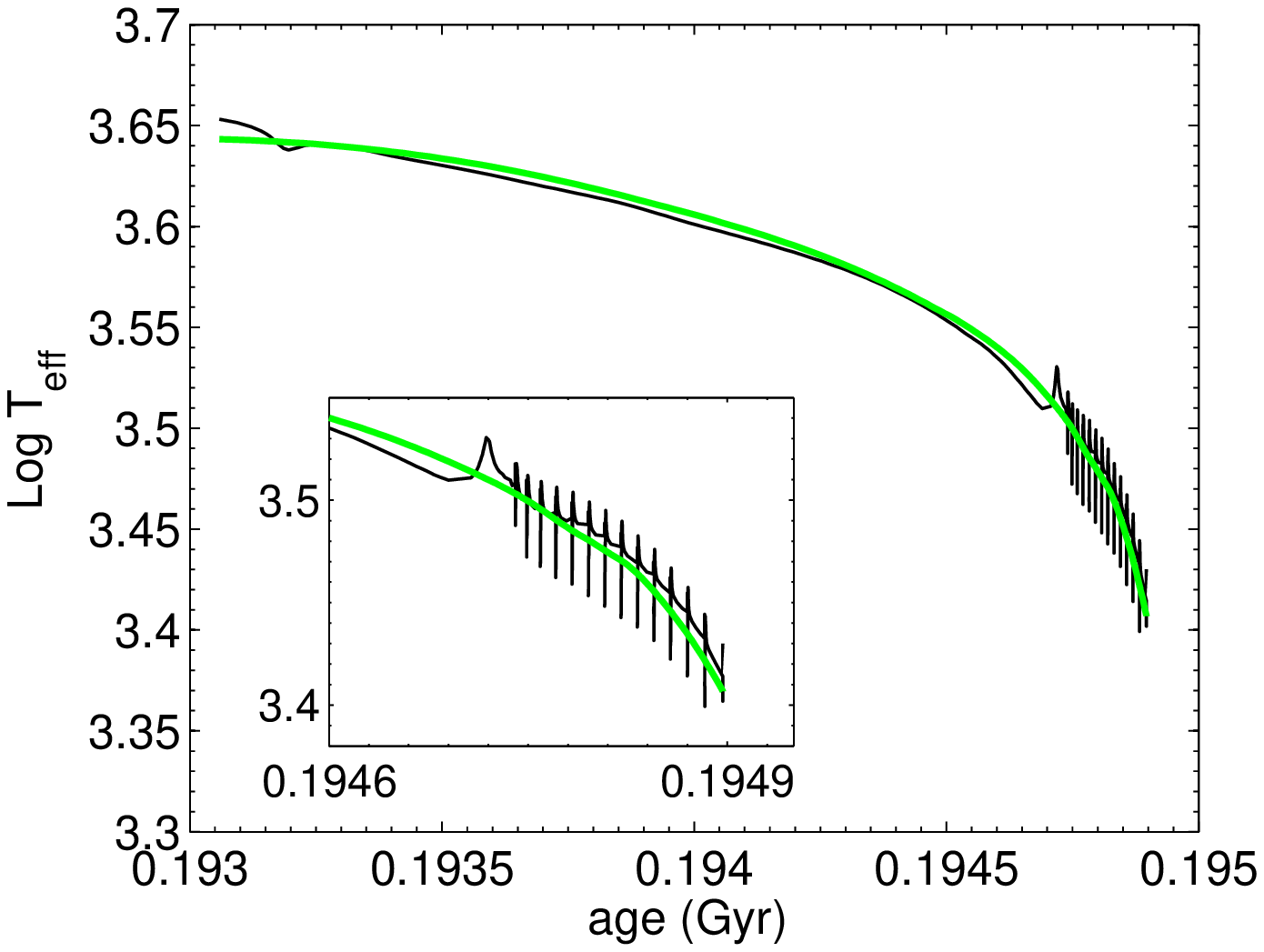}}
\caption{Evolutionary track for the AGB phase of a M=4 M$_{\odot}$, $Z$=0.02 stellar model.
The top panel show the temporal variation of the luminosity, the bottom panel the evolution of the effective temperature.
We display, superimposed on the track, the smooth approximations we have adopted.}
\label{tracks_temp}
\end{center}
\end{figure}

\section{The GARSTEC AGB models}\label{star_models}
This section describes briefly the key aspects of the AGB
phase, and summarizes \citet{Weiss2009} model prescriptions
for mass and opacities. Our new libraries of SEDs for
dust-enshrouded AGB stars are based on these stellar models and make use of the same mass-loss rates and the same opacities.\\

\noindent \textbf{\textsf{AGB stars in a nutshell}}. AGB stars
are found in the high luminosity and low-temperature region of the
HRD. They have evolved through core H- and He-burning, to develop
an electron degenerate CO-core. The luminosity is produced by
alternate H-shell and
He-shell burnings during the TP-AGB phase \citep[see the classical review][]{IbenRenz1983}. In brief, the He-burning shell becomes
thermally unstable (mild He-burning flash) every $\approx$10$^5$ yrs,
depending on the core mass. The energy provided by the thermal pulse
drives the He-burning convective zone inside the He-rich inter-shell region, and
He nucleosynthesis products are mixed inside this region. The stars expands
and the  H-shell is pushed to cooler regions, where it is
almost extinguished. At this stage the lower boundary of the convective
envelope can move inwards (in
mass) to regions previously mixed by the flash-driven convective
zone. This phenomenon is known as third dredge-up (TDU) and is
the responsible for enriching the surface with $^{12}$C and other
products of He-burning. Following the TDU,
the star contracts and the  H-shell is re-ignited,
providing most of the surface luminosity for the next inter-pulse period.
This cycle \textit{inter-pulse}-\textit{thermal
  pulse}-\textit{dredge-up} can occur many times during the AGB phase,
depending on the initial stellar mass, composition, and
in particular on the mass-loss rate. In intermediate-mass AGB
stars (M $\gtrsim$ 4M$_{\odot}$) the convective envelope can dip into
the top of the H-shell when it is active, and nuclear H-burning can occur at the base of
the convective envelope. This event is called \textit{envelope
  burning} (EB) or hot bottom burning, and can dramatically change the surface
composition. Indeed, the convective turn-over time of the envelope is
$\approx$ 1 year, hence the whole envelope will be processed in a few
thousand times over one inter-pulse
period. As a consequence, an AGB star of  suitable mass can
evolve from an oxygen-rich giant to a carbon-rich star ([C/O] $\geq
1$) due to the third dredge-up, and back to an oxygen-rich surface composition,
due to CN-burning in the envelope.
The new AGB models  by \citet{Weiss2009} include the latest physical inputs as
far as the treatment of C-enrichment of the envelope due to the TDU and related opacities are concerned.
These latter determine the surface  temperature of the models and the dust-driven mass-loss rates,
in turn affecting  the transition to the post AGB stages \citep{Marigo2002}.

\subsection{Mass-Loss and Opacities} \label{MassLoss}
\textbf{
\textsf{Mass-Loss}}. The AGB evolution is characterized by strong mass-loss due
to stellar winds. Mass-loss is one of the driving mechanisms of AGB
evolution as it determines how and when the TP-AGB phase ends, and
 what yields can be expected from these stars. It will also affect the nuclear burning
 at the bottom of the convective envelope.  The mass-loss rate  for the RGB  and pre-AGB
 evolution is the  \citet{Reimers1975} relation,
\begin{equation}
\dot{M}=4 \times 10^{-13} \frac{(L/L_{\odot})(R/R_{\odot})}{(M/M_{\odot})}\eta_{R}
\label{eq_reimers}
\end{equation}
with $\eta_R=0.45$. The rate is in   M$_{\odot}$/yr. This is
consistent with the  mass-loss rate adopted by \citet[][and companion papers]{Bertelli1994}.
If and when,
along the  TP-AGB and later stages, observed mass-loss rates are higher than predicted by  Eq.~(\ref{eq_reimers}),
the following prescription is adopted:  for carbon-rich chemical compositions (in which
nearly all oxygen is bound in CO, and the excess carbon gives rise to carbon-based molecules and dust)
the  mass-loss rate by \cite{Wachter2002}  is used
\begin{equation}
\small
 \dot{M}_{\rm AGB}=  3.98\times10^{-15} \left(\frac{L}{L_{\odot}}\right)^{2.47}
 \left(\frac{T_{\rm eff}}{2600 K}\right)^{-6.81} \left(\frac{M}{M_{\odot}}\right)^{-1.95}
\label{eq_Crich}
\end{equation}
\noindent whereas  for oxygen-rich stars ([C/0] $<$ 1), the empirical fitting formula by
\cite{vanLoon2005}, obtained from dust-enshrouded oxygen-rich AGB stars, is considered.
\begin{equation}
\small
\dot{M}_{\rm AGB}= 1.38\times10^{-10} \left(\frac{L}{L_{\odot}}\right)^{1.05}
 \left(\frac{T_{\rm eff}}{3500 K}\right)^{-6.3}
\label{eq_Orich}
\end{equation}
\noindent
As a star leaves the AGB,  its $T_{\rm eff}$ increases; by using
hydro-simulations of dusty envelopes around evolving post-AGB stars,
\citet{Steffen2007} show that strong mass-loss should occur for
temperatures higher than  $T_{\rm eff} \simeq 5000\, K$ or $\simeq
6000\, K$. This trend is reproduced by keeping the AGB-wind mass-loss
rates until the pulsation period $P$ has dropped to 150 days.  As  the
beginning of the post AGB phase is usually taken at $P=100$ days, an
interpolation is  needed to connect the end of the AGB and the
start of the post AGB phases. From there on, \citet{Weiss2009} employ the larger rate of
either Eq.~(\ref{eq_reimers}) or the radiation-driven wind mass
  loss rate (M$_{\odot}$/yr):
\begin{equation}
\small
\dot{M}_{CSPN}=-1.29 \times 10^{-15}\left( \frac{L} {L_{\odot}} \right)^{1.86}.
\end{equation}
\\

\noindent \textbf{\textsf{Opacities}}. The C-enhancement of the
stellar envelopes due to the TDU, is treated by using
opacity tables with varying [C/O]-ratio, and theoretical
mass-loss rates for carbon stars.  More precisely,  OPAL tables
for atomic Rosseland opacities by \citet{Iglesias1996} were obtained from
the OPAL-website\footnote[2]{http://physci.llnl.gov/Research/OPAL},
whereas for low temperatures new tables  with molecular opacities
were generated following the prescriptions by
\citet{Ferguson2005}. In all cases, the chemical compositions of low-
and high-temperature tables are the same, and tables from the different
sources are  combined \citep{Weiss2008}. The spectra of M-, S- and C-type
giant stars show the presence of different types of molecules, whose abundances are regulated by the [C/O]-ratio.
The spectra of O-rich stars ([C/O]-ratio  $\lesssim 1$) show strong bands of TiO, VO, H$_{2}$O,
 whereas  C-rich stars with [C/O] $>$ 1 display C$_{2}$, CN, SiC,  some HCN, and C$_{2}$H$_{2}$ bands.
Different tabulations of Rosseland opacities at low temperature must be prepared in advance at varying
the [C/O]-ratio, for different combinations of  X, Y, and Z. The dependence of the opacity on the [C/O]-ratio at given total
metallicity cannot be easily foreseen.

\begin{figure*}
\begin{center}
\subfigure
{\includegraphics[width=0.33\textwidth]{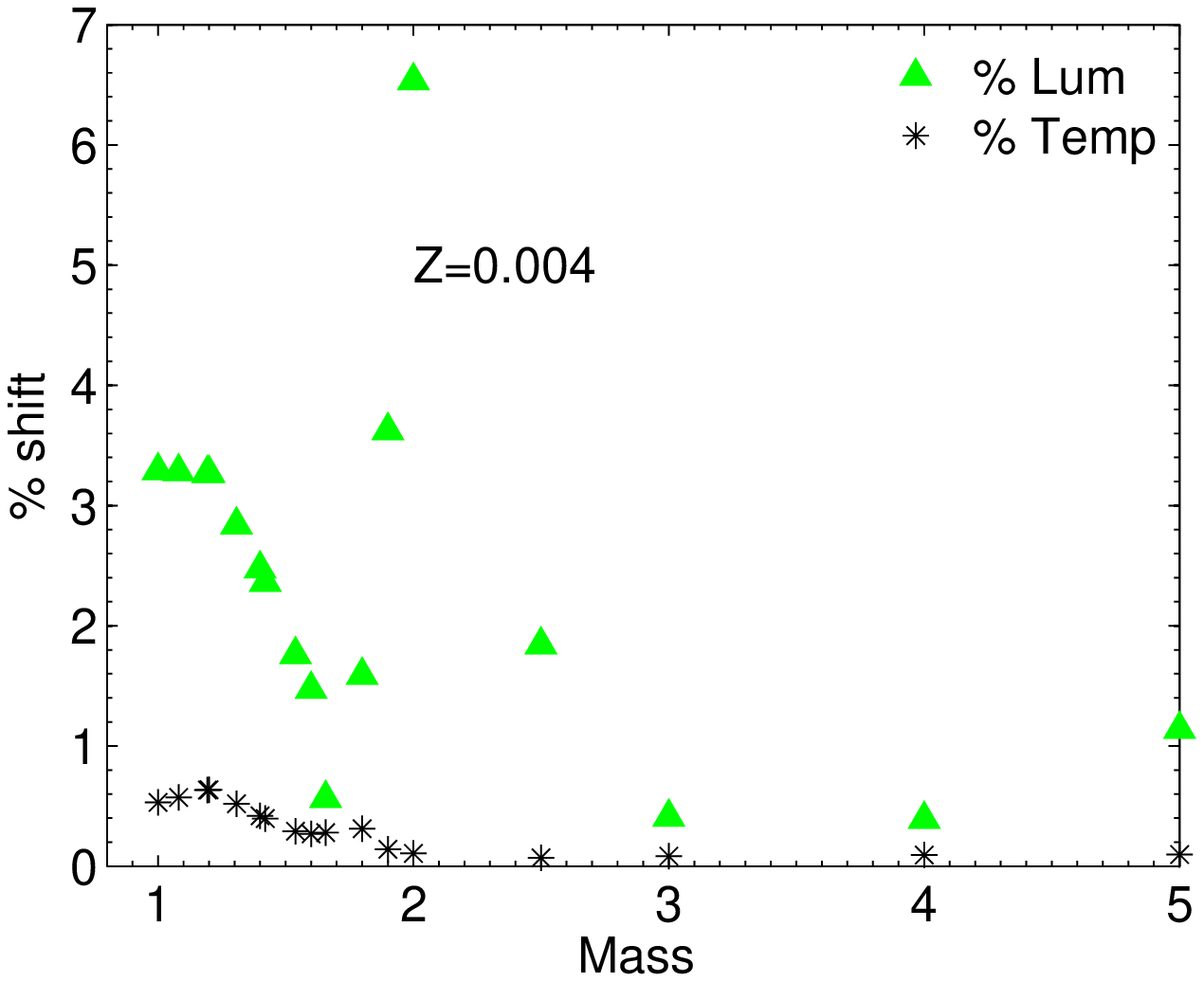}}
{\includegraphics[width=0.33\textwidth]{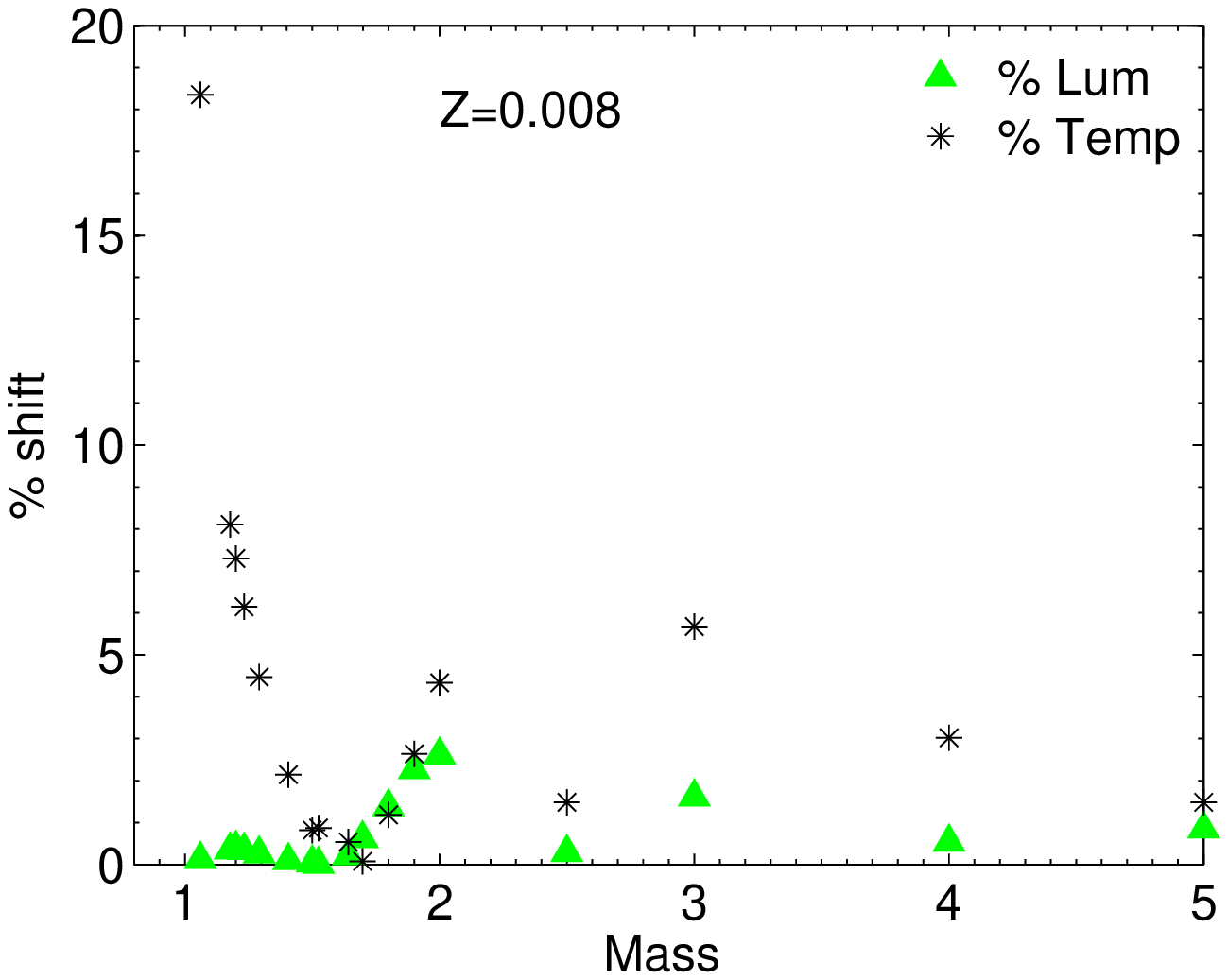}}
{\includegraphics[width=0.33\textwidth]{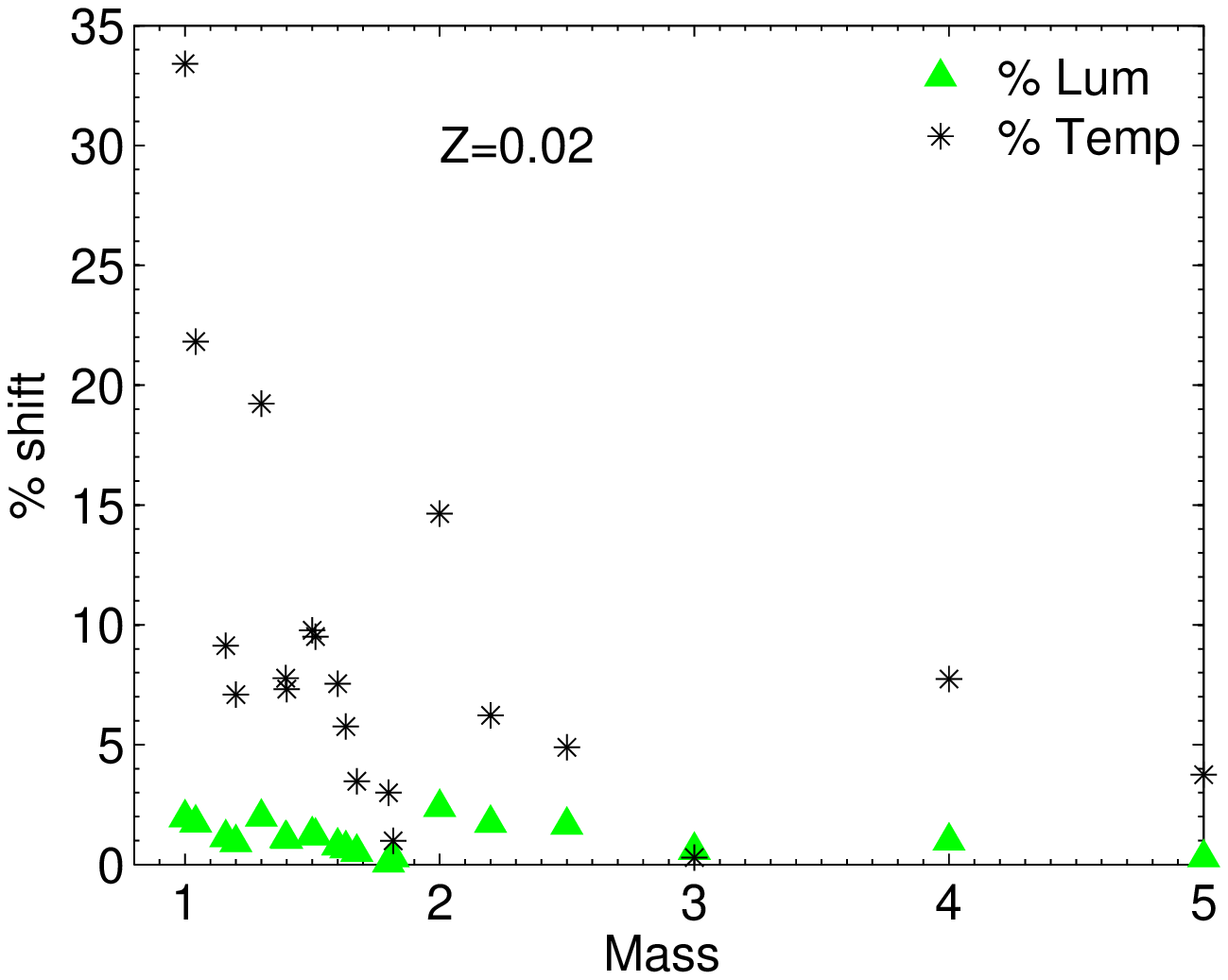}}
\caption{Relative variation $\Delta f/f$ of effective temperature (black stars)
  and luminosity (green triangles) as a function of the initial stellar mass,
  between the \citet{Bertelli1994}
  and \citet{Weiss2009} tracks at the end of the E-AGB phase. Results for three metallicities are
  displayed: $Z\simeq Z_{\rm SMC}$=0.004, $Z\simeq Z_{\rm LMC}$=0.008
  (typical of LMC)  and $Z \simeq Z_{\rm SN}$=0.02 (Solar Neighbourhood).}
\label{shift_temp}
\end{center}
\end{figure*}

\subsection{Smoothing the AGB phase}\label{smoothing}

 Although the TP-AGB phase is characterized by periodic
oscillations (a manifestation of the thermal pulses)
of the luminosity and effective temperature of the star,
there is a steady increase of the mean luminosity and a
decrease of the mean effective temperature. The typical trend of the two
quantities is shown in Fig.~\ref{tracks_temp} for a 4M$_{\odot}$ star
with $Z$=0.02. The inclusion of this oscillatory phase in isochrones and
SSPs would be a cumbersome affair from a numerical point of
view, with no real advantage compared to adopting the mean
luminosity and effective temperature, simply because the oscillations
take place on a very short time scale (essentially, the inter-pulse
time scale of the thermally pulsing He-burning shell in the deep
  interior of the star). Therefore, the standard procedure for
including the AGB phase envisages a smoothing of the luminosity/effective temperature
evolution, and makes use of  the resulting mean values \citep{Bertelli1994,Girardi2002,Bertelli2008}.
To appreciate the reasons for this approximation, some additional comments are necessary:
\begin{description}
  \item[-] In principle it is possible, but in practice it is
    numerically very cumbersome, to interpolate between the
    oscillating $L/L_\odot-T_{\rm eff}$ paths of stars of different
    mass.  Since the AGB  phase is short-lived, the interpolation
      between  pulses would  require short age-differences,
        corresponding to almost infinitesimal mass differences along an isochrone.
  \item[-] Star clusters have a small number of AGB stars, as expected according to the short
	duration of the double shell H-He nuclear burning phase. Therefore, both colour-colour diagrams and
CMDs of real clusters cannot reveal photometric signatures of the pulses. In case of very rich assemblies of stars
-- like field objects in a galaxy -- that sample rich populations of AGB stars, one could in principle
detect signatures of the oscillations associated to the thermal pulses, if all objects
were of the same mass. In practice, AGB stars in a galaxy span a large range of
masses, and their paths on the CMD would overlap, to produce a stream of AGB stars of different mass
(and probably chemical composition as well) at different stages of their AGB evolution.
\end{description}
\noindent  Based on these considerations, the thermal pulses
  have been replaced by smoothed quantities in all evolutionary
  sequences that include the AGB phase. The procedure can be summarized as follows:
\begin{description}
\item[i)] For each evolutionary sequence of fixed mass (and chemical composition) that includes the AGB phase, we have determined
the start, duration and end of all the evolutionary phases of interest, to carefully select the TP-AGB stage;
\item[ii)] As discussed in \citet{Weiss2009}, nearly all
  evolutionary sequences under consideration are followed to the
  end of the AGB phase, but for the
  highest masses (typically 5 and 6 M$_{\odot}$),
  because of numerical difficulties. In such cases, considering
  the rate of mass-loss and the  mass of the remaining envelope of the last model, an estimate
	of the number of missing pulses until the end of the TP-AGB is provided by \citet{Weiss2009}.
	We use this estimate to evaluate correctly the total TP-AGB
	lifetime for the few evolutionary sequences where this is required.
\item[iii)] Using the MATLAB environment, we plot for each star the
  current mass (M/M$_{\odot}$), age (yr), mass-loss rate
  $\dot{\textrm{M}}$, luminosity $\log$(L/L$_{\odot}$),
  effective temperature $\log$ T$_{\rm eff}$, gravity $\log$ g,
  central hydrogen mass fraction X$_{c}$ and central helium Y$_{c}$, the
  core mass  within the H- and He-burning shells, M$_{c1}$ and
  M$_{c2}$,  and the surface abundances of C$_{s}$ and O$_{s}$, both as function of the age
	and/or mass as appropriate. Making use of \textit{cftool} (\textit{Curve Fitting Toolbox})
	and  \textit{Smooth Options Loess} (Locally weighted scatter plot smoothing)
	we try to reproduce each of the above quantities by means of analytical fits. The method uses linear
	least-squares fit and second-order polynomials. The span parameter, that is the number of
	data points used to compute each smoothed value, is suitably varied. In locally weighted
	smoothing methods like \textit{Loess}, if the span parameter is less than 1, it can be
	interpreted as the percentage of the total number of data points. For all the physical variables
	that do not oscillate,  smoothing is not required and the span can be varied in such a way that
	the shape and the form of the original data are preserved\footnote{See   \textit{http://www.mathworks.it/help/techdoc/index.html}
	for the MATLAB documentation and   more details.}
\item[iv)] Once the smoothing procedure has been applied, we determine the start and
the end of the E-AGB and TP-AGB phases, and the  oxygen-rich to carbon-rich transition.
This is required for the interpolation between different values of the initial mass, to account correctly for
the carbon-rich and oxygen-rich stages.
\item[v)] In order to include these new models of AGB stars in old
  isochrones and SSPs, we need to extend \citet{Weiss2009} evolutionary
  models to mass as low as $0.6\,M_{\odot}$ (at
  least). As already recalled, \citet{Weiss2009} data set extend
  only down to $1\, M_{\odot}$. To this aim, we gently extrapolate the
  \citet{Weiss2009} stellar models down to $0.8\, M_{\odot}$ trying to
  scale consistently all the physical variables (luminosity, T$_{\rm
    eff}$, time-scales) obtained for  $1\,M_{\odot}$. We follow a
  numerical technique similar to the one used for the smoothing
  procedure. For even lower masses, that never reach the carbon-rich phase, a simple description is
	sufficient, and we follow \citet{Bertelli1994} and \citet{Piovan2003}.
\item[vi)] Finally, we match the TP-AGB part of each sequence derived from the  \citet{Weiss2009} models
to the end of the E-AGB phase of the corresponding (same initial mass and chemical composition) evolutionary
tracks from  \citet{Bertelli1994}. Some details of this are given below.
\end{description}

\begin{table*}
\begin{center}
\caption{
Fraction of a SSP total stellar mass at birth, contained in several stellar mass ranges (see text for the definitions) as prescribed by several IMFs.
The normalization constants are set to one.
Column (1) lists the chosen IMF; (2) \& (3) the corresponding lower and upper mass
integration limits; (4) fraction contained in stars with mass larger than 1$M_\odot$;
(5):  fraction contained in stars with M$<$1M$_{\odot}$, which do not contribute to the
chemical enrichment; (6) \& (7) fraction contained in stars
that contribute to the star-dust budget via the AGB and SN channel; (8) the total SSP mass $M_{SSP}$ at birth, in solar units.}
\label{tab_imf}
\vspace{1mm}
\begin{tabular*}{146.5mm}{l l l c c c c c c}
\hline
\noalign{\smallskip}
IMF                     &   & $M_{l}$  & $M_{u}$  & $\zeta_{>1}$ & $\zeta_{<1}$ & $\zeta_{1,6}$     &   $\zeta_{>6}$ & $M_{SSP}$  \\
\hline
\noalign{\smallskip}
  (1) & & (2) & (3) & (4) & (5) & (6) & (7) & (8) \\
\hline
\noalign{\smallskip}
Salpeter                    & IMF$_{\rm Salp}$     &  0.10    & 100    & 0.392   &0.6075  & 0.2285  & 0.1640 & 5.826  \\
Larson Solar Neighbourhood  & IMF$_{\rm Lar-SN}$   &  0.01    & 100    & 0.439   &0.5614  & 0.3130  & 0.1256 & 2.306  \\
Larson  (Milky Way Disc)    & IMF$_{\rm Lar-MW}$   &  0.01    & 100    & 0.653   &0.3470  & 0.3568  & 0.2962 & 3.154  \\
Kennicutt                   & IMF$_{\rm Kenn}$     &  0.10    & 100    & 0.590   &0.4094  & 0.3883  & 0.2023 & 3.048  \\
Kroupa (original)           & IMF$_{\rm Kro-Ori}$  &  0.10    & 100    & 0.405   &0.5948  & 0.3016  & 0.1036 & 3.385  \\
Chabrier                    & IMF$_{\rm Cha}$      &  0.01    & 100    & 0.545   &0.4550  & 0.3517  & 0.1933 & 0.025  \\
Arimoto                     & IMF$_{\rm Ari}$      &  0.01    & 100    & 0.500   &0.5000  & 0.1945  & 0.3055 & 9.210  \\
Kroupa 2002-2007            & IMF$_{\rm Kro-27}$   &  0.01    & 100    & 0.380   &0.6198  & 0.2830  & 0.0972 & 3.134  \\
Scalo                       & IMF$_{\rm Sca}$      &  0.10    & 100    & 0.320   &0.6802  & 0.2339  & 0.0859 & 4.977  \\
\hline
\end{tabular*}
\end{center}
\end{table*}

\subsection{Matching GARSTEC to Padua models}\label{cut_paste}

Both GARSTEC and Padua
models, beside the same assumptions for the mass-loss
rates until to the start of TP-AGB phase, similar sources and treatment
of the opacities, same metal mixture
\citep{Grevesse1993}, and many other common physical ingredients,
are calculated with numerical codes that are descendants of
the G\"{o}ttingen code developed by \citet{Hofmeister1964}. \\
This makes easier the match between evolutionary models from the main sequence to the E-AGB
 phase calculated by the Padua group, and those for
the TP-AGB phase calculated by \citet{Weiss2009}.\\
\indent We carefully checked that, with some exceptions, the shifts we must apply to
\citet{Weiss2009} TP-AGB models to match the \citet{Bertelli1994} E-AGB endpoints are acceptable.
We have scaled luminosity, effective temperature, core
mass, and envelope mass of the GARSTEC tracks to match those of the E-AGB stage of \citet{Bertelli1994} models.
 The zero point of the age of GARSTEC  AGB models is also rescaled to match that of \citet{Bertelli1994}
E-AGB endpoint.
Figure \ref{shift_temp} displays the required $\log$ T$_{\rm eff}$
and $\log$(L/L$_{\odot}$) shifts, for the initial masses under consideration.
Recalling that the luminosity of stellar models is far less affected by theoretical uncertainties than the effective temperature,
we analyze the match of the two sets of tracks
by means of the relative variations of luminosity and effective temperature, defined in
the following way: let $L_f$ and $T_{\rm eff,f}$ be the luminosity/effective temperature of the final model of the E-AGB phase and
$L_i$ and $T_{\rm eff,i}$ the counterparts for the initial model of the TP-AGB phase, the relative luminosity shift is given by
\[ \Delta f/f = \left|\log (L_i /L_\odot) - \log (L_f/L_\odot)\right| / \log (L_i/L_\odot) .\]
For the effective temperature it is more convenient to normalize the shift to the total length of the TP-AGB phase projected
onto the $T_{\rm eff}$-axis, i.e.
\[ \Delta f / f = \left|\log T_{\rm eff,i}  - \log T_{\rm eff,f}\right| /(\log T_{\rm eff,l} -\log T_{\rm eff,i} ), \]
where $T_{\rm eff,l}$ is the temperature of the last TP-AGB point.
Figure~\ref{shift_temp} displays results for three values of the metallicity, namely
$Z$=0.004 (typical of the SMC), $Z$=0.008 (typical of the LMC),
and $Z$=0.02 (approximately solar). Models for other
metallicities behave in the same way. For low-mass stars, whose lifetime
on the TP-AGB is very short, shifts comparable
to the total temperature interval of the TP-AGB phase are possible. This is
evident in Fig.~\ref{shift_temp}: the low-mass, high-metallicity models need the largest 
shifts in temperature. This is due to the strong sensitivity
of the envelope size (hence effective temperature) to the mean opacity and
to the amount of mass lost in the previous phases -- see Eq.~(\ref{eq_reimers}) -- which governs the mass-loss during the RGB
and pre-AGB evolution. Indeed, even though GARSTEC and Padua
models include very similar recipes for the mass-loss and the
opacities, the evolutionary time spent in the pre-AGB phases may
still vary because of other different input physics of the models
like, e.g., nuclear reaction rates, thus causing a different size of the envelope. The luminosity 
is much more stable because it is generated deep inside the star.
For the other mass ranges and metallicities involved in
the TP-AGB, the shifts are $\lesssim 5\%$, thus introducing an unavoidable, but small, uncertainty 
in the region of the HR diagram covered by TP-AGB stars.
In particular, for the highest masses, the shifts are just a small
fraction of the full TP-AGB extension.

\subsection{The Initial Mass Function} \label{IMF_description}

To calculate spectro-photometric properties of SSPs (SED,
magnitudes, colours and luminosity functions) it is necessary to consider an
IMF. There are several popular prescriptions in the literature. A few of them are listed
in Table~\ref{tab_imf}.  For the purposes of our study all IMFs are assumed to
 be constant in time and space. The IMFs in our list are:  Salpeter \citep[][IMF$_{\rm Sal}$]{Salpeter1955},
 Larson \citep[][IMF$_{\rm Lar-MW}$, IMF$_{\rm Lar-SN}$]{Larson1998}, - with different parameters for the Milky
 Way disk and for the solar neighbourhood \citep{Portinari2004} - Kennicutt \citep[][IMF$_{\rm Kenn}$]{Kennicutt1998},
the original IMF by Kroupa \citep[][IMF$_{\rm Kro-Ori}$]{Kroupa1998}, a revised and more recent version of this IMF by
Kroupa \citep[][IMF$_{\rm Krou-27}$]{Kroupa2007}, Chabrier \citep[][IMF$_{\rm Cha}$]{Chabrier2003}, Scalo
\citep[][IMF$_{\rm Sca}$]{Scalo1986}, and Arimoto \& Yoshii \citep[][IMF$_{\rm Ari}$]{Arimoto1987}.
We refer either to the original sources or to to \citet{Piovan2011II} for a detailed explanation of the main features of these IMFs.

The IMFs are expressed as the number of stars per mass interval, $dN=\Phi(M)dM$, and require a normalization, for
$\Phi(M)$ contains an arbitrary constant.
This can be accomplished in different ways. In view of the calculation of integrated spectral
energy distributions, magnitudes and colours of  the SSPs, we  introduce here  the concept of (zero age) SSP mass, given by

\begin{equation}
 M_\mathrm{SSP} = \int_{M_l}^{M_u} \Phi (M) M dM
\label{imf_norm}
\end{equation}

\noindent i.e. the total mass contained in a SSP with lower stellar mass limit M$_l$ and upper mass limit M$_u$, and we set the
constant entering $\Phi(M)$ equal to one.
The photometric properties of a SSP of given age and chemical composition
will refer to a given SSP mass. By doing this, one can easily scale the SSP monochromatic
flux to any population of stars of arbitrary total mass.

For the purposes of the discussion below, we denote with $\zeta_1$ the
the fraction of the population total stellar mass at birth, contained in stars whose lifetime is shorter than the age of the Universe,
-- and therefore able to chemically pollute the interstellar medium -- given by

\begin{equation}
  \zeta_{1} = { \int_{1}^{M_u} \Phi(M) M dM  \over \int_{M_l}^{M_u} \Phi(M) M dM }
\end{equation}

where $M_l$ and $M_u$ have the same meaning as before. In a similar way, we define the mass fraction of the stars contributing to the dust-budget
via the AGB channel ($1\, M_\odot \leq M \leq 6\, M_\odot$) denoted by $\zeta_{1,6}$, and the mass fraction of stars that contribute to the dust budget
via the core collapse supernovae channel, ($ M > 6\, M_\odot$) denoted by $\zeta_{>6}$.
Table~\ref{tab_imf}
summarize the mass ranges where the various IMFs are defined, the mass fractions of stars defined above, and the corresponding total SSP mass at birth,
for a normalization constant equal to one.

\begin{figure}
\begin{center}
{\includegraphics[width=0.40\textwidth]{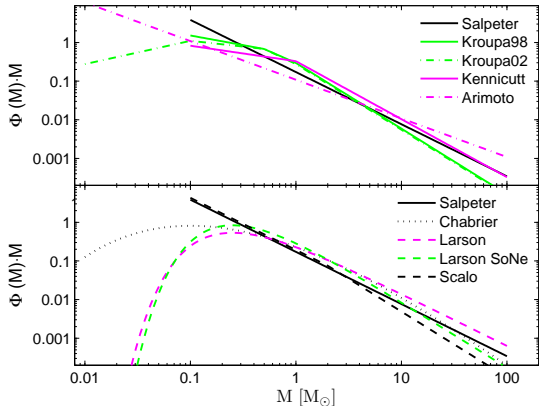}}
\caption{Fractional contribution to the total SSP mass budget of stars of different masses, as 
predicted by the labelled IMFs over the range where they are defined
(see text for details).
The widely used IMF by \citet{Salpeter1955} is shown
in both panels for the sake of comparison. Stellar masses are in solar units and all the IMFs
are in this case normalized to a total SSP mass equal to 1$M_\odot$.}
\label{imf_plot}
\end{center}
\end{figure}

Figure~\ref{imf_plot} shows the mass dependence
of the different IMFs, and implicitly the mass
interval covered by stars going through the TP-AGB and WD phases or
ending in a SN explosion and thus contributing to the star-dust budget.

\begin{figure*}
\begin{center}
{\includegraphics[width=0.33\textwidth]{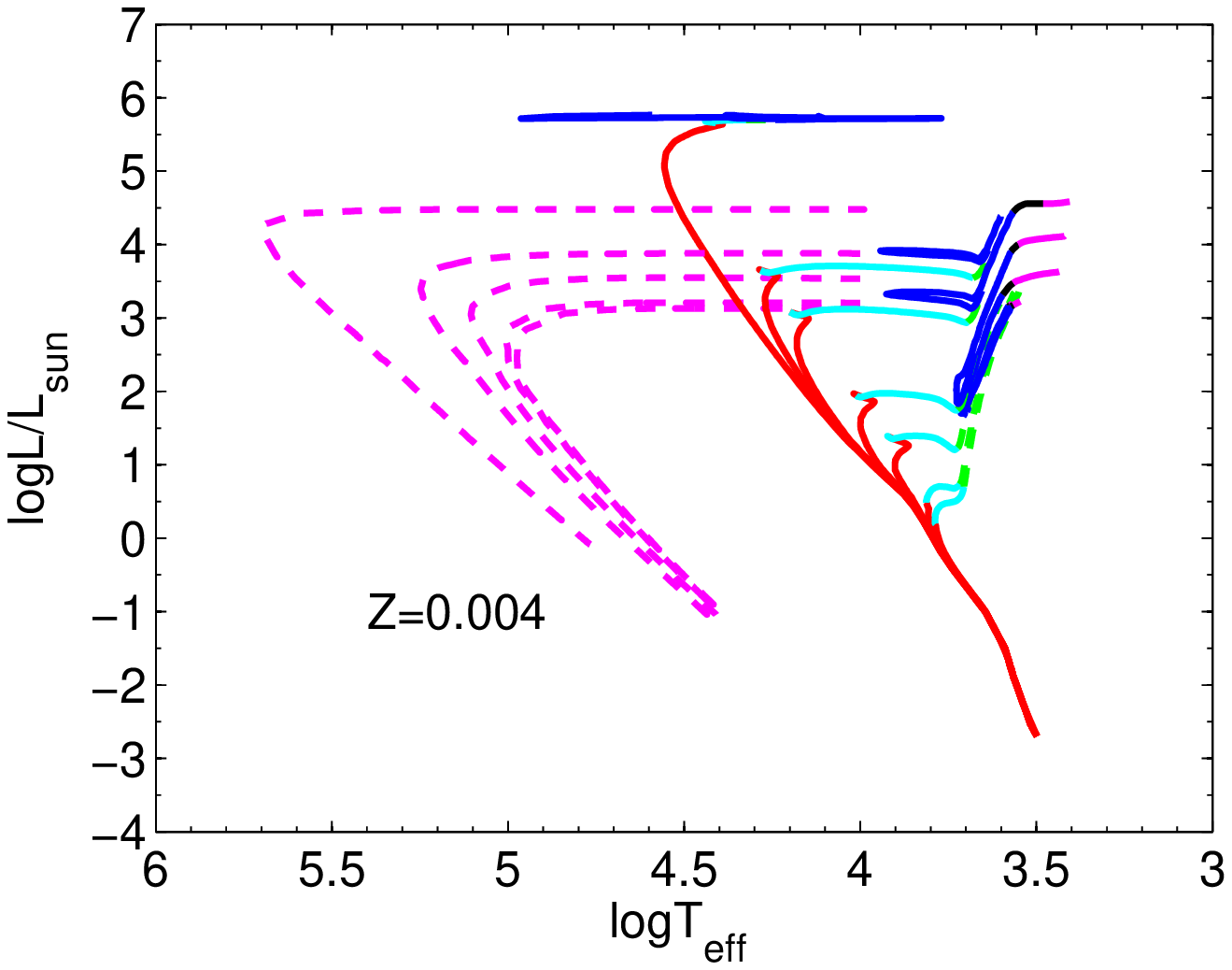}}
{\includegraphics[width=0.33\textwidth]{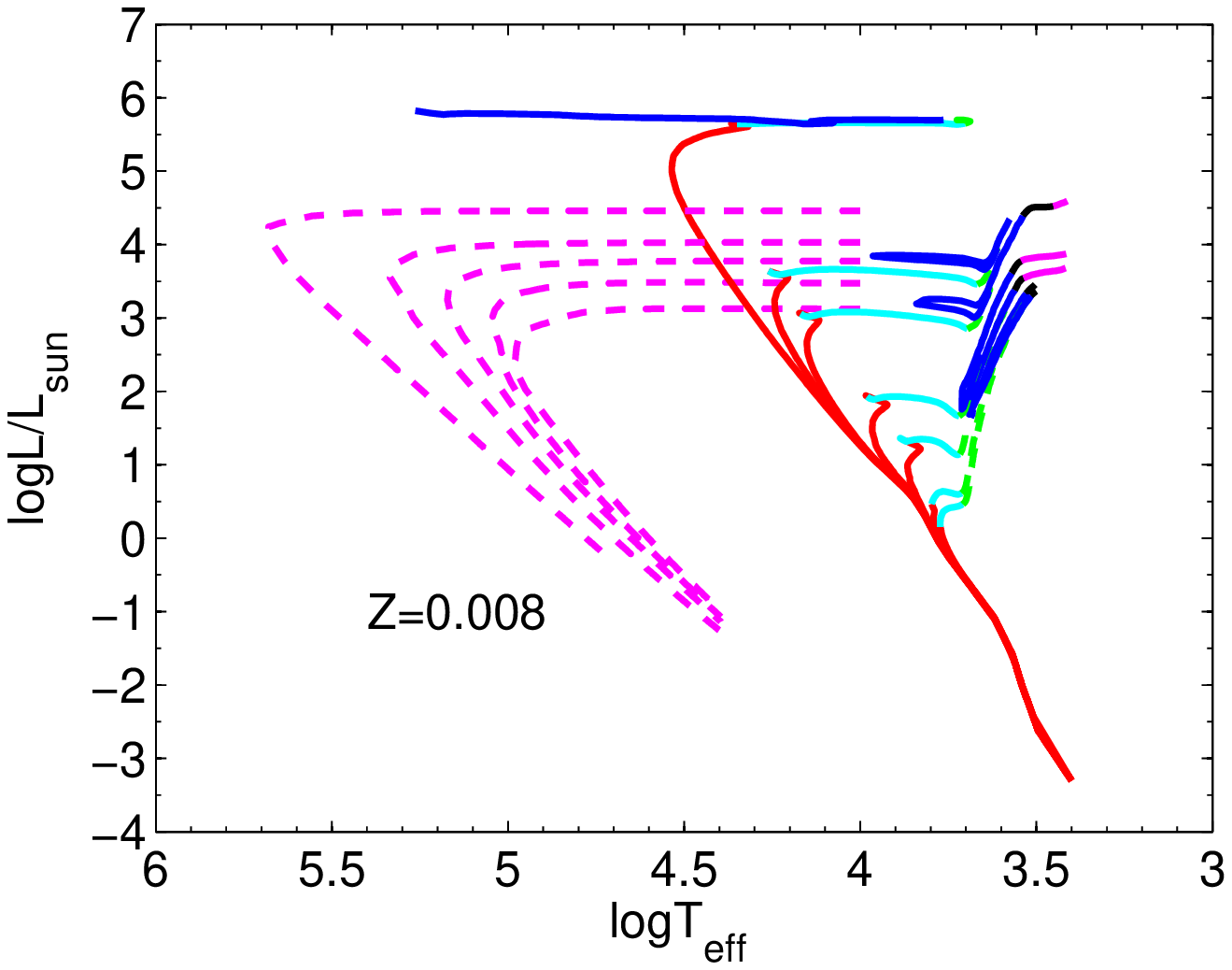}}
{\includegraphics[width=0.33\textwidth]{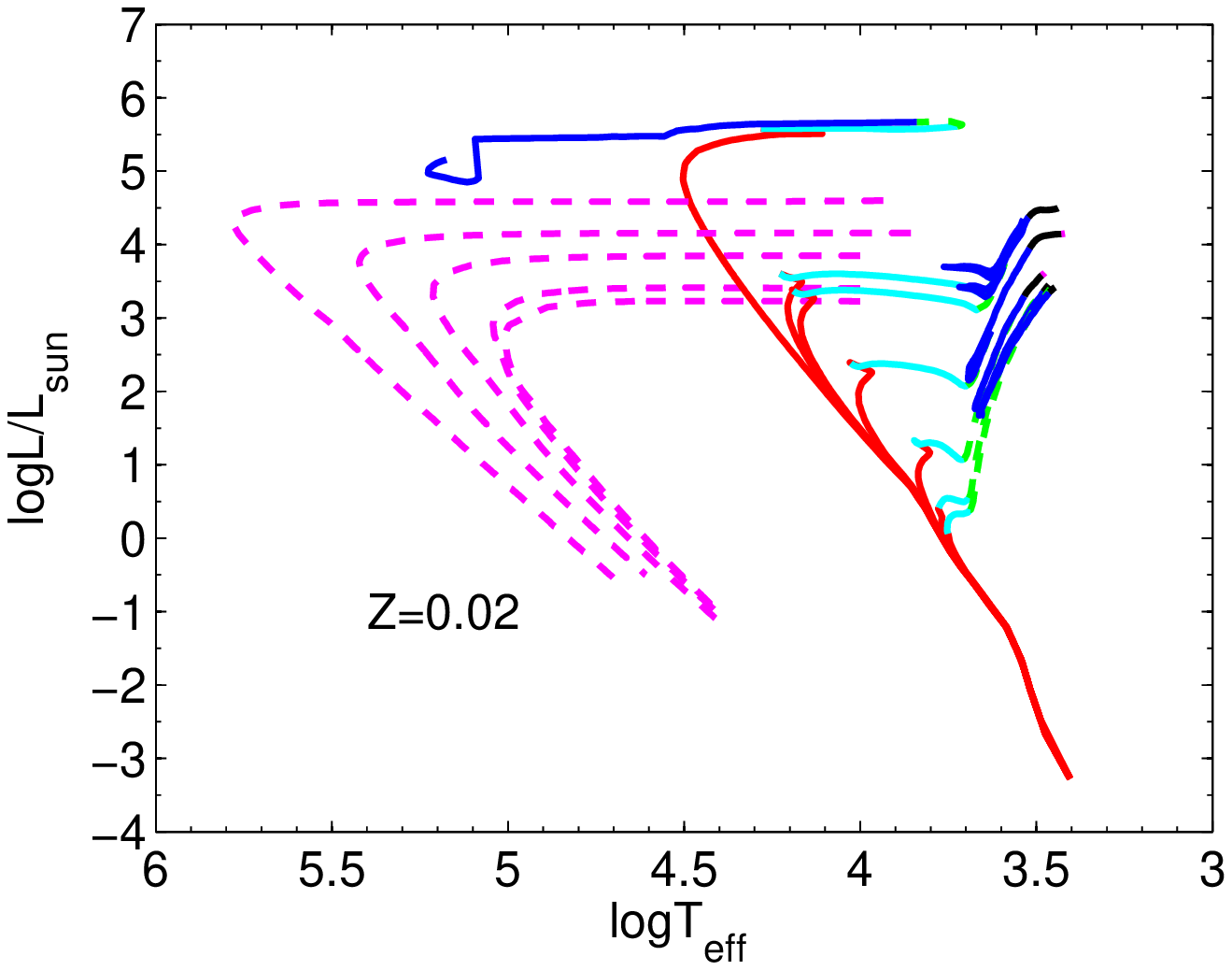}}
\caption{Isochrones with the new AGB models, from the zero age main sequence
to the stage of PN formation or central carbon ignition, depending on the initial stellar mass. Three
metallicities are shown: $Z$=0.004, typical of the Small Magellanic Cloud (left), $Z$=0.008, typical
of the Large Magellanic Cloud (middle), and $Z$=0.02, typical of the Solar neighbourhood (right).
The isochrones are plotted for a few selected ages between 5 Myr and 15 Gyr.}
\label{iso_same}
\end{center}

\end{figure*}

\section{The new isochrones: results}\label{new_isocs}

We present here the sets of isochrones obtained
with the new TP-AGB models. Each set contains isochrones
for more than fifty age values, ranging from  $\sim$3.0
Myr to 15 Gyr. The age range for the development of an AGB
varies with metallicity according to

\begin{figure*}
\begin{center}
\subfigure
{\includegraphics[width=0.33\textwidth]{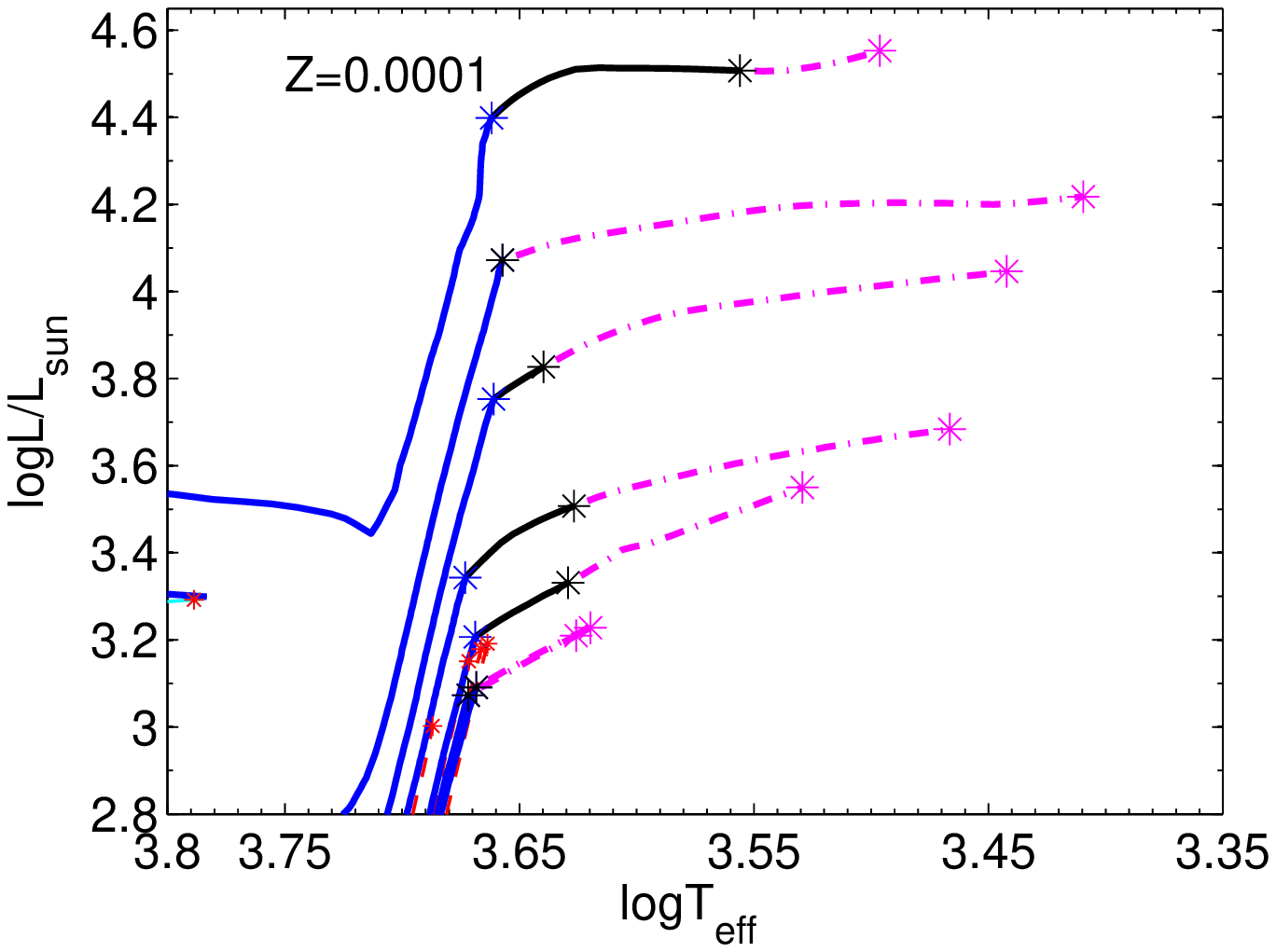}
 \includegraphics[width=0.33\textwidth]{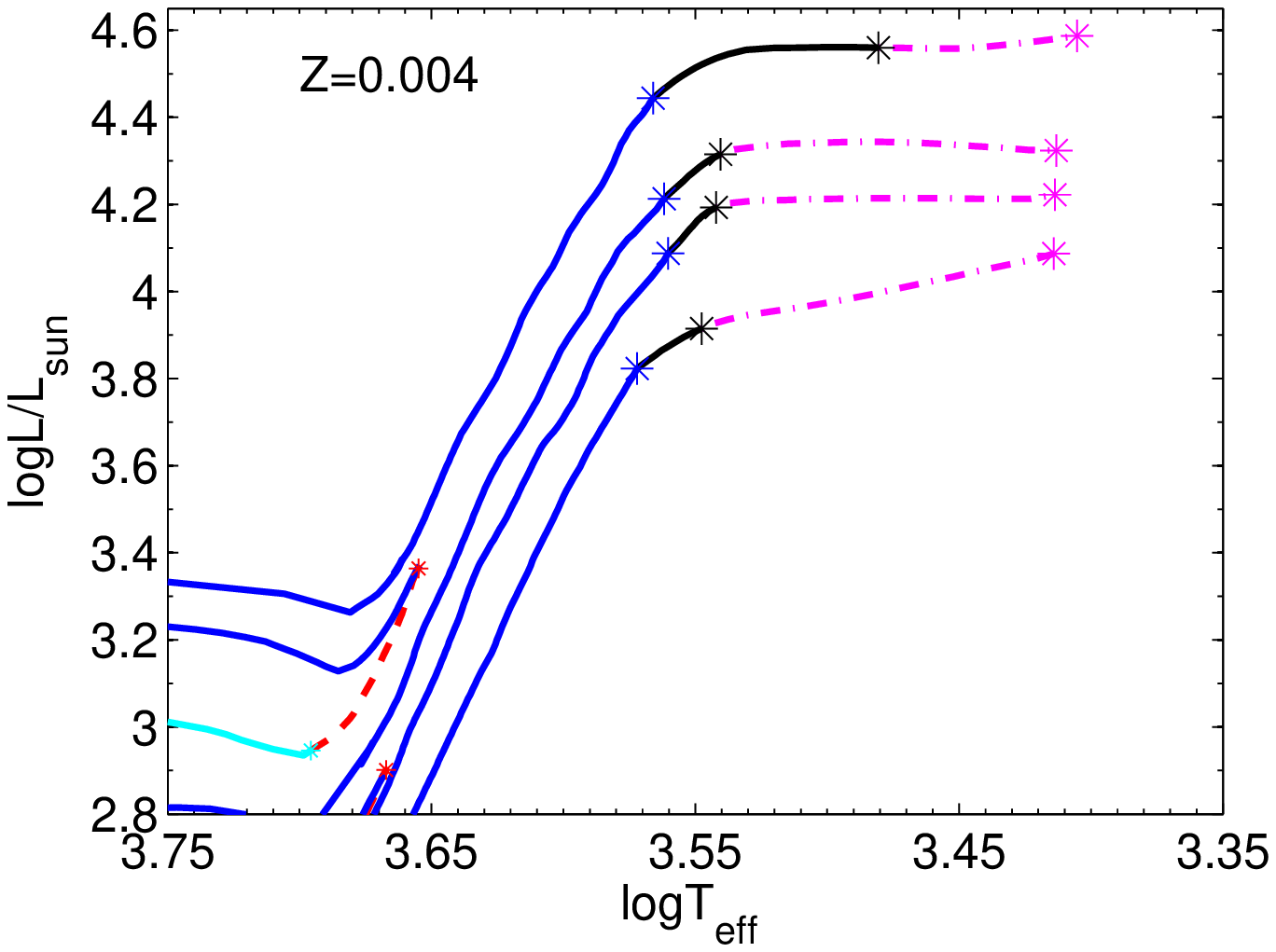}
 \includegraphics[width=0.33\textwidth]{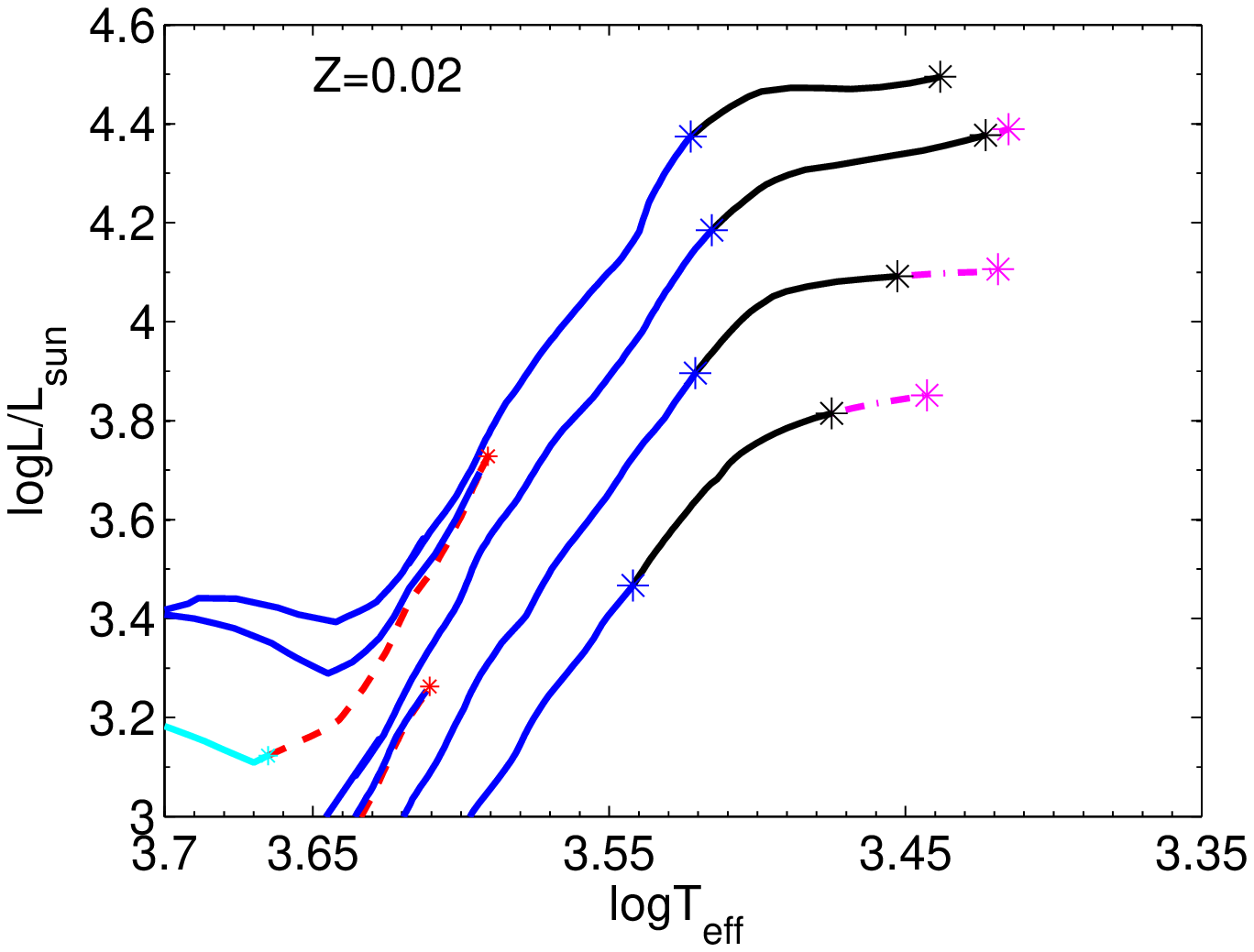}}
\subfigure
{\includegraphics[width=0.33\textwidth]{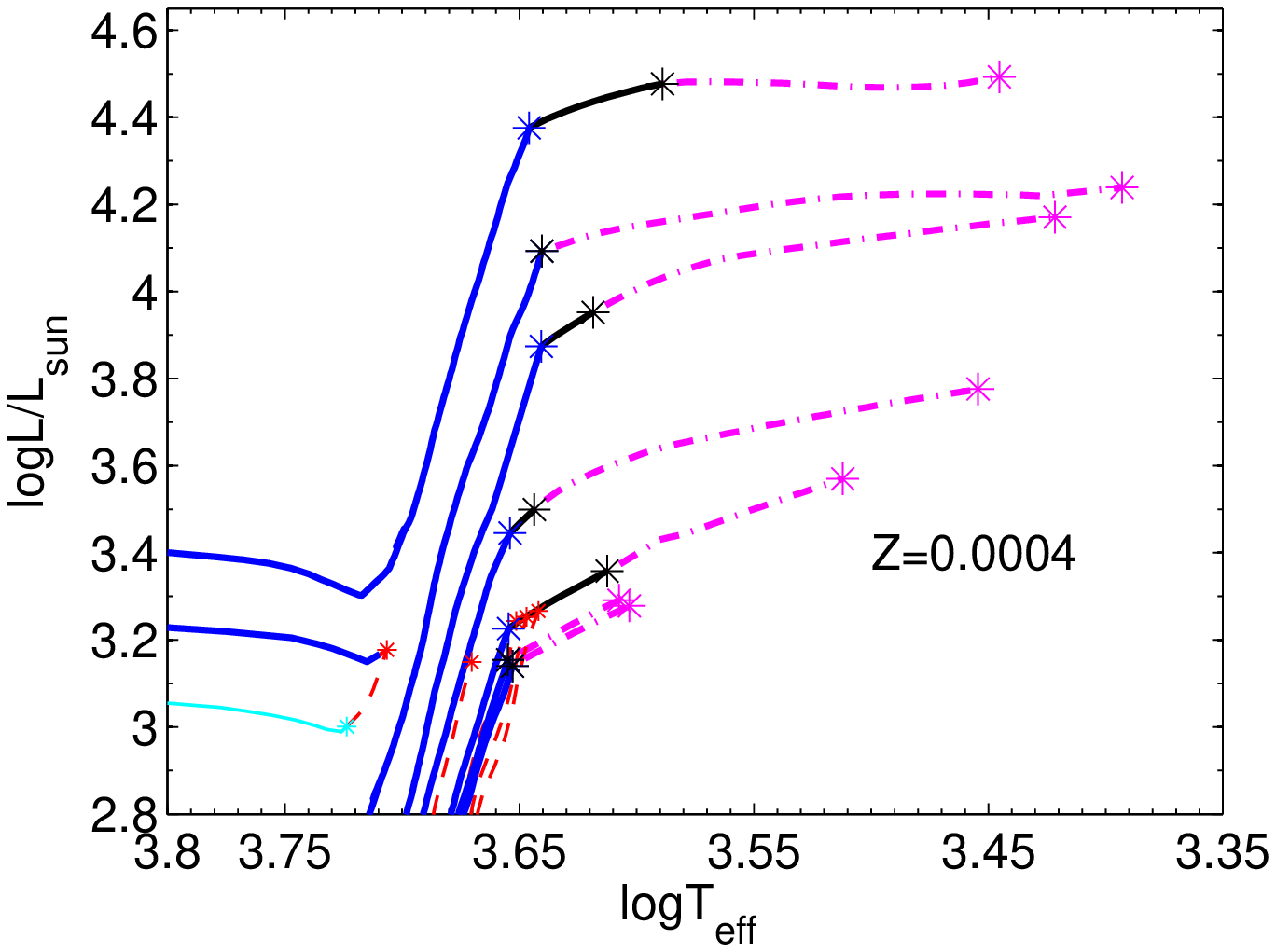}
 \includegraphics[width=0.33\textwidth]{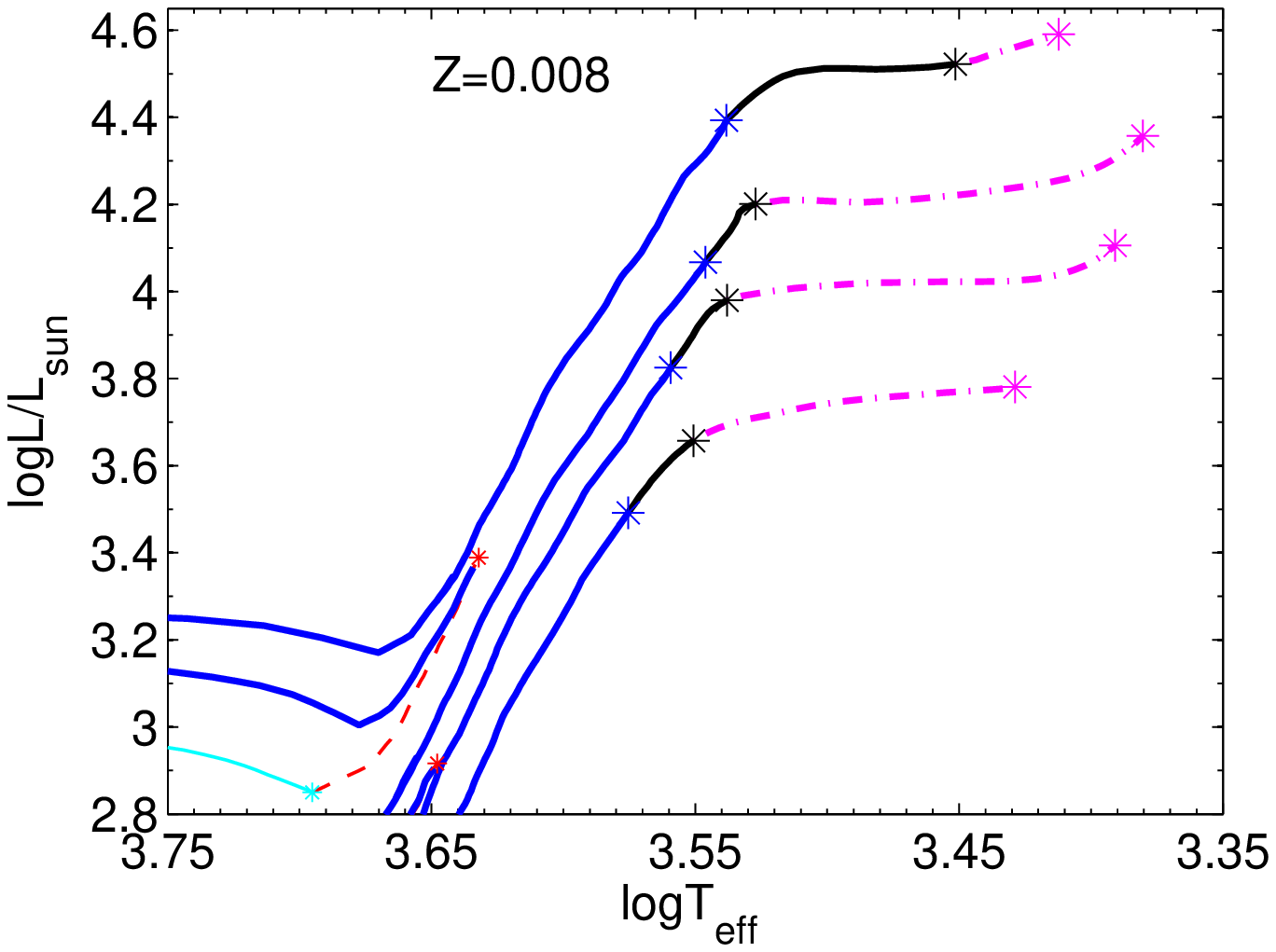}
 \includegraphics[width=0.33\textwidth]{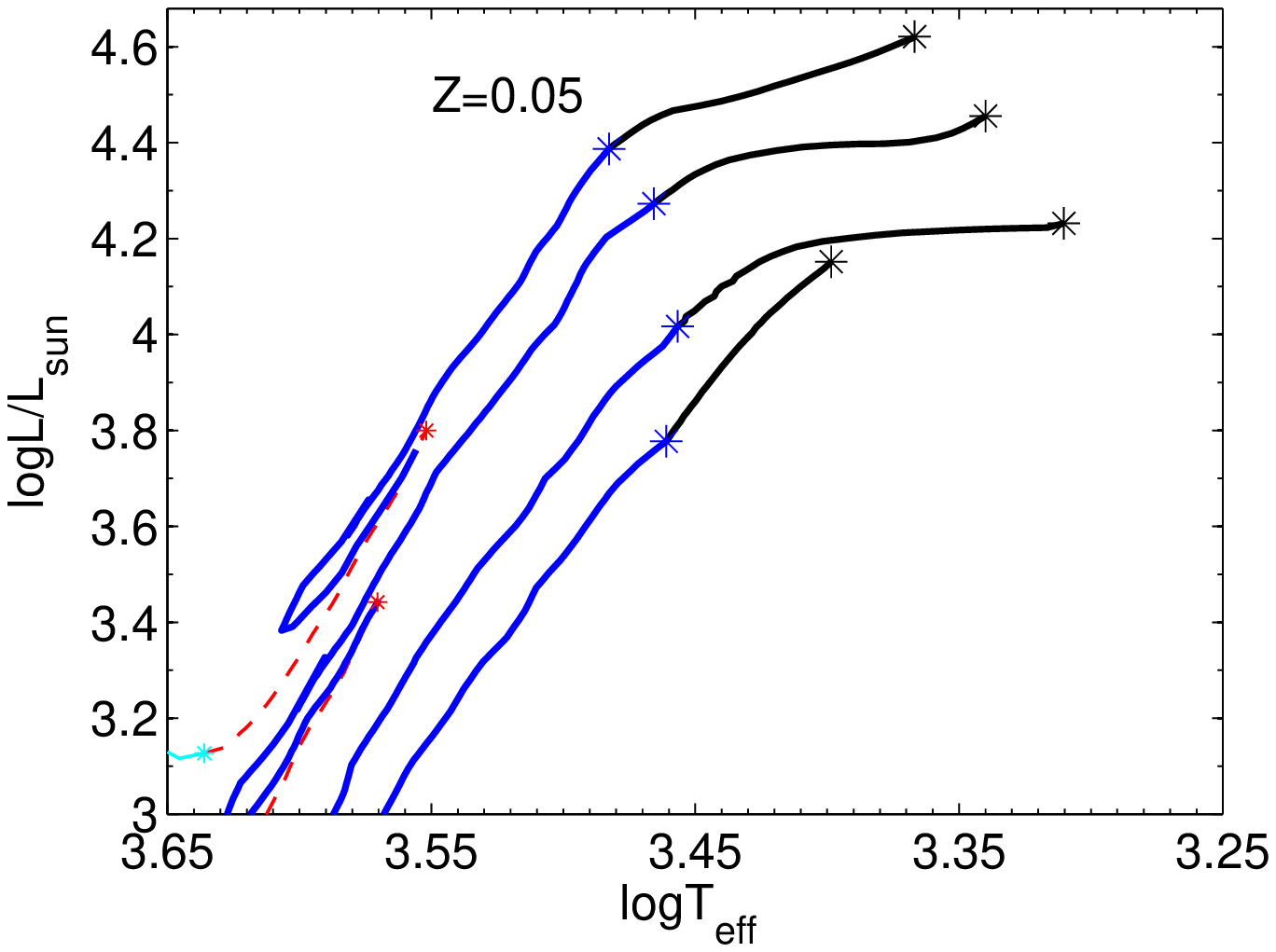}
     }
 \caption{Isochrones in the theoretical HRD, centred on the E-AGB and TP-AGB phases, for the labelled
metallicities. They are organized in three groups, from left to right.
 \textbf{Left Panels}: Very low metallicities, $Z$=0.0001 and $Z$=0.0004 respectively.
 \textbf{Central Panels}: as in the left panels, but for $Z$=0.004 and
 $Z$=0.008.  \textbf{Right Panels}: as in the previous panels, but
 for $Z$=0.02 (solar value) and $Z$=0.05. Along each isochrone the end
 of the E-AGB phase is marked by the blue star. The TP-AGB phase is in
 turn drawn as a solid black line when the envelope is
 oxygen-rich, and as a dot-dashed magenta line when if carbon-rich. }
\label{agb_123}
\end{center}
\end{figure*}

\begin{description}
\item[-] $Z$=0.050:\,\,\,  $7.78 \leq \log t \leq  10.18$;

\item[-] $Z$=0.020:\,\,\,  $7.90 \leq \log t \leq  10.18$;

\item[-] $Z$=0.008:\,\,\,  $8.10 \leq \log t \leq  10.18$;

\item[-] $Z$=0.004:\,\,\,  $8.10 \leq \log t \leq  10.18$;

\item[-] $Z$=0.0004:\, $8.10 \leq \log t \leq  10.18$;

\item[-] $Z$=0.0001:\, $8.00 \leq \log t \leq  10.18$.
\end{description}

where $t$ is in yr.
All the isochrones are calculated with the IMF$_{Salp}$:
indeed, varying the IMF  would affect only the way the different mass
bins along an isochrone are populated, i.e.\ the so-called normalized
luminosity function. The effect of changing  the IMF becomes more
evident when calculating SEDs of SSPs (see below).\\
\indent Figure~\ref{iso_same} shows  a few selected isochrones
for metallicities $Z$=0.004 (typical  for the SMC),
$Z$=0.008 (typical for the LMC), and $Z$=0.02 (typical  for
the Sun and the solar vicinity) respectively. All other metallicities have similar HRDs. 
Important differences with \citet{Bertelli1994} arise obviously along the AGB phase, as shown
in Fig.~\ref{agb_123}. The AGB phase for oxygen-rich envelopes
is displayed with black, solid lines, whereas the carbon-rich case with
[C/O]$>$1 is displayed with magenta dot-dashed lines. The
beginning and end of each evolutionary phase is marked with a
little star. Thanks to the new low temperature opacities \citep{Weiss2009},  the
isochrones now  extend towards lower temperatures than in the
old models. The enrichment of the C-abundance at the surface of TP-AGB
stars, accompanied by an important  reduction of the effective
temperature and the formation of a shell of dust surrounding the
star (see below)  are important steps forward,
that amply justify our efforts to calculate a
library of stellar spectra for O- and C-rich dust-enshrouded AGB stars.

Looking at the grids of isochrones for different metallicities, the following considerations can be made:

-\textsf {Solar and super-solar metallicities: $Z \geq 0.02$}. These stars are normally
oxygen-rich at the surface, even if a late transition to the carbon-rich phase may take place
due to the final dredge-up events, in agreement with observations \citep[e.g. see ][\, for more details]{Vanloon1998,Vanloon1999}.
For solar metallicity, the transition occurs only in isochrones of intermediate ages  and  at
very low $T_{\rm eff}$, during the final stages of the TP-AGB
phase. In contrast, isochrones of super-solar metallicity show only oxygen-rich material at the
 surface. As expected, the youngest isochrones are the most extended in the HRD during the
AGB phase. The TDU does not occur in the oldest isochrones of both
metallicities, and the TP-AGB phase is much shorter than the E-AGB phase.\\
\indent -\textsf {Sub-solar metallicities:  $0.004 \leq Z < 0.02$}. These stars show an extended
carbon-rich phase, even at rather young ages.
This is due to the onset of the ON cycle, that converts O into N, increasing the [C/O]-ratio above
1 \citep{Ventura2002,Marigo2008}. Furthermore, the carbon-enrichment at the surface
starts at higher effective temperatures (compared to solar metallicity isochrones), because the
lower molecular concentrations in the atmospheres \citep{Marigo2008}.\\
\indent -\textsf{Low metallicities: $Z < 0.004$}. All trends
described for isochrones of moderate metallicities become  more evident.
The transition to a carbon-rich envelope starts at even higher
effective temperatures and the majority of the isochrones show almost exclusively the C-star phase. Only
few isochrones of intermediate ages have an oxygen-rich phase. Our results fairly agree with those 
by \citet{Marigo2008} even though some marginal differences can be noticed. The agreement is
ultimately due to the fact that both include opacities that depend on the [C/O]-ratio.
This is confirmed by the nearly identical effective temperatures of the AGB models, and the similar behaviour of the oxygen-rich
and carbon-rich stages with the metallicity.

\section{The dust-free SSPs }\label{dust_free_SSP}

The most elementary population of stars is the so-called
\textit{Single} (or \textit{Simple}) \textit{Stellar Population} made
of stars born at the same time in a burst of star formation activity
of negligible duration, and with the same chemical composition. SSPs
are the basic tool to understand the spectro-photometric properties of
more complex systems like galaxies, which can be considered as
linear combinations of SSPs with different composition and age, each of
 them weighted by the corresponding rate of star formation.\\
\indent The integrated monochromatic flux  $ssp_{\lambda}(\tau^{},Z)$ of a SSP of any age and metallicity is given by

\begin{equation}
ssp_{\lambda}(\tau^{},Z)=\int_{M_{l}}^{M_{u}}\Phi(M)f_{\lambda}(M,\tau^{},Z) dM
\label{eq_ssp}
\end{equation}

\noindent where $f_{\lambda}$ is the monochromatic flux of a star of mass $M$, metallicity $Z$ and age  $\tau$. $\Phi(M)$
is the IMF, expressed as the number of stars per mass interval $dM$. The integrated  $ssp_{\lambda}(\tau, Z)$
refers to an ideal SSP of total mass  $M_\mathrm{SSP}$ (expressed in solar units). The integrated bolometric luminosity
is then calculated by integrating $ssp_{\lambda}(\tau^{}, Z)$  over the whole wavelength range:

\begin{equation}
L_{SSP}(\tau^{},Z)=\int_{0}^{\infty} ssp_{\lambda}(\tau^{},Z) d\lambda
\label{eq_lumssp}
\end{equation}

In more detail, the steps to calculate the SED of a SSP are as follows:

(i) for a fixed age and metallicity, the corresponding isochrone in the
HRD is divided into elementary intervals small enough to ensure that
luminosity, gravity, and $T_{\rm eff}$ are nearly constant.
In practice, the isochrone is approximated  by a series of
virtual stars, to which we assign a spectrum;

(ii) in each interval the stellar mass spans a range
$\Delta M$ fixed by the evolutionary  speed. The
number of stars assigned to each interval is proportional to the integral of the IMF over the range $\Delta M$ (the
differential luminosity function);

(iii) finally, the contribution to the integrated flux (at each wavelength)
by each elementary interval is weighted by the number of stars and their luminosity;

(iv) the spectra of the virtual stars are taken from suitable spectral libraries,
as a function of effective temperature, gravity, and chemical composition.
We employed the spectral library by \citet{Lejeune1998}, based upon
the Kurucz (1995) release of theoretical spectra, with several important implementations.
For $T_{\rm eff}<3500$ $K$ the
spectra of dwarf stars by \citet{Allard1995} are included, whilst the spectra by \citet{Fluks1994}
and \citet{Bessell1989, Bessell1991} are considered for giant stars.
Following \citet{Bressan1994}, for $T_{\rm eff}> 50000$ $K$
the library has been extended using black body spectra.

We have calculated grids of dust-free SSP-SEDs of different ages, for the six values
of metallicity,  and the nine  different IMFs of
  Tab.~\ref{tab_imf}, and derived
magnitudes and colours for different photometric systems.

\begin{figure*}
\begin{center}
\subfigure
{\includegraphics[width=0.33\textwidth]{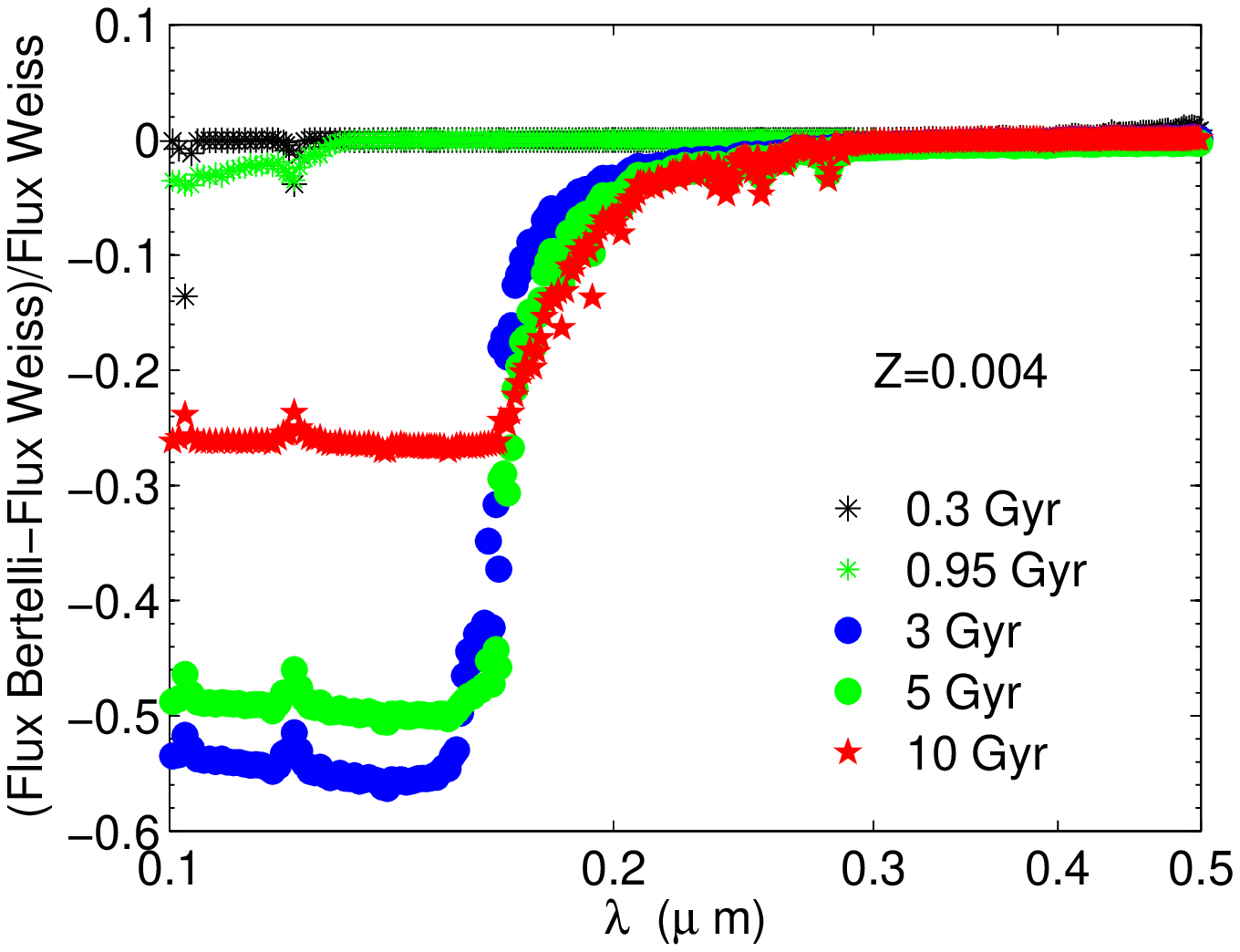}
 \includegraphics[width=0.33\textwidth]{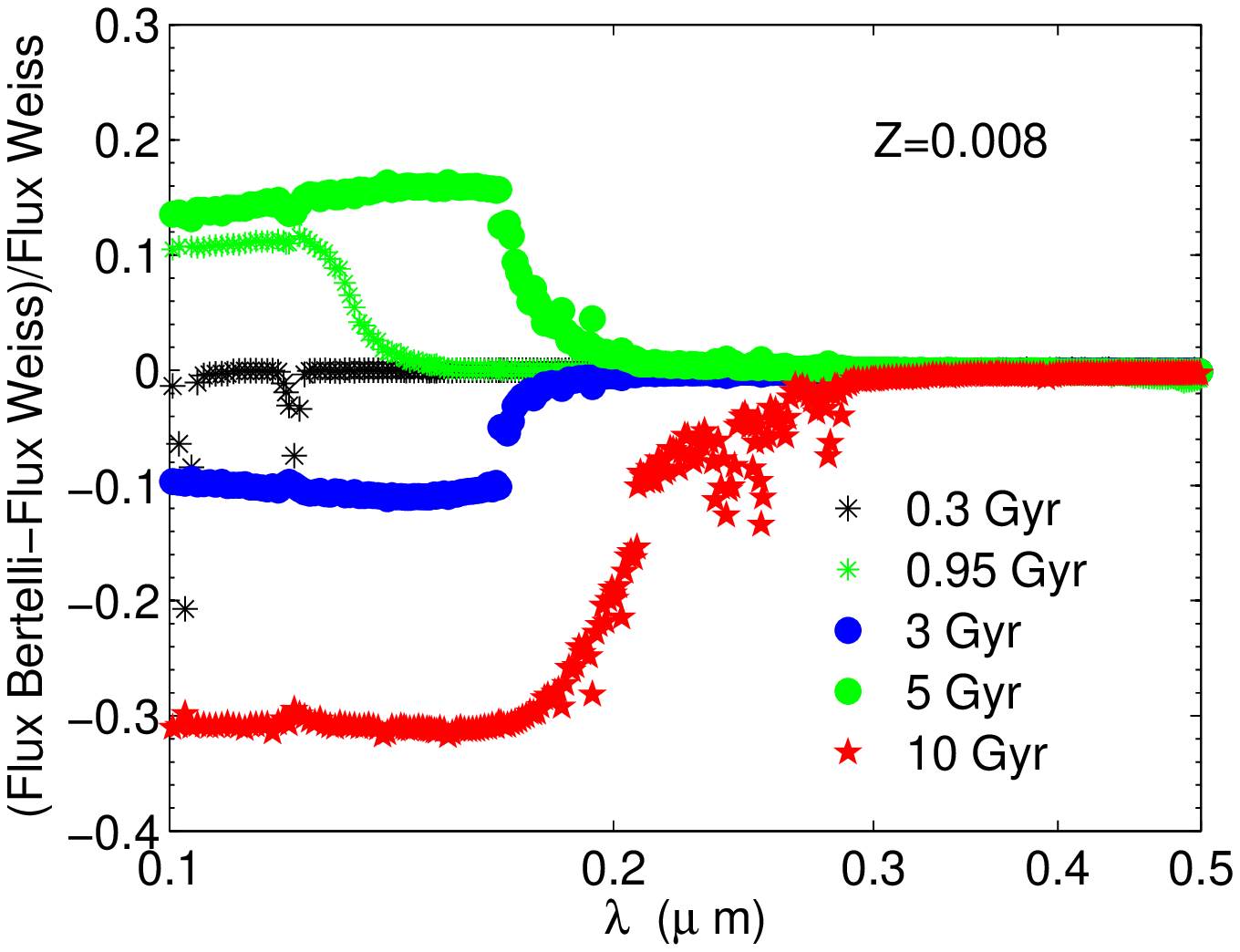}
 \includegraphics[width=0.33\textwidth]{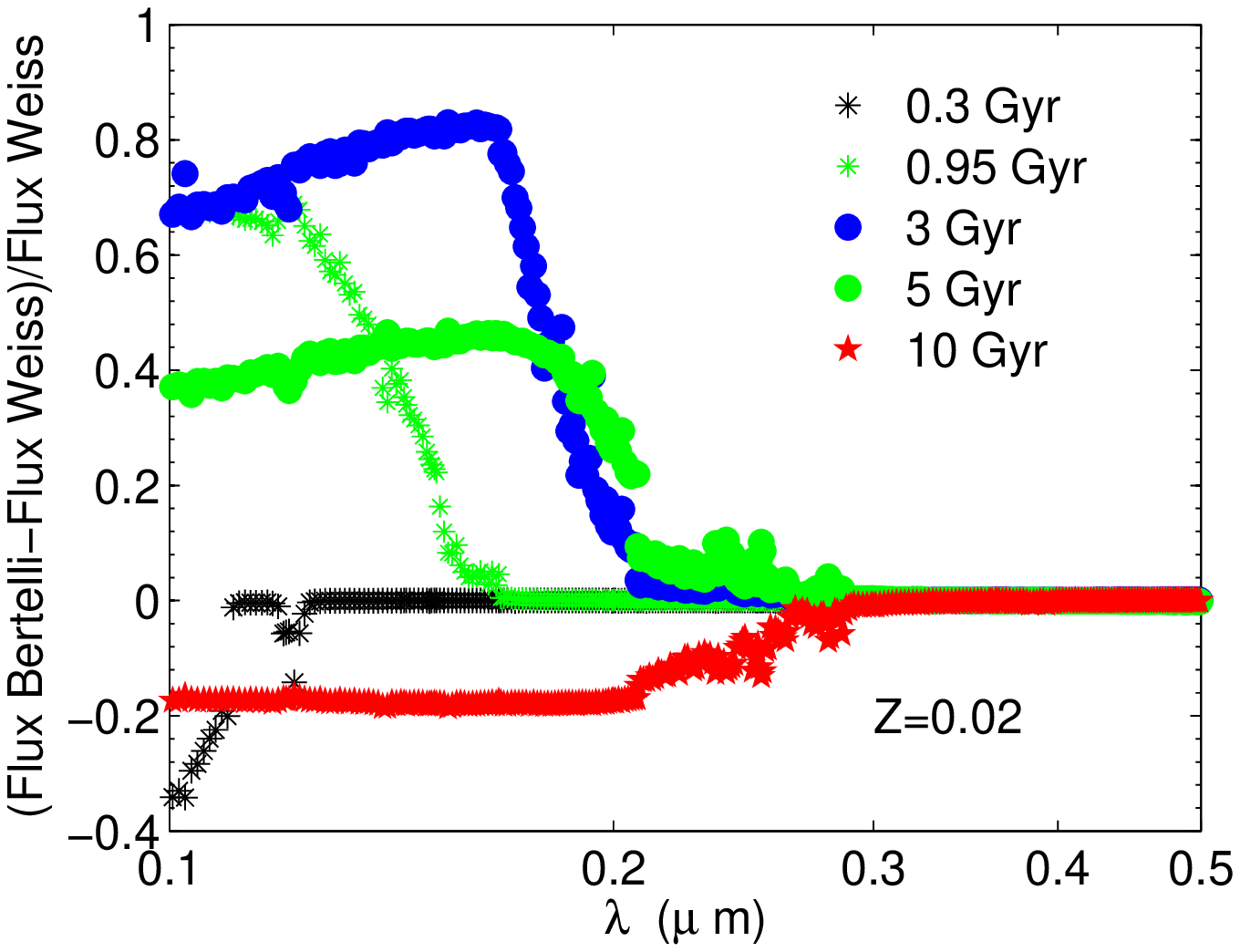}}
\subfigure
{\includegraphics[width=0.33\textwidth]{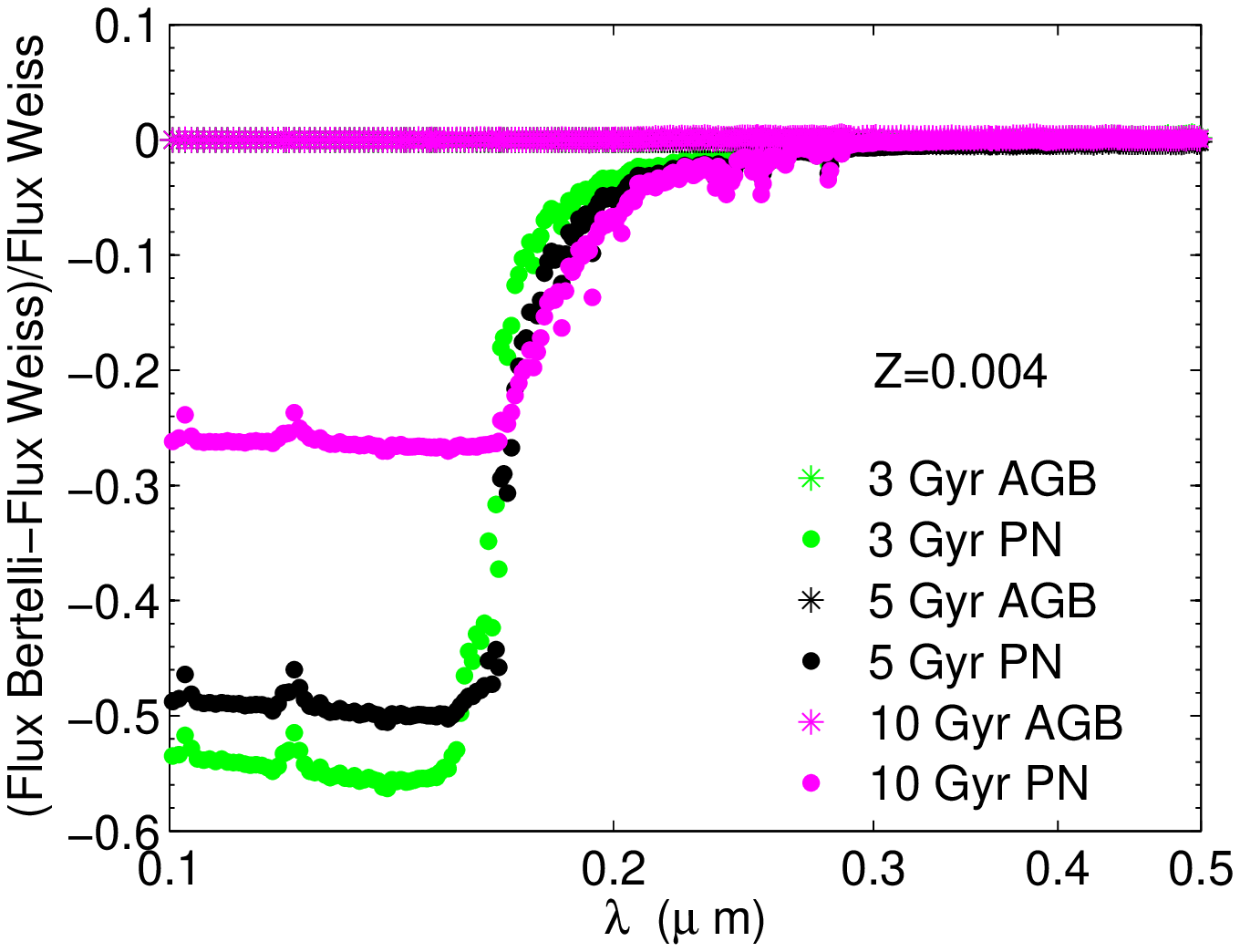}
 \includegraphics[width=0.33\textwidth]{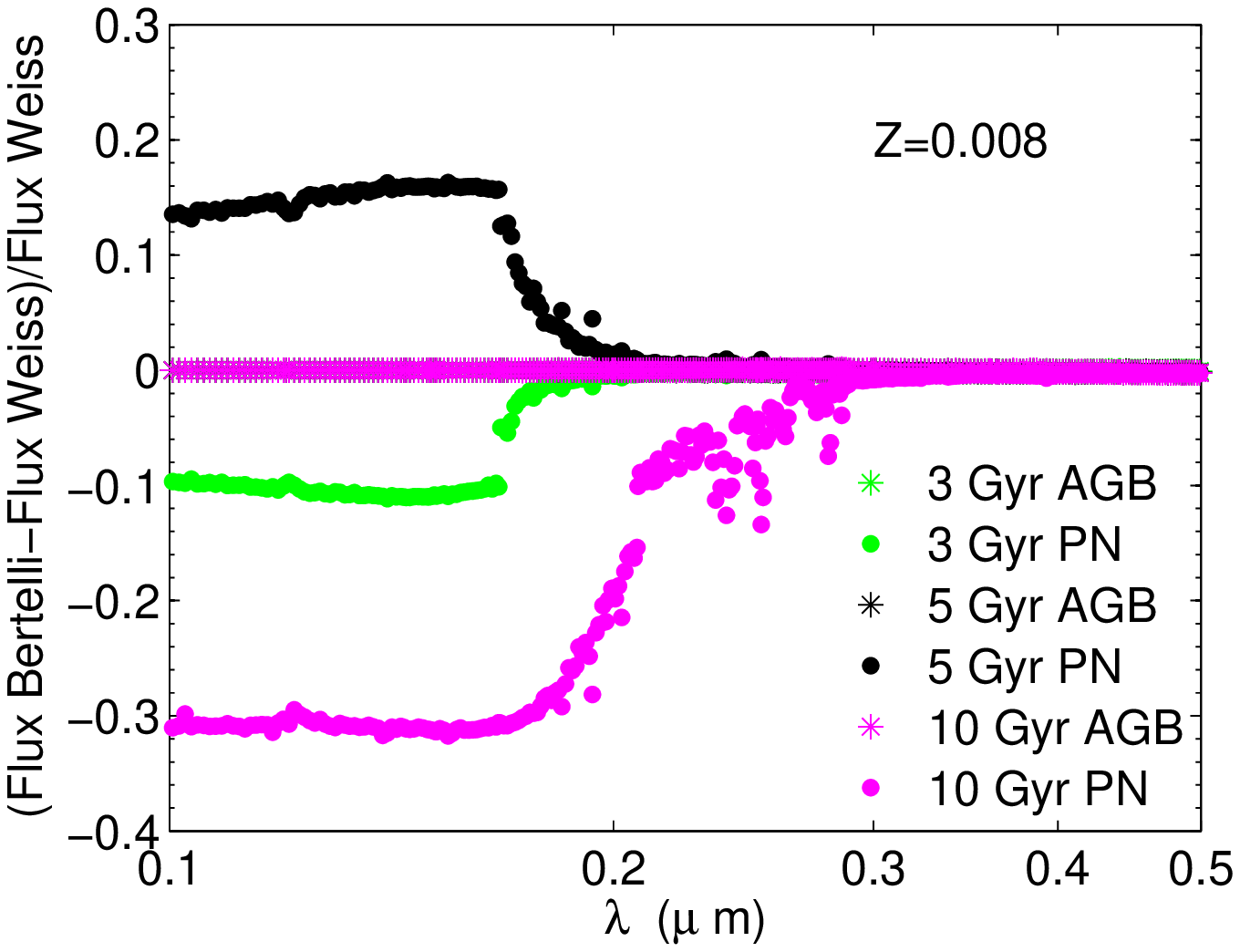}
 \includegraphics[width=0.33\textwidth]{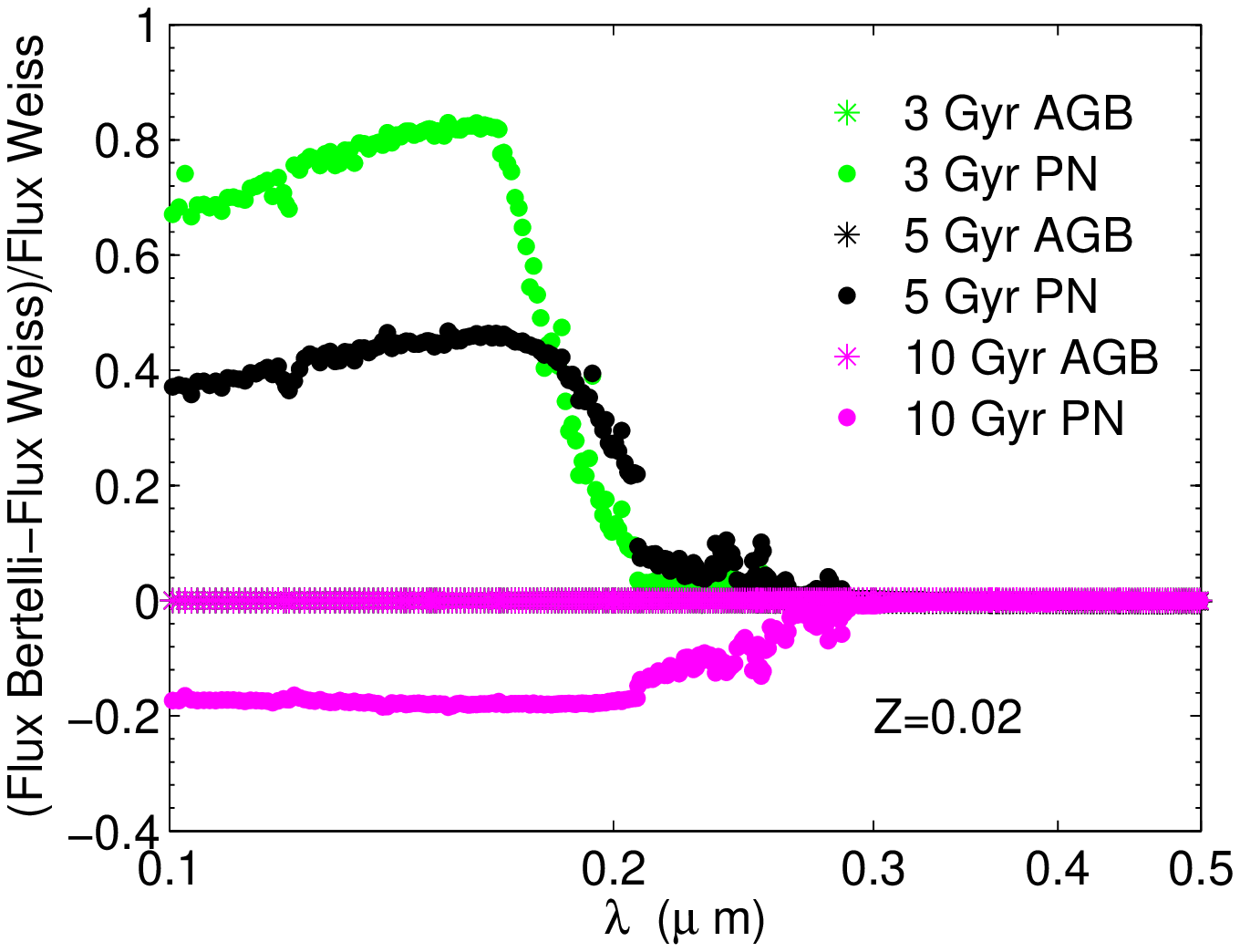}}
\caption{\textbf{Upper panels}: comparison between the integrated flux of our new SSPs and the
\textit{old} SSPs, as a function of \ $\lambda$, for the labelled ages and metallicities.
 \textbf{Lower panels}: as the upper panels, but for both integrated and cumulative fluxes to the end of the AGB,
 for fewer selected ages.
 Left panels are for $Z$=0.004, central panels
  for $Z$=0.008, and right panels for  $Z$=0.02.}
\label{residui_123}
\end{center}
\end{figure*}

\subsection{A comparison with the old dust-free SSPs}

We compare in this section the old SSPs computed by
\citet{Bertelli1994} with our new database. The \textit{only improvement} is the
TP-AGB phase, based on the new models by  \citet{Weiss2009}.
We start  by defining at each wavelength $\lambda$ a residual flux ratio
                 \[FR_\lambda = [F_\lambda(Bertelli) - F_\lambda(Weiss)]/F_\lambda(Weiss), \]
where $F_\lambda(Bertelli)$ is the monochromatic SSP flux of the SED calculated with the \citet{Bertelli1994}
AGB models and $F_\lambda(Weiss)$ is the counterpart with the \citet{Weiss2009} AGB models.
The results are presented in Fig. \ref{residui_123}, for $Z$=0.004,
$Z$=0.008, and $Z$=0.02, respectively. The top panels display the
total monochromatic flux for five selected ages, moving from young
ages where the AGB phase is well developed, to old ages where the AGB
is of much less importance. We define as \textit{cumulative
    flux} the monochromatic flux integrated between the zero age main sequence
   and a given advanced evolutionary phase, like, i.e.,
  the tip of the RGB or the end of the TP-AGB. The bottom
  panels show, for ages of 3, 5 and 10 Gyr, the
  cumulative flux to the end of the AGB phase and the total flux, that includes the post-AGB PN and
  WD phases.
There are two regions of the SED where we expect differences,
even when dust is not introduced:
(1) the near-IR region affected by cool stars and (2) the UV region,
because different AGB lifetimes lead to differences in the PN
phase. Indeed, Fig.~\ref{residui_123} reveals significant differences
between old and new SSPs in the UV region (say up to 0.3
$\umu$m).
These are likely caused by the different assumptions made by
\citet{Bertelli1994} and \citet{Weiss2009} for the  mass-loss rate
during the TP-AGB phase. The old  SSPs make use of  the
\citet{Vassiliadis1993} prescription; the new models include the mass-loss
rates by either \citet{Wachter2002}
or \citet{vanLoon2005}, depending on the surface chemical compositions of the
models. For a given initial mass, different mass-loss rates
produce, when the TP-AGB phase is over, remnants with
different core masses and, in turn, different PNs. This is clear when
looking at the bottom panels of Fig.~\ref{residui_123}. The cumulative
fluxes to the end of the AGB phase do not result in any
  visible residuals. Instead, the inclusion of the PN phase changes the residuals by as much as 30\%. As expected,
this effect increases at decreasing ages: higher mass stars experience
a larger mass-loss rate and produce remnants (cores) of smaller mass
and hotter surface temperatures. Finally, there is a
systematic trend of the ratio $FR_\lambda$ in the UV, when passing from low to high
metallicity (see Fig.~\ref{residui_123}).

\begin{figure*}
\begin{center}
{\includegraphics[width=0.325\textwidth]{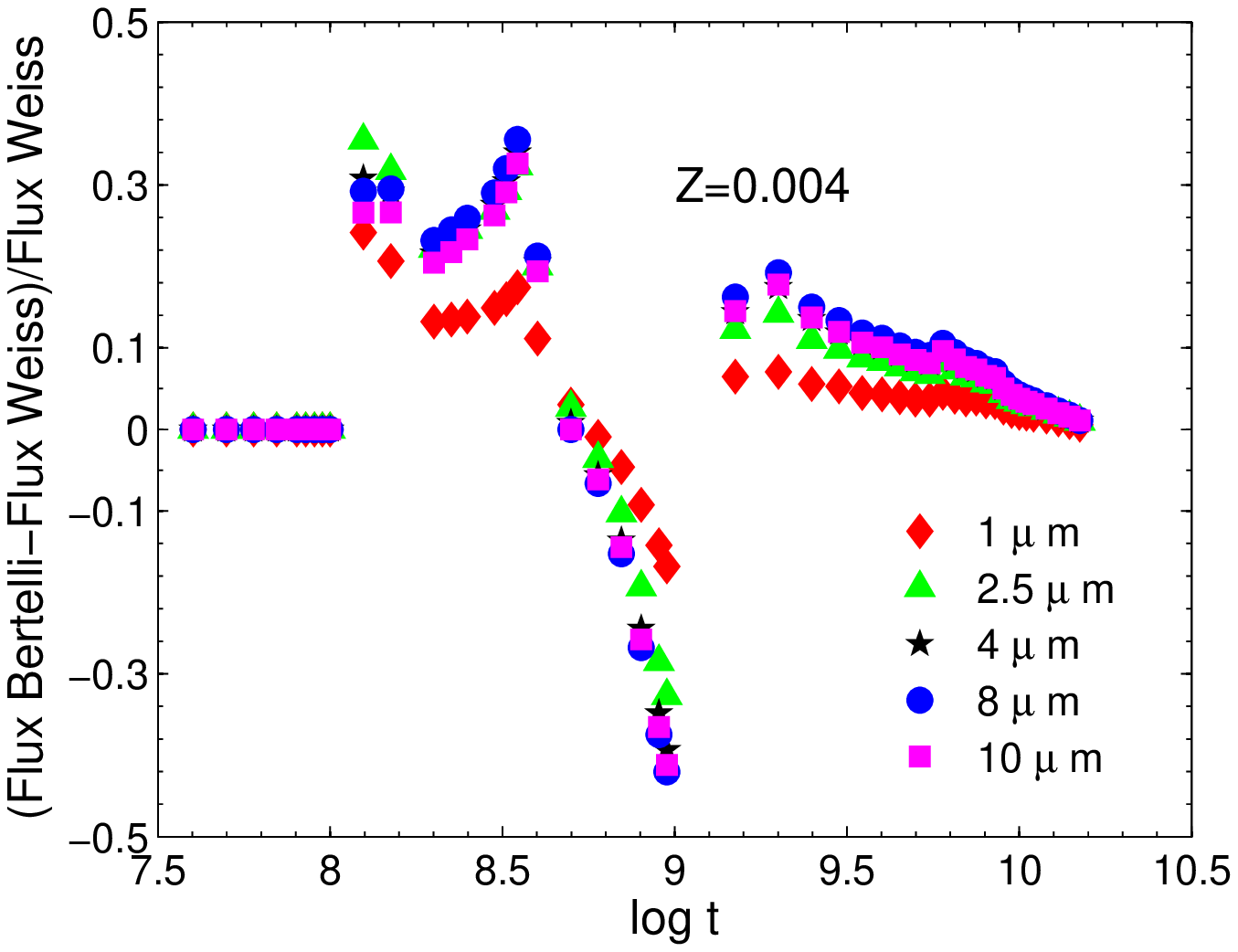}
 \includegraphics[width=0.325\textwidth]{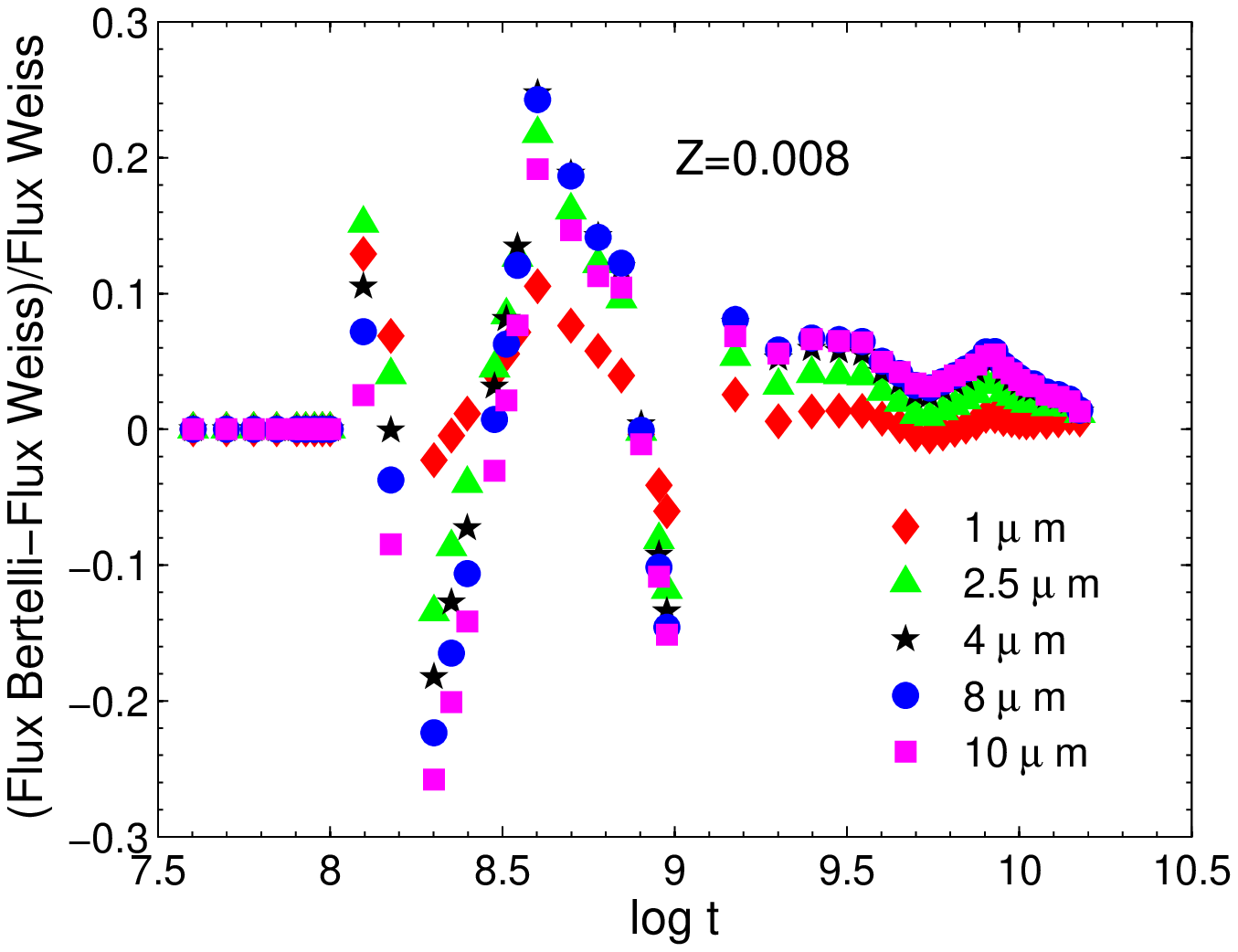}
 \includegraphics[width=0.325\textwidth]{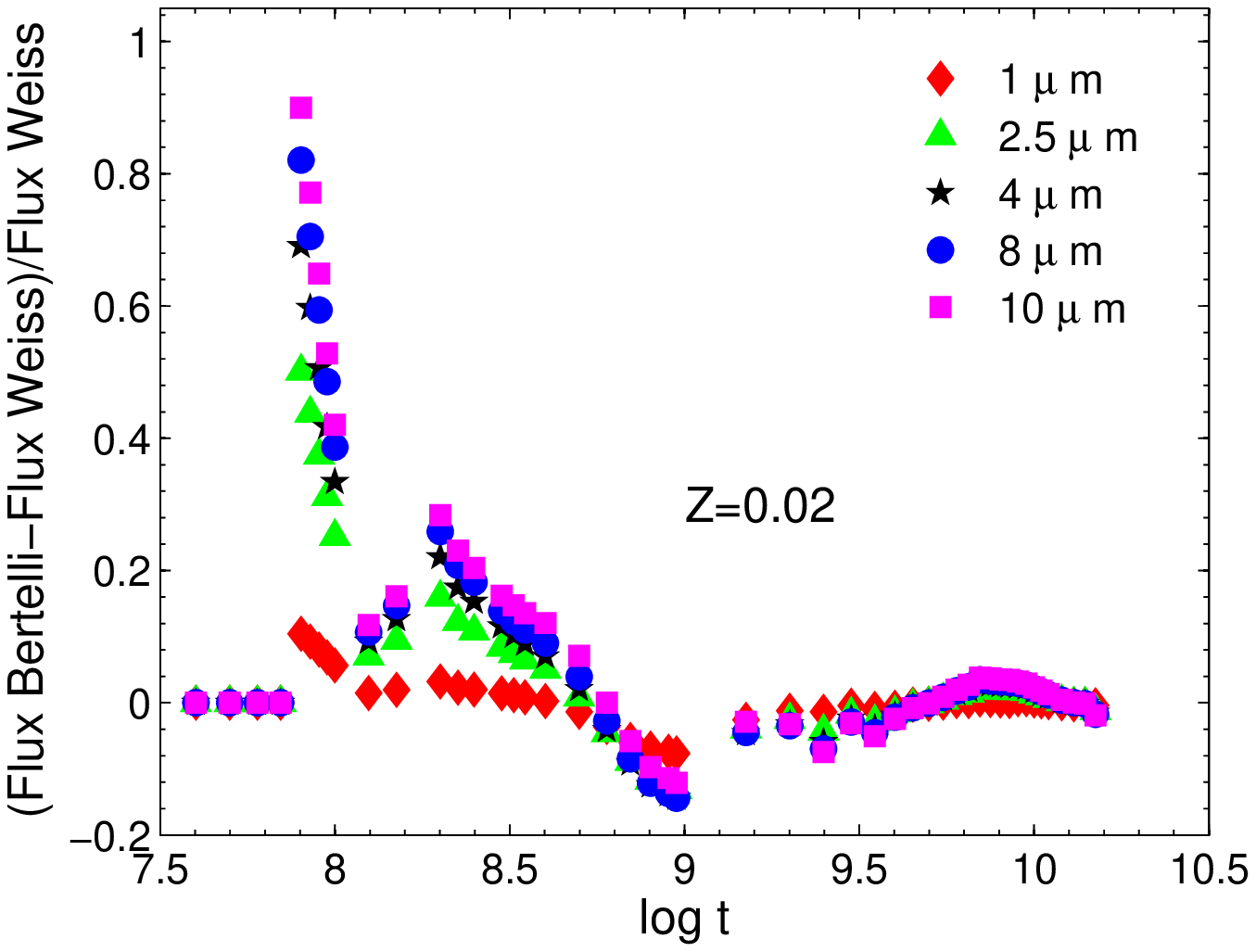} }
\caption{As in the upper panels of Fig. \ref{residui_123},
  but in this case we consider the residual flux ratios as a function of
  age for the labelled reference wavelengths $\lambda$. Three
  metallicities are shown:  $Z$=0.004 (\textbf{left panel}), $Z$=0.008
  (\textbf{middle panel}) and $Z$=0.02 (\textbf{right panel}). The unit of time $t$ is yr.}
\label{residui_4}
\end{center}
\end{figure*}

Other major differences appear in the IR spectral region, where
AGB stars emit most of their  light. This is shown by
Fig. \ref{residui_4}, that displays the ratios FR$_\lambda$ as a
function of age at selected near-IR wavelengths,
for different metallicities ($Z$=0.004, 0.008, and
0.02). Given that we neglect the effects of circumstellar dust
shells around the AGB stars, we expect that the cool M and C models
emit most of the flux in the range $\sim$ 1-4 $\umu$m (dust
would shift the emission towards longer wavelengths). The agreement
 between the two sets of SSPs is very good before the onset of
the AGB phase and for old ages, where the differences amount
to only a few percent. As expected, when the
AGB phase sets in at $\log t\sim$8, differences are much larger. They   can be ultimately ascribed to the
different prescriptions for the TP-AGB phase in the Padua and GARSTEC
models.

\section{SSPs with circumstellar dust around AGB stars} \label{AGBShell}

It has long been known that low and intermediate-mass AGB stars
are amongst the main contributors to the ISM dust content. The previous
evolutionary phases are not as important as dust
factories: dust formation in RGB and E-AGB stars is poorly efficient
because of the unfavorable wind properties and the low mass-loss rate
\citep{Gail2009}. As for the calculation of SEDs,  magnitudes,
  and colours, the presence of dust shells
is usually -- with just a few exceptions \citep[see for
  instance][]{Bressan1998,Mouhcine2002c,Piovan2003,Marigo2008} --
 not taken into account.

As TP-AGB stars are expected to form significant amounts of dust and therefore suffer
self-obscuration and re-processing of their photospheric radiation,
the effect of dust on their SEDs cannot be ignored \citep{Piovan2003}.\\
\indent Dust formation in AGB stars has been modelled with increased accuracy over the years
\citep{Gail1984,Gail1985,Gail1987,Dominik1993,Gail1999,Ferrarotti2002,Ferrarotti2006,Gail2009}, and we are now
in the position to calculate the amount of newly formed dust in M-stars,
S-stars and C-stars (a sequence of growing [C/O]-ratio). This ratio
determines the composition of dust formed in the outflows
\citep{Piovan2003,Ferrarotti2006,Gail2009}. The oxygen-rich M-stars
([C/O]$<1$)  produce dust grains mainly formed by refractory elements
(generically named silicates), like pyroxenes and olivines, oxides and
iron dust. Carbon-rich  stars ([C/O]$>1$)  produce carbon-dust;
$\textrm{SiC}$ and maybe iron dust can condensate. In S-stars ([C/O]$
\approx 1$)   quartz and iron dust should form
\citep{Ferrarotti2002}. However, the carbon-rich or oxygen-rich phases
dominate, and for example the contribution of $\textrm{SiC}$ produced
during the S-star phase can be neglected, compared to the $\textrm{SiC}$
produced during the C-star phase.

\subsection{Modelling a dusty envelope}\label{par:Dusty}

The problem of the radiative transfer in the dusty shells that form around AGB stars has been
addressed by many authors \citep[see the classical review by][and references therein]{Habing1996}.
The best approach would clearly be to couple the equations describing the radiative transfer through
the dusty envelope, with the hydro-dynamical equations for the motion of the two components, dust and gas,
taking into account the interplay between gas, dust, and radiation pressure. For the purposes of this work,
 it is however enough to limit ourselves to solve the problem of the radiative transfer through the envelope
\citep{Ivezic1997,RowanRobinson1980}. Indeed, our purpose is to build a library of dusty SEDs to
determine the effects of dust around AGB stars, not to study the dynamical behaviour of the outflows.

\subsubsection{The optical depth}\label{AGB_tau}

As our aim is to determine the SED of AGB stars after it has been
filtered by circumstellar dust shells, we can use DUSTY, the
classical code for radiative transfer \citep{Ivezic1997}. The
original version of the code cannot handle large
wavelength grids without becoming computationally very demanding. To cope
with this, we suitably modified the public version 2.06 of DUSTY
to handle a much larger grid at still reasonable computational cost. Our full grid is built by adding the Kurucz-Lejeune wavelength grid
plus the all the wavelengths characterizing the tabulated optical
properties and features of the dust. For the sake of simplicity, we
assume spherical
symmetry. The key parameter needed to solve the radiative
transfer problem is the
optical depth $\tau_{\lambda}$ of the shell, defined as
follows:
\begin{equation}
\tau_{\lambda} = \int_{r_{in}}^{r_{out}}d\tau _{\lambda}\left(r\right) =
\int_{r_{in}}^{r_{out}}k_{\lambda} \left(r \right) \rho_{d}
\left( r \right) dr
\label{taulambda}
\end{equation}

\noindent where $k_{\lambda,d}$ is the overall dust extinction coefficient per mass unit and $\rho_{d}$
 is the dust mass density. They both depend on the radial distance $r$ from the central source.
 The integral is evaluated over the thickness of the shell, from the innermost to the outermost radius. If we
now apply the continuity equation for the gas and dust \citep{Schutte1989,Piovan2003}
we can recast the optical depth of Eq.~\ref{taulambda} as

\begin{equation}
\tau_{\lambda} = \int_{r_{in}}^{r_{out}}k_{\lambda} \left(r \right)
\frac{\dot{M} \left(r \right) \delta\left(r\right)}{ 4 \pi r^{2} v_{d} \left( r
\right)}  dr
\label{taulambdabis}
\end{equation}

\noindent where $\delta$ is the dust-to-gas ratio in the shell. To proceed further, the mass-loss rate $\dot{M(r)}$,
the expansion velocity of the dust $v_{d}(r)$, the extinction coefficient $k_{\lambda}(r)$, and the dust-to-gas ratio
$\delta$ together with their radial dependence must be specified. Common assumptions are the following
\citep{Groenewegen1993b,Bressan1998,Piovan2003,Groenewegen2006,Marigo2008}: at any given time, the rate of mass-loss and the velocity are constant and do not depend on $r$.
The same holds for the optical properties of the dust and the dust-to-gas
ratio. The radial dependence is neglected. With these simplifications and assuming that $r_{out}\gg r_{in}$ and
$r_{in} \sim r_{c}$ we have

\begin{equation}
\tau_{\lambda} = \frac{\delta \dot{M} k_{\lambda}}{4 \pi
v_{\infty}r_{c}}
\label{tauter}
\end{equation}

\noindent where $v_{\infty}$ is the wind terminal velocity and
$r_c$ the condensation radius or the innermost distance from which
dust starts to absorb the stellar radiation.  A safe approximation
  is that $v_{d} \left( r \right)\sim v_{\infty}$ because of the
small drift between gas and dust \citep{Groenewegen1993b}. The
extinction coefficient per unit mass $k_{\lambda}$ is in general given by:

\begin{equation}\label{klambda}
k_{ \lambda} = \frac{\sum_{i} n_{i} \sigma_{i}(a,\lambda)}{\rho_{d}}=
\frac{\sum_{i} n_{i} \pi a^{2}Q_{i}(a,\lambda)}{\rho_{d}}
\end{equation}

\noindent where the summation is extended over all types of grains in
the envelope, and $\sigma_{i}$ and $n_{i}$ are the cross
section and the number of grains  per unit volume of the
\textit{i-th} dust type, respectively. For the sake of simplicity only
one typical dimension $a$ of the grains is assumed.
  The total mass density of the grains for unit volume $\rho_{d}$, is

\begin{equation}
\rho_{d}=\frac{4}{3}\pi a^{3}\sum_{i}n_{i}\rho_{i}
\end{equation}

\noindent where $\rho_{i}$ is the mass density of a grain of dust type $i$, assumed to be spherical. Finally, we get:

\begin{equation}
\tau_{\lambda} = \frac{3 \delta \dot{M}}{16 \pi
v_{\infty}r_{c}}\frac{\sum_{i} n_{i} Q_{i}(a,\lambda)/a}{\sum_{i} n_{i}\rho_{i}}
\label{tauquater}
\end{equation}

\noindent Starting from Eq.~\ref{tauquater}, introducing a single type of grain and properly normalizing the
 various quantities, it is possible to recover the expression by \citet{Groenewegen2006} for the optical depth.
 The inner radius of the shell can be derived from the conservation of the total luminosity
$L = 4 \pi R_{*}^{2} \sigma T_{\rm eff}^{4} = 4 \pi r_{c}^{2} \sigma T_{d}^{4}$, thus obtaining

\begin{equation}\label{taufive}
\tau_{\lambda} = A_{d}\delta \dot{M} v_{\infty}^{-1} L^{-1/2}
\end{equation}

\noindent where $A_{d}$ depends on the adopted mixture of dust
  \citep{Marigo2008}:

\begin{equation}
A_{d} = \frac{3}{8}T_{d}^{2}\left(\frac{\sigma}{\pi}\right)
\frac{\sum_{i} n_{i} Q_{i}(a,\lambda)/a}{\sum_{i} n_{i}\rho_{i}}
\end{equation}

\noindent Different kinds of dust would imply different condensation
temperatures $T_{d}$, thus leading to different radii $r_{c}$. For the
sake of simplicity and due to the DUSTY requirements,
only a single condensation temperature will be used, even in case of a multi-component dust shell.\\
\indent We need now to connect the quantities defining the optical depth of the shell with
the parameters of the AGB models. We
can take the surface bolometric luminosity $L/L_\odot$, the effective temperature  T$_{\rm eff}$, the mass-loss
rate $\dot{M}$, the metallicity $Z$, the [C/O]-ratio, and the
chemical composition of the star at the surface. To get the terminal
velocity of the wind, \citet{Bressan1998} and \citet{Piovan2003}
adopted the simple recipes by \citet{Vassiliadis1993} and
\citet{Habing1994}. In this paper, we  employ the formulation of
dusty winds  by
\citet{Elitzur2001}  as also done recently  by \citet{Marigo2008},

\begin{equation}\label{vinfty}
v_{\infty}=\left(A\dot{M}_{-6}\right)^{1/3}\cdot
\left(1+B\frac{\dot{M}_{-6}^{4/3}}{L_{4}}\right)^{-1/2}
\end{equation}

\noindent where the velocity is in km s$^{-1}$, the mass-loss rate $\dot{M}_{-6}$
in units of $10^{-6}$ M$_{\odot}$yr$^{-1}$, and finally the AGB star luminosity L$_{4}$  in units of
10$^{4}$L$_{\odot}$.
The two parameters A and B are defined as  in \citet{Elitzur2001}:

\begin{equation}\label{A_Elitzur}
A= 3.08 \times 10^{5} T^{4}_{c3}Q_{*}\sigma_{22}^{2}\Psi_{0}^{-1}
\end{equation}

\begin{equation}\label{B_Elitzur}
B=\left(2.28\frac{Q_{*}^{1/2}\Psi_{0}^{1/4}}{Q_{V}^{3/4}\sigma_{22}^{1/2}T_{c3}}\right)^{-4/3}
\end{equation}

\noindent The meaning of the various parameters contained in the functions
$A$ and $B$ is as follows. First, T$_{c3}$ is the dust condensation
temperature in units of $10^{3}$;  literature values range from 800 K to 1500 K
\citep{RowanRobinson1982,David1990,Suh1999,Suh2000,LorenzMartins2000,LorenzMartins2001,Suh2002}.
Our choice is in the range between 1000~K and 1500~K, depending on
the dust mixture and the [C/O]-ratio (see below for more details),
in agreement with most of the literature and with similar works
on dusty AGBs by \citet{Groenewegen2006} and \citet{Marigo2008}.
Then, Q$_{*}$ is the  mean of the quantity $Q(\lambda,a)$, averaged over
the Planck function B$\left(\lambda,\textrm{T}_{\textrm{eff}}\right)$:

\begin{equation}\label{Qstar}
Q_{*}=\frac{\pi}{\sigma T^{4}_{\textrm{eff}}}\int Q\left(a,\lambda\right)B
\left(\lambda,\textrm{T}_{\textrm{eff}}\right)d\lambda
\end{equation}

\noindent where $Q\left(a,\lambda\right)$ is the sum of the absorption and scattering radiation
pressure efficiencies, assuming isotropic scattering. The cross section $\sigma_{22}$ is defined
by the following relation with the gas cross section:

\begin{equation}\label{sigma22}
\sigma_{g}=\sigma_{\textrm{22}} \cdot 10^{-22} \textrm{cm}^{2}
\end{equation}

\noindent where

\begin{equation}\label{sigmag}
\sigma_{g}=\pi a^{2}\frac{\sum_{i}n_{i}}{\sum_{i}n_{i,g}}=\frac{3}{4}
\frac{A_{g}m_{H}}{a\overline{\rho}}\delta .
\end{equation}

\noindent  Here $n_{i,g}$ is the gas number density, $A_{g}\simeq
4/(4X_{H}+X_{He})$  the mean molecular weight of the gas
\citep{Marigo2008}, m$_{H}$  the atomic mass unit and
$\overline{\rho}$  the average mass density of the grains calculated
      for the actual mixture of dust, given by $\overline{\rho}=\sum_{i}n_{i}\rho_{i}/\sum_{i}n_{i}$.
			The parameter $\Psi_{0}$ is defined in \citet{Elitzur2001} as:

\begin{equation}\label{Psi0}
\Psi_{0}=\frac{Q_{P}\left(T_{\rm eff}\right)}{Q_{P}\left(T_{d}\right)}
\end{equation}

\noindent where the subscript $P$ means an average of the
absorption efficiency over the Planck function, similar to the average that defines Q$_{*}$ in Eq.~(\ref{Qstar}).
It must be underlined that \citet{Elitzur2001} assume the temperature of the star to be fixed at 2500 K:
in our case we will take into account the variation of T$_{\rm eff}$, by considering every time
 the temperature of the current stellar model. Finally, the last parameter in Eqs.~(\ref{A_Elitzur}) and (\ref{B_Elitzur})
is Q$_{V}$, the absorption efficiency at optical wavelengths.

\subsubsection{Mass-loss}\label{AGB_MassLoss}

\indent
It is currently widely accepted and supported by hydro-dynamical
calculations, that large amplitude pulsations are required to
accelerate the mass outflow from the stellar surface of AGB stars to
regions where the gas cools enough so that refractory elements can
condense into dust. Once dust grains are formed, they transfer
energy and momentum from the stellar radiation field to the
gas by collisions, so that the flow velocity may grow enough to exceed the escape
velocity \citep{Gilman1972}.  This stellar wind increases with
time until the so-called super-wind regime is reached: the star
quickly evolves into a PN, with the whole envelope being stripped off. The remnant is
a bare CO core that evolves to high effective temperatures. We
have already reported on the  mass-loss rates adopted for the various
evolutionary phases from the RGB to the formation of PN stars. They
are  also used here for the sake of consistency between stellar
models and their dusty envelopes.
The only point to note is that a minimum mass-loss is required to form
enough dust to be able to accelerate the gas beyond the escape
velocity \citep{Elitzur2001}. The minimum mass-loss is:

\begin{equation}\label{MinimumMassLoss}
\dot{M}_{min}=3\times 10^{-9}\frac{M^{2}}{Q_{*}\sigma_{22}^{2}L_{4}T_{k3}^{1/2}}
\end{equation}

\noindent where T$_{k3}$ is the kinetic temperature at the inner
boundary of the shell, that we simply set to
$\textrm{T}_{k3}\approx \textrm{T}_{c3}$. It may easily occur that
envelopes are optically thin and $\dot{\textrm{M}}\lesssim
\dot{\textrm{M}}_{min}$. In this case  dust is formed, but according
to \citet{Elitzur2001} it cannot  sustain  the wind. When this happens,
we apply  the recipe proposed by \citet{Marigo2008} to
evaluate the expansion velocity by means of $\dot{\textrm{M}}_{min}$,
and get an estimate for $v_{\infty}$ to insert in the expression for
$\tau_{\lambda}$.

\begin{figure*}
\subfigure{
\includegraphics[width=0.45\textwidth]{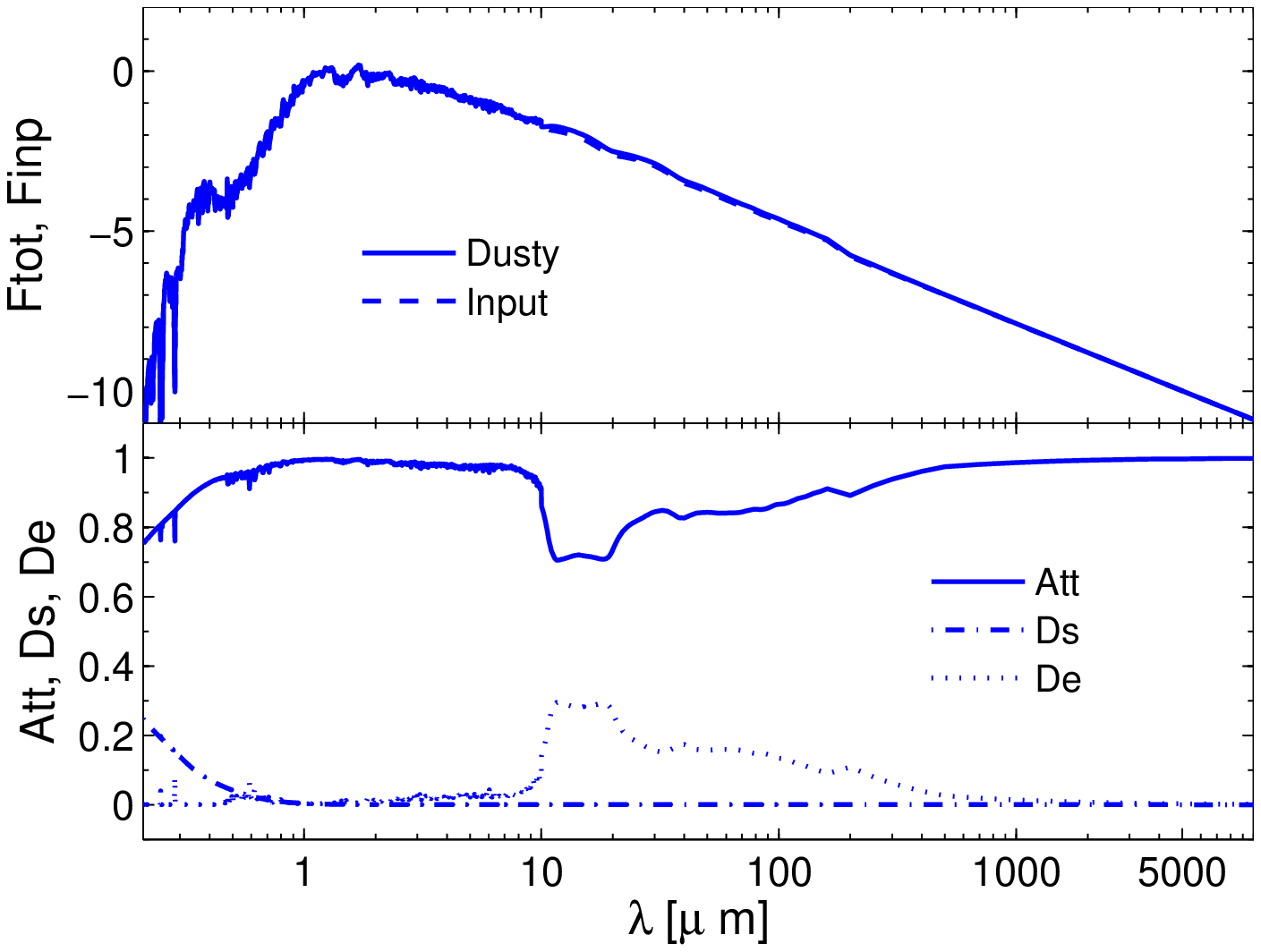}  
\includegraphics[width=0.45\textwidth]{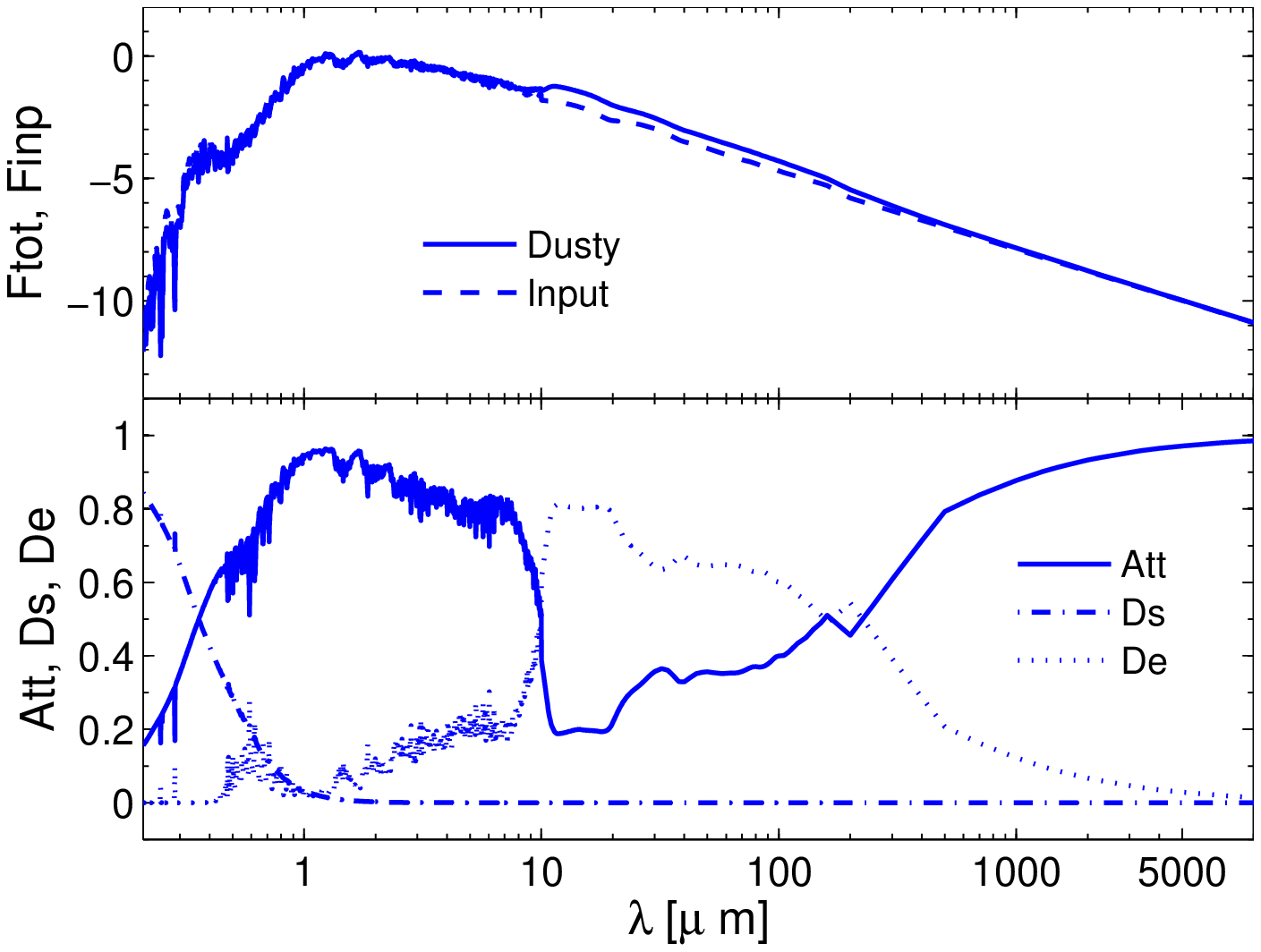}}
\subfigure{
\includegraphics[width=0.45\textwidth]{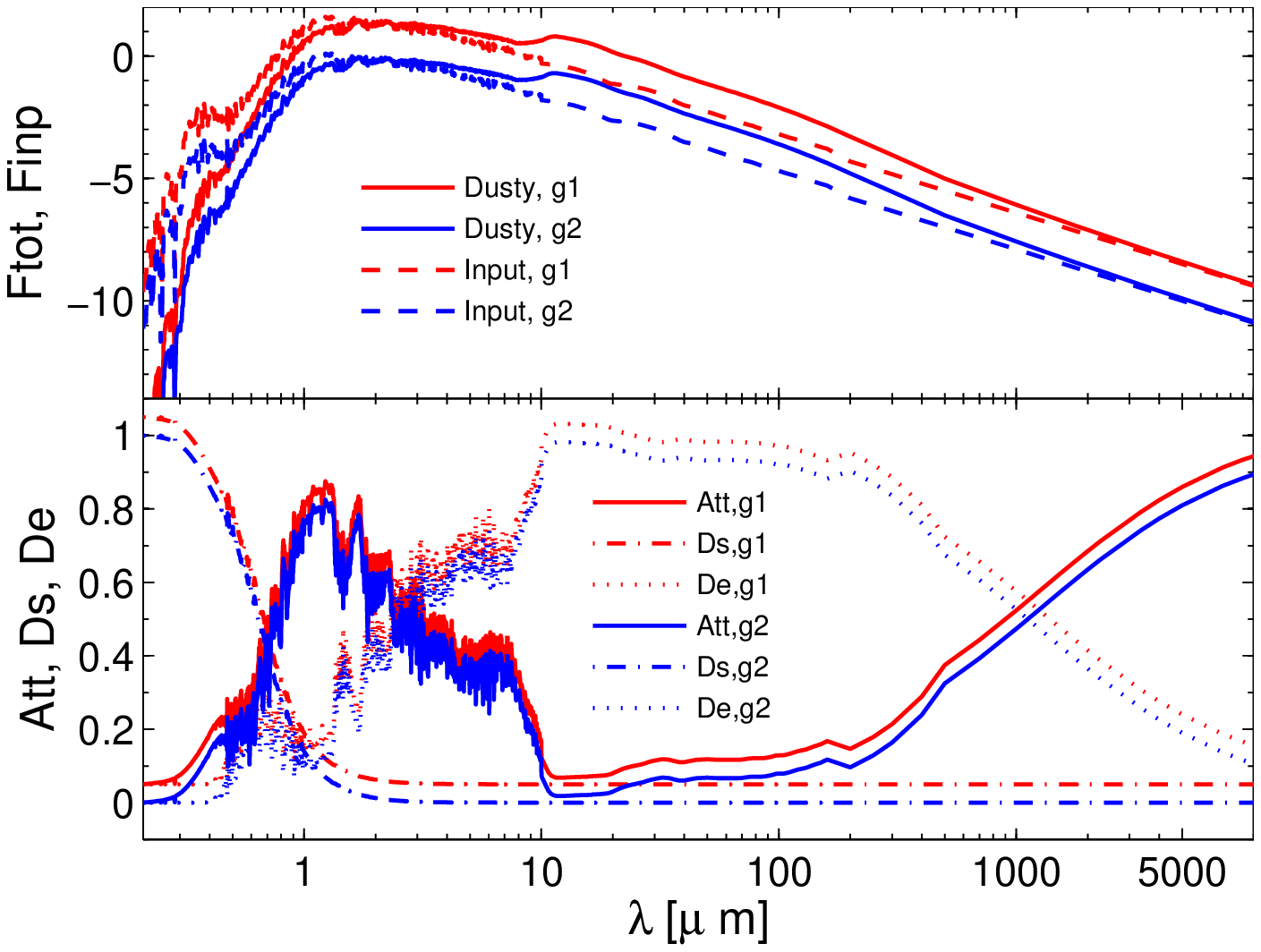}
\includegraphics[width=0.45\textwidth]{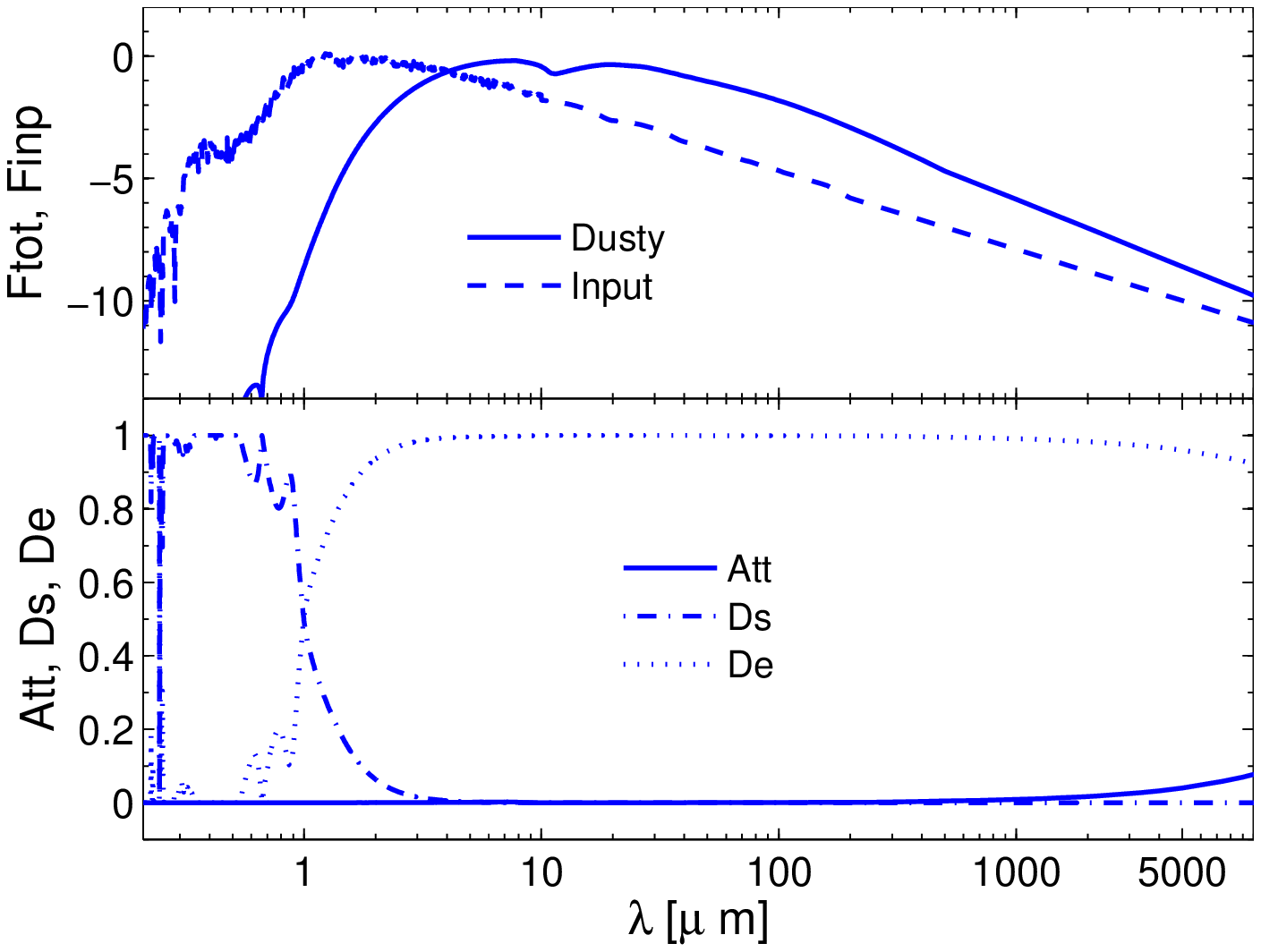}
}
\caption{Dust-enshrouded spectra for AGB stars obtained with our modified version of the radiative transfer code DUSTY.
The input parameters are: M-type AGB stars with T$_{\rm eff}$=2500 and L=3000~L$_{\odot}$,
oxygen-rich surface composition  with 60$\%$ Al$_{2}$O$_{3}$ and 40$\%$ silicates.
The SEDs for four values of the optical depth are shown:
$\tau$=0.0224 (\textbf{upper-left panel}),
$\tau$=0.2083 (\textbf{upper-right panel}),
$\tau$=1.306 (\textbf{lower-left panel}),
and $\tau$=30.0 (\textbf{lower-right panel}).
The lower-left panel shows the results for two different gravities of the input spectra. More details are given in the text.
} \label{MstarsTau_123}
\end{figure*}

\subsubsection{Dust-to-gas ratio}\label{AGB_DustToGas}

\indent Another important parameter of Eq.~(\ref{tauquater}) is the
dust-to-gas ratio $\delta$. In \citet{Piovan2003} the dust-to-gas
ratio was obtained by simply inverting a relation between velocity,
luminosity and dust-to-gas ratio based upon the results by
\citet{Habing1994}.\\
\indent Over the years, increasingly
refined models of AGB stars have simulated the process of dust formation
in the envelope \citep{Gail1984,Gail1985,Gail1987,Dominik1993,Gail1999,
Ferrarotti2001,Ferrarotti2002,Ferrarotti2003}. In \citet{Ferrarotti2006} dust formation
is described through the concept of key species \citep{Kozasa1987} and detailed tables of
yields of dust for oxygen-rich and carbon-rich stars are
presented. The dust grains considered by   \citet{Ferrarotti2006} are
pyroxenes, olivines, quartz and iron dust for oxygen-rich M-stars,
quartz and iron dust for S-stars,  and finally silicon carbide and
carbonaceous grains for carbon-rich C-stars. For each one of them,
according to the initial metal distribution adopted by \citet{Weiss2009}, the
key element will be silicon, iron or carbon, depending on the grain
type. Indeed, only the abundances of C and O may change during the AGB
evolution due to  TDU and  E-HB, whereas the abundances of Mg, Si, S
and Fe remain unchanged. Introducing the key elements and the
equations of continuity for the two-fluids medium made of gas and
dust, the dust-to-gas ratio can be expressed as
\citep{Ferrarotti2003}:

\begin{equation}\label{dusttogas}
\delta=\frac{\dot{M}_{d}}{\dot{M}-\dot{M}_{d}}=\frac{\sum_{i}\dot{M}X_{i}\frac{\displaystyle A_{d,i}}{\displaystyle n_{d,i}A_{i}}f_{d,i}}{\dot{M}-\sum_{i}\dot{M}X_{i}\frac{\displaystyle A_{d,i}}{\displaystyle n_{d,i}A_{i}}f_{d,i}}
\end{equation}

\noindent where the summation is over all dust compounds. Simply,
$\dot{\textrm{M}}\textrm{X}_{i}$ is the abundance of the i-th
key-element in the wind and
$\dot{\textrm{M}}\textrm{X}_{i}\textrm{f}_{d,i}$  the fraction of the
key element condensed into dust. Dividing by
$\textrm{n}_{d,i}\textrm{A}_{i}\textrm{m}_{\textrm{H}}$, where
$\textrm{n}_{d,i}$ is the number of atoms of the key elements required
to  form one dust unit, and A$_{i}$ the atomic weight of the i-th
element, we get the number of dust units. Finally, multiplying A$_{d,i}\textrm{m}_{\textrm{H}}$
by the mass of one dust unit, we obtain the
total mass of the $i-th$ dust compound. We then divide the AGB
evolution  into three regions, corresponding to different
[C/O]-ratios. Following  \citet{Ferrarotti2006}, we  define two
critical carbon abundances:
$\epsilon_{C,1}=\epsilon_{O}-2\epsilon_{Si}$ and
$\epsilon_{C,2}=\epsilon_{O}-\epsilon_{Si}+\epsilon_{S}$, where
$\epsilon=X/A$ is the abundance in mol g$^{-1}$. The two corresponding
critical [C/O]-ratios are  $\left(\textrm{[C/O]}\right)_{1}=0.9$ and
$\left(\textrm{[C/O]}\right)_{2}=0.97$. \citet{Ferrarotti2003} groups the  stars along the AGB into three classes: M-stars, S-stars and C-stars .

\noindent \textbf{\textsf{Oxygen-rich M-stars}}.
The spectra of oxygen-rich, M-type AGB stars show two typical
features at 10 $\umu$m and 18 $\umu$m, either in absorption or in emission, depending on the optical depth
of the surrounding envelope. These features are usually attributed to stretching and bending modes of
$\textrm{Si-O}$ bonds and $\textrm{O-Si-O}$ groups, and probe the existence of silicate grains in the
circumstellar shell. Because of the strong bond between O and C in the carbon monoxide,
it is believed that all C is blocked into CO molecules and none is available for the formation of dust
grain with other chemical species of low abundance. In contrast, the fraction of O not engaged in CO
reacts with other elements such as $\textrm{Mg}$ and $\textrm{Si}$, and forms various types of compounds.
Iron dust can accrete onto the envelope as well. By applying Eq.~\ref{dusttogas} to the specific case we get:

\begin{equation}\label{Ostarsdust}
\dot{M}_{d}=\dot{M}\left(X_{\textrm{Si}}\frac{A_{\textrm{sil}}}{A_{\textrm{Si}}}f_{\textrm{sil}}+
X_{\textrm{Fe}}\frac{A_{\textrm{iro}}}{A_{\textrm{Fe}}}f_{\textrm{iro}}\right)
\end{equation}

\noindent where X$_{\textrm{Si}}$ and X$_{\textrm{Fe}}$ are the mass
fractions of the key elements involved (iron for iron dust grains and
silicon for silicates), A$_{\textrm{Si}}$ and A$_{\textrm{Fe}}$  the
atomic weights, and A$_{\textrm{sil}}$ and A$_{\textrm{iro}}$  the
mass numbers of one typical unit of dust for silicates and iron
respectively \citep{Zhukovska2008}. According to the dust types
  considered in \citet{Ferrarotti2006} we have that silicates includes
olivines/pyroxenes/quartz: $f_{\textrm{sil}}=f_{\textrm{ol}}+f_{\textrm{pyr}}+f_{\textrm{qu}}$
and the mean molecular weight of the mixture of silicates is
$A_{\textrm{sil}}=\left(A_{\textrm{ol}}f_{\textrm{ol}}+A_{\textrm{pyr}}f_{\textrm{pyr}}+A_{\textrm{qu}}
f_{\textrm{qu}}\right)/f_{\textrm{sil}}$. The total fraction of silicates is calculated following \citep{Ferrarotti2003}:

\begin{equation}\label{fsilicates}
f_{\textrm{sil}}=0.8\frac{\dot{M}}{\dot{M}+5 \times 10^{-6}}\sqrt{\frac{\epsilon_{\textrm{C,1}}-\epsilon_{\textrm{C}}}{\epsilon_{\textrm{C,1}}}}
\end{equation}

\noindent where we still need to specify $f_{\textrm{ol}}$, $f_{\textrm{pyr}}$ and $f_{\textrm{qu}}$.
According to \citet{Ferrarotti2001}, the mixtures depend on the ratio between the abundances
of Mg and Si that is about 1.06 for solar abundances \citep{Zhukovska2008}. For a typical M-star:
$f_{\textrm{ol}}/f_{\textrm{pyr}}$=4 and $f_{\textrm{ol}}/f_{\textrm{qu}}$=22 \citep{Marigo2008}. Finally, for the iron dust:

\begin{equation}\label{firon}
f_{\textrm{iro}}=0.5\frac{\dot{M}}{\dot{M}+5 \times 10^{-6}}
\end{equation}

\noindent \textbf{\textsf{S-stars}}. S-stars fall into the range
$0.90\leq \textrm{[C/O]} \leq 0.97$. With the  scarce oxygen available,
only  iron-dominated dust mixtures are possible. The situation is described  by Eqs.~(\ref{Ostarsdust})
 and (\ref{firon}). Once more, the silicates are grouped with the same ratios as for M-stars \citep{Ferrarotti2002},
 whereas the condensation fraction is lower than predicted by Eq.~(\ref{fsilicates}):

\begin{equation}\label{fsilicatesS}
f_{\textrm{sil}}=0.1\frac{\dot{M}}{\dot{M}+5 \times 10^{-6}}
\end{equation}

\noindent \textbf{\textsf{Carbon-rich C-stars}}. According to the scheme adopted,
we consider a carbon-rich environment of dust
formation when [C/O] $\geq 0.97$. When this occurs, the
formation of oxygen-rich dust ceases, replaced by carbon-rich compounds, and the C-star phase.
Thereinafter, the continuous formation of carbon-rich dust
makes the envelopes of these stars increasingly optically thick. By
losing mass at very high rates, these stars get enshrouded by thick envelopes
that absorb and scatter the UV-optical radiation to the IR and radio
wavelengths. According to \citet{Ferrarotti2006}, two types of dust are  present: carbonaceous grains, that are
the natural product of a carbon-rich environment, and silicon carbide (SiC). Indeed, almost all these
stars show an emission feature at 11.3 $\umu$m due to SiC, whose presence was predicted by \citet{Gilman1969}
and observationally confirmed by \citet{Hackwell1972}. Applying Eq.~(\ref{dusttogas}) to the C-stars  we get:

\begin{equation}\label{Cstarsdust}
\dot{M}_{d}=\dot{M}\left(X_{\textrm{C}}\frac{\textrm{A}_{\textrm{SiC}}}
{\textrm{A}_{\textrm{C}}}f_{\textrm{SiC}}+\textrm{X}_{\textrm{C}}f_{\textrm{car}}\right)
\end{equation}

\noindent with the obvious meaning of the symbols. The terms $f_{\textrm{car}}$ and $f_{\textrm{SiC}}$ are
evaluated following \citet{Ferrarotti2003}.  For $f_{\textrm{SiC}}$ we used Eq.~(\ref{firon}), while

\begin{equation}\label{fC}
f_{\textrm{car}}=0.5\frac{\dot{\textrm{M}}}{\dot{\textrm{M}}+5 \times 10^{-6}}\left(\frac{\epsilon_{\textrm{C}}-\epsilon_{\textrm{O}}}{\epsilon_{\textrm{O}}}\right)
\end{equation}

\noindent Once the dust-to-gas ratio is specified, we have all
parameters  entering Eq.~(\ref{tauquater}) for the optical depth. We then
proceed in the following way: given an AGB star (or an elementary interval of
the AGB isochrone)  with  (L,T$_{\rm eff}$) and the corresponding surface  element abundances, we calculate the
SED of the resulting dust-enshrouded object.  In brief, for every
AGB model or  evolutionary track/isochrone elementary interval we need:
L, T$_{\rm eff}$, $\dot{\textrm{M}}$, the [C/O]-ratio and the element
abundances at the surface, X$_{i}$. This fixes the optical depth
$\tau$ at the surface and  the mixture of dust formed in the envelope, that
in turn determine the extinction coefficients Q to be used in the
radiative transfer problem.  We can thus calculate the final SED to compare with observations.

\begin{figure}
\centering
\subfigure{
\includegraphics[width=0.4\textwidth]{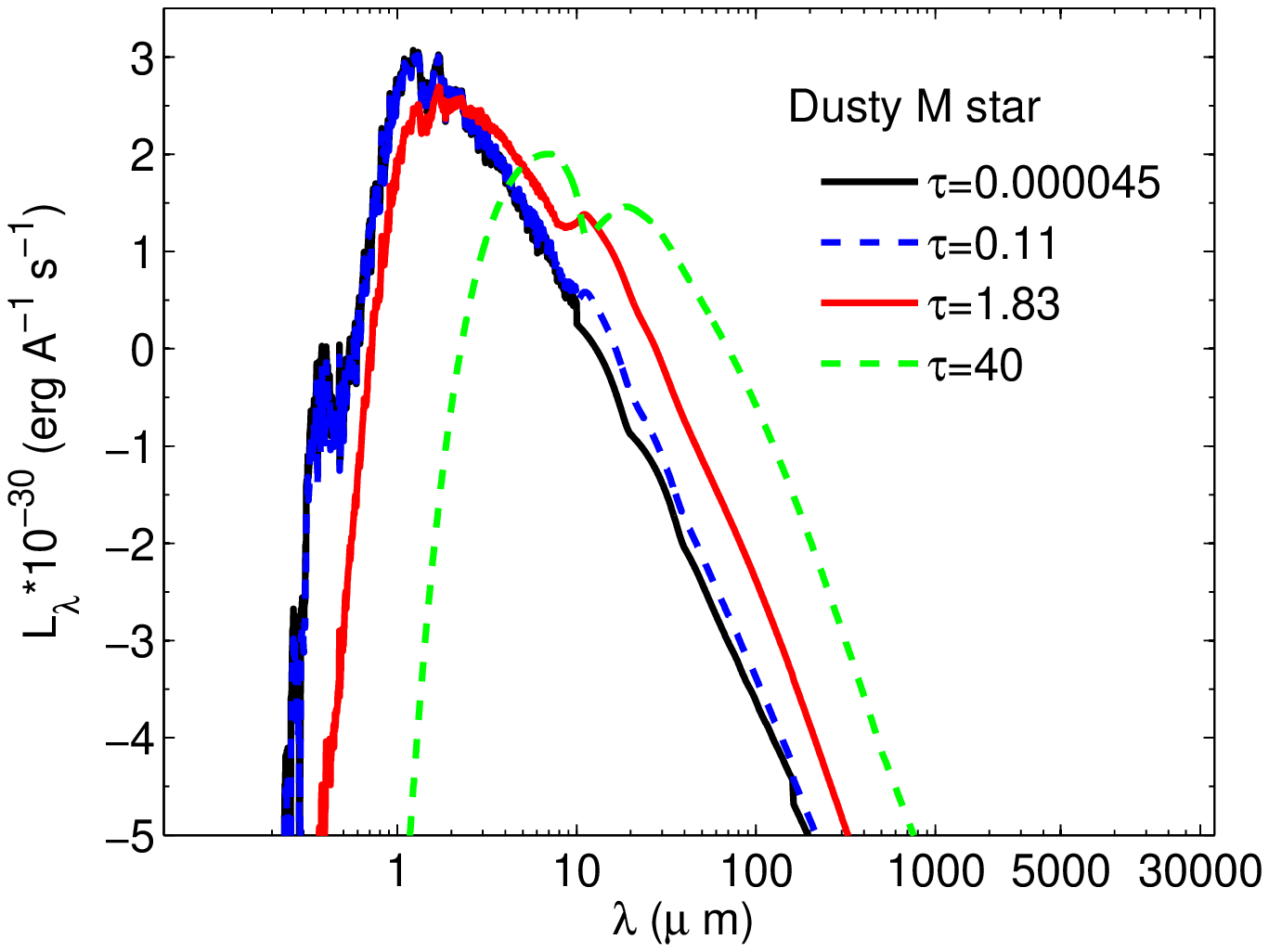}}
\subfigure{
\includegraphics[width=0.4\textwidth]{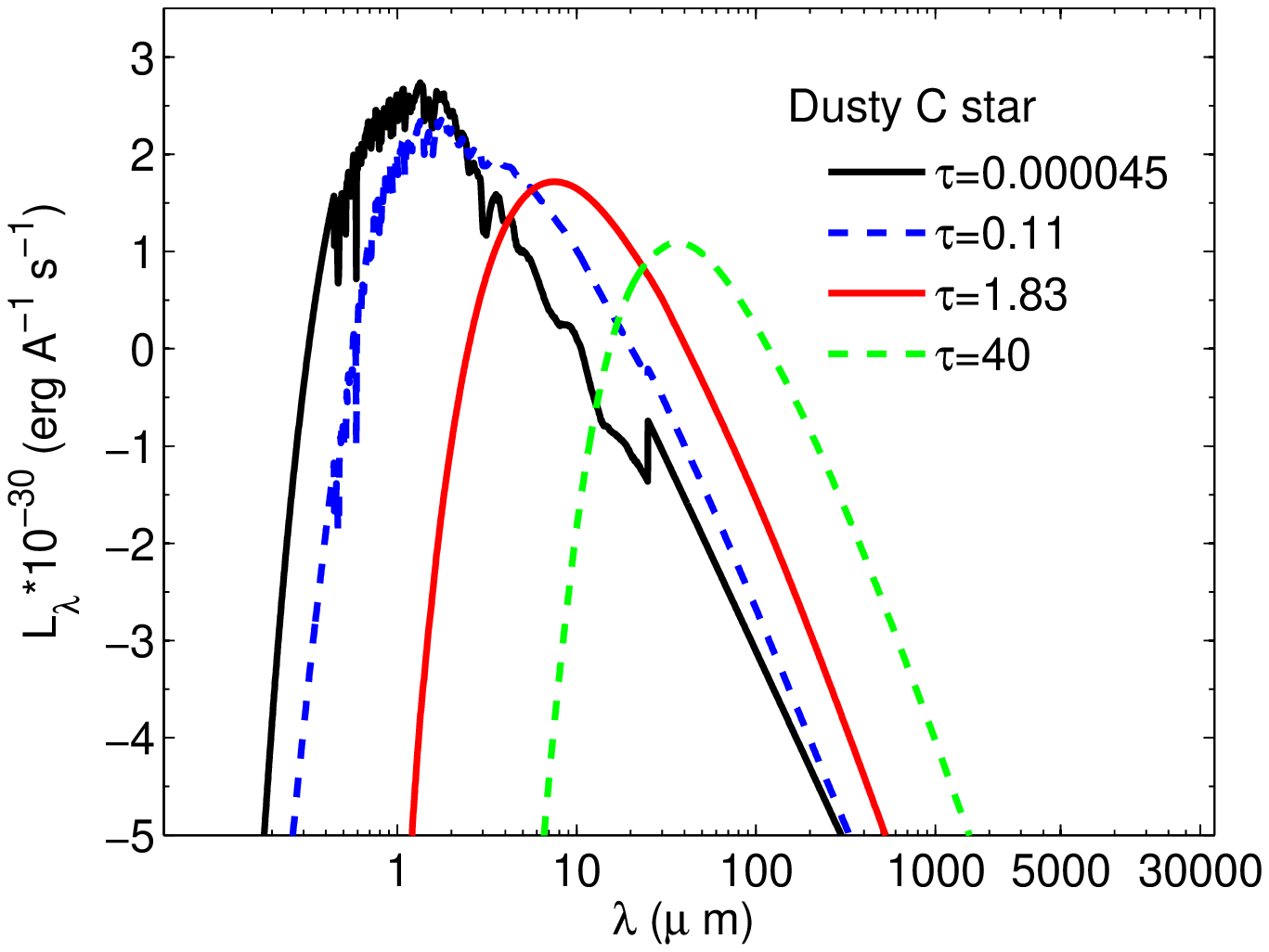}}
\caption{Dust-enshrouded  AGB spectra  for various optical depths for oxygen-rich M-stars
(\textbf{top panel}) and carbon-rich C-stars  (\textbf{bottom panel}).} \label{DustyStars}
\end{figure}

\begin{figure}
\begin{center}
{ \includegraphics[width=0.4\textwidth]{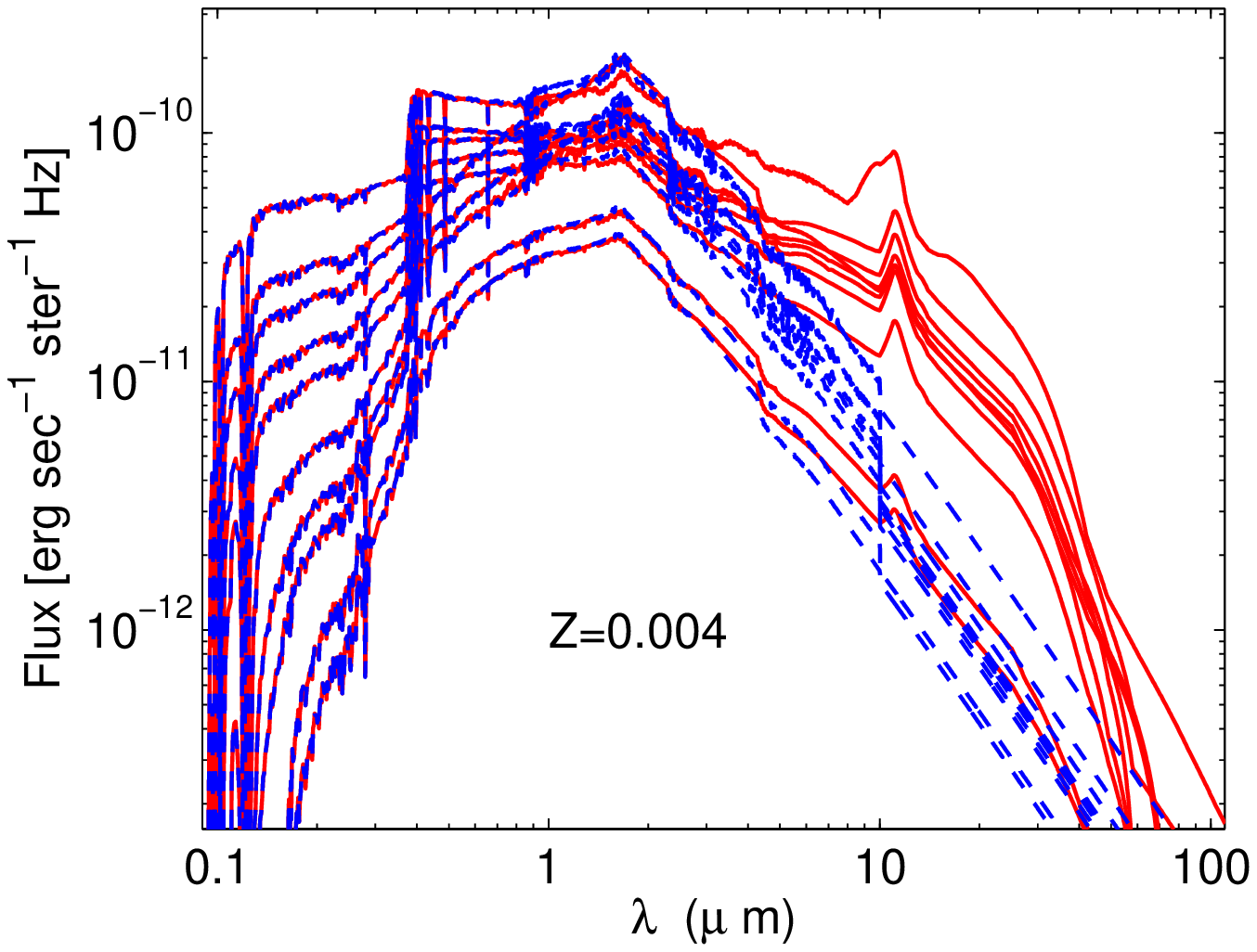} }
{ \includegraphics[width=0.4\textwidth]{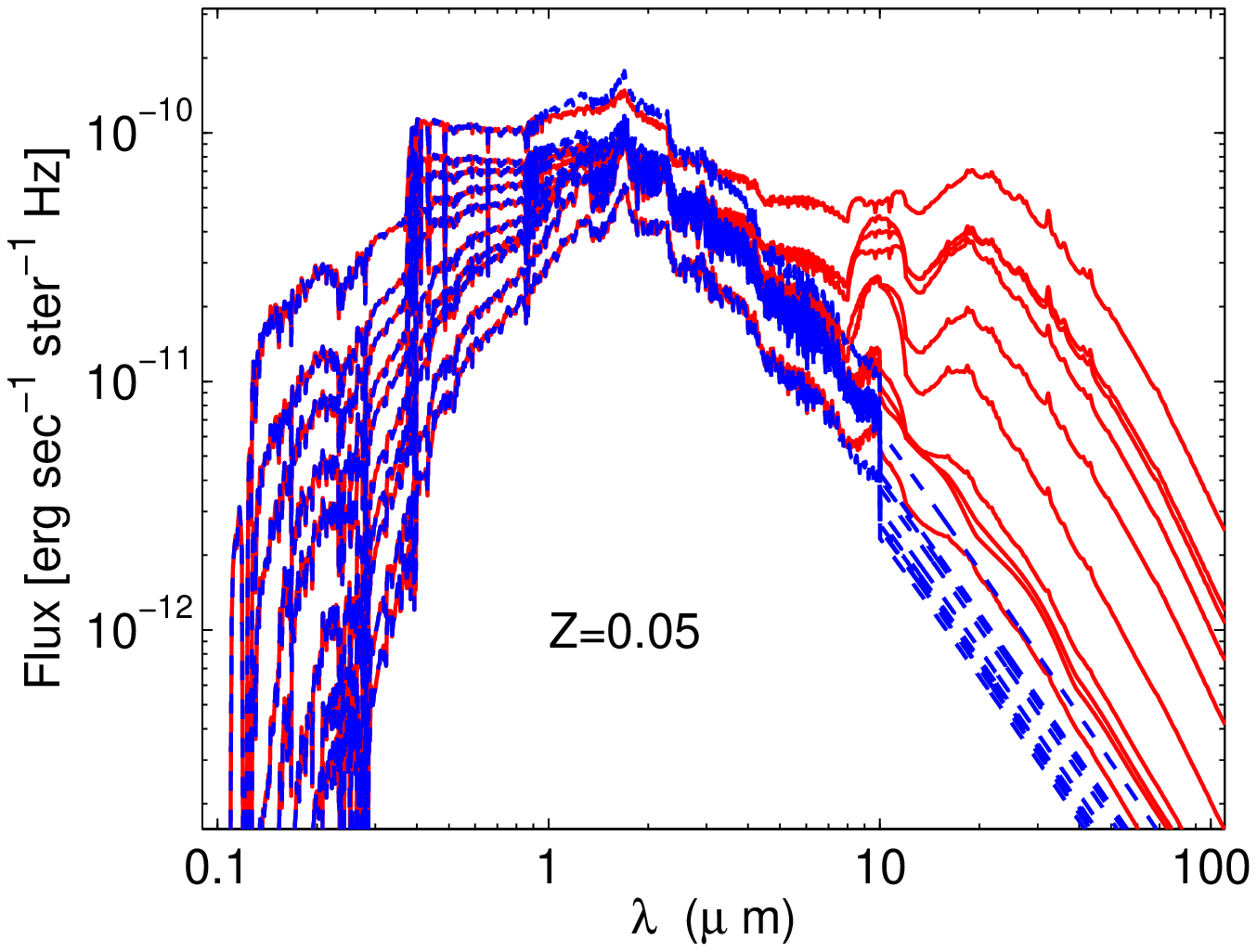}  }
\caption{SEDs  ($F_{\nu}$ vs. $\lambda$) for the SSPs with
 ages from 0.1 to 2 Gyr.  The case with dusty circumstellar envelopes around
 AGB stars are displayed as red solid
 lines, while results without dust are plotted with blue dotted
 lines.  Ages range from the oldest (bottom) to the youngest (top)
 values (2.0, 1.5, 0.95, 0.8, 0.6, 0.4, 0.325, 0.25, 0.15, 0.1 Gyr).
The metallicity is $Z$=0.004 (\textbf{top panel}) and  $Z$=0.05 (\textbf{bottom
panel}).}
\label{SSP_flux_123}
\end{center}
\end{figure}

\begin{figure}
\begin{center}
{
 \includegraphics[width=0.39\textwidth]{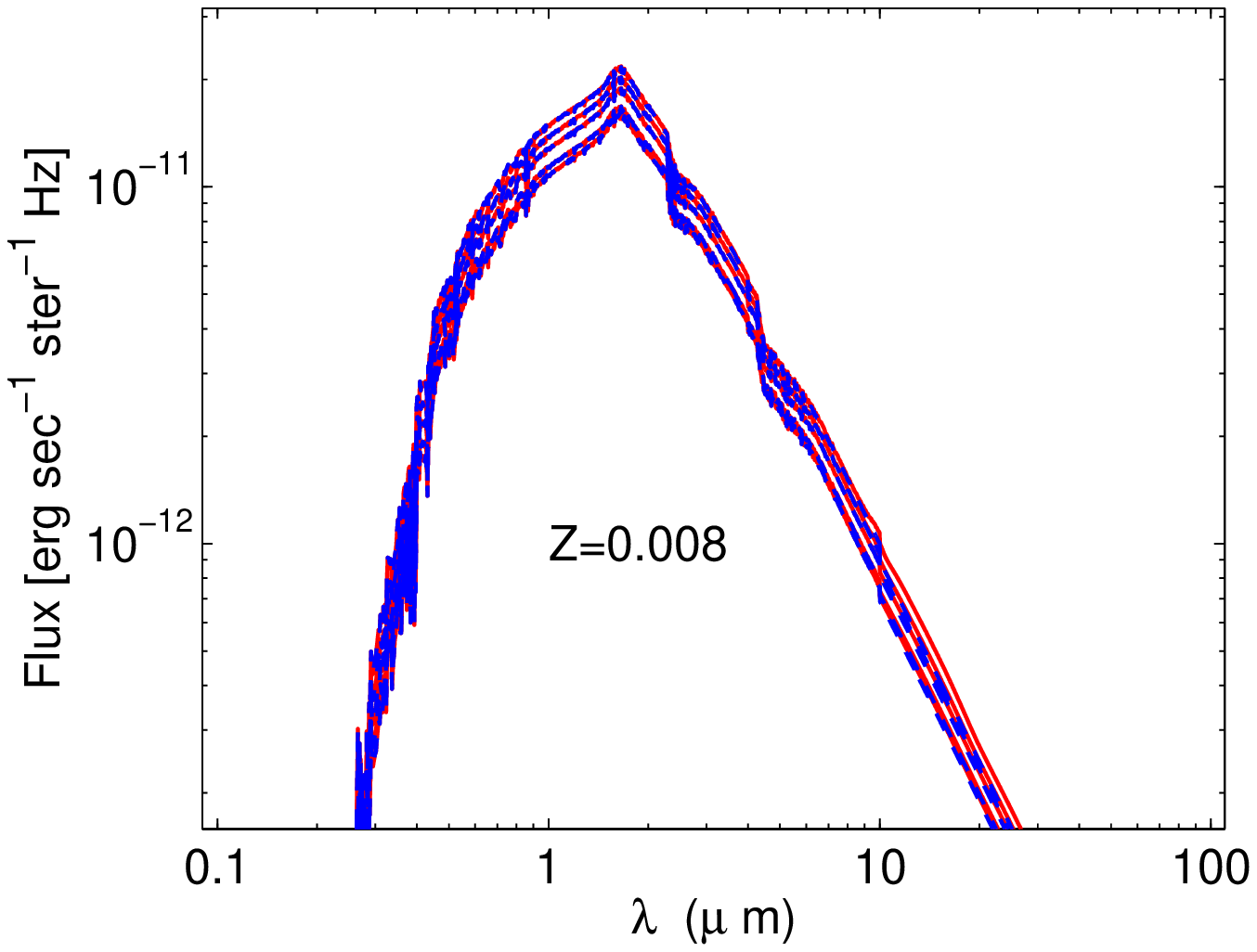}}
{\includegraphics[width=0.37\textwidth]{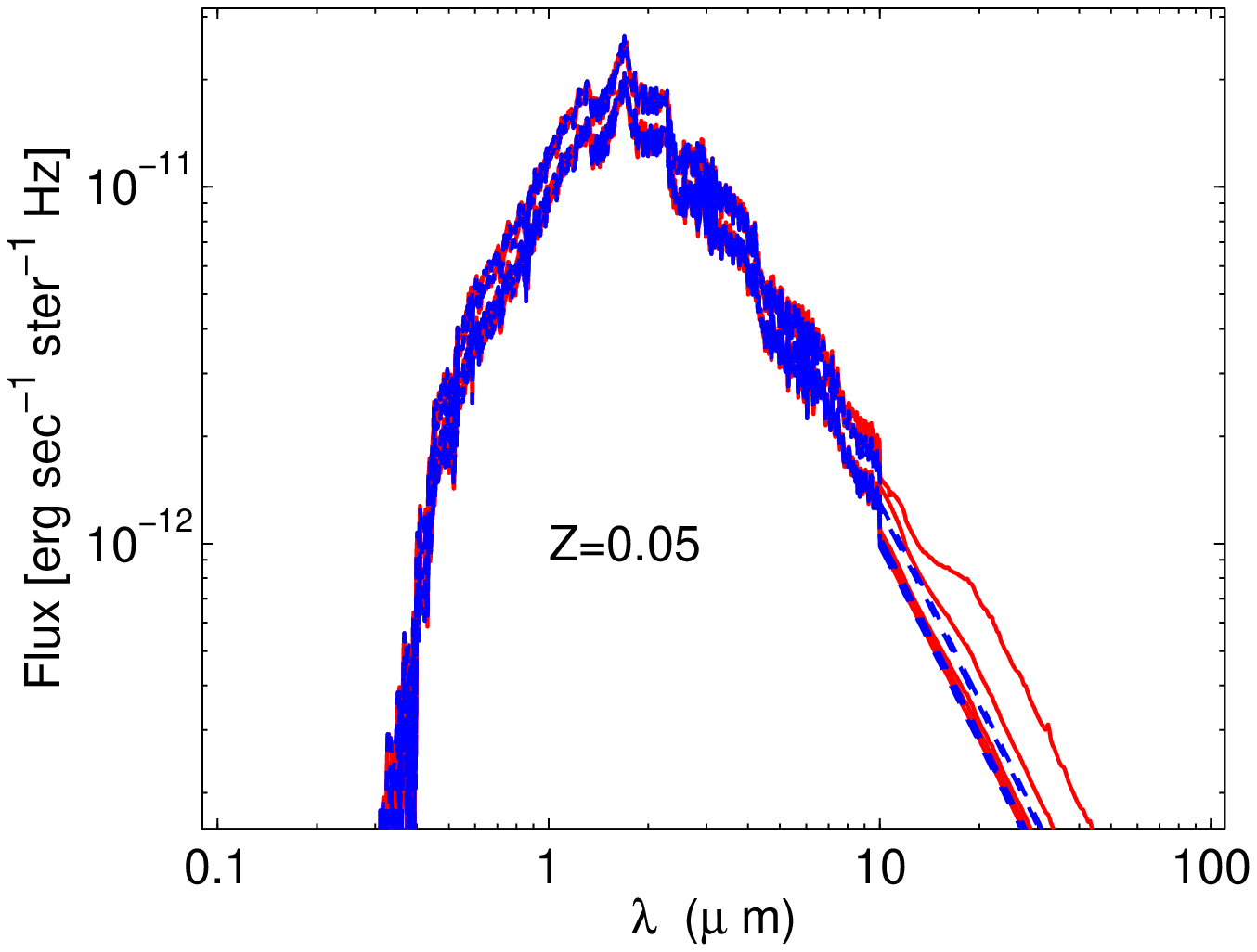}}
\caption{SEDs ($F_{\nu}$ vs. $\lambda$) for the SSPs with $Z$=0.008
  (\textbf{top panel})  and $Z$=0.05 (\textbf{bottom panel}).  Red solid lines
    correspond to models including dusty circumstellar envelopes,
  and  blue dotted lines to SSPs without dust. From
  bottom to top the displayed ages are: 6, 7, 8, 9 and 10 Gyr.}
\label{SSP_flux1b}
\end{center}
\end{figure}

\section{Theoretical spectra of O- and C-rich stars }\label{dustyspectra}

Our goal is to calculate spectra modified by the effect of the dust shells
around the AGB stars. The ideal approach would be to
generate for each AGB model the corresponding SED and use it to derive
magnitudes and colours in a given photometric system. However, this way
of proceeding that was occasionally adopted by  \citet{Piovan2003} is
very time-consuming. It requires solving the radiative transfer
problem on a star-by-star basis: it can be applied only if the number
of models is small. In the  present study, we follow a
  different approach. We first set up two libraries of dust-enshrouded
  AGB spectra, one for O-rich and the other for C-rich objects,
  that cover the full parameter space spanned by our AGB models.
  Interpolations among the library SEDs will provide the spectro-photometric
 properties of the AGB section of our isochrones. Each library contains 600
spectra. The parameters  have been grouped according to:
\begin{enumerate}
\item[-] \textbf{\textsf{The optical depth}}. $\tau$ is derived from
  Eq.~(\ref{tauquater}) using the appropriate physical parameters that describe the
  central star and the surrounding dust shell. For each group (C-stars and
  M-stars) we calculate 25 optical depths, going from 0.000045 to 40
  \citep{Groenewegen2006}, at a suitable reference wavelength of the
  MIR, for the chemical mixture that forms the dust.
\item[-]  \textbf{\textsf{The SED of the central star embedded in the
    dust-shell}}.  The total luminosity is not
  required, a normalized flux $\lambda \cdot \textrm{F}_{\lambda}$
  in some arbitrary units being sufficient. The following SEDs for
  the central stars are adopted: for the oxygen-rich stars we use the
  SEDs of the \citet{Lejeune1997} library, that includes also
  semi-empirical spectra of cool M-stars by \citet{Fluks1994};
  for the C-stars we select a suitable number of SEDs from the
  \citet{Aringer2009} models of dust-free C-stars. For the  M-stars,
  we adopt six values of the temperature (2500, 2800, 3000, 3200,
  3500, and 4000 K), but no specification is made for the
  gravity, because the sample of \citet{Fluks1994} contains empirical
  spectra. For the library of C-stars we adopt six values of
  temperature (2400, 2700, 3000, 3200, 3400, and 3900 K), see also
  \citet{Aringer2009}. We  consider two values for the
        [C/O]-ratio, namely [C/O] = 1.05 and [C/O] = 2. Finally, for the
        input mass, gravity, and metallicity we  use  M = 2
        M$_{\odot}$,  $\log$ g = 0.0, and Z=Z$_\odot$.
\item[-] \textbf{\textsf{The composition of the dust in the outer envelope}}. \\
 \textit{(i) C-stars}. Several types of dust grains in carbon-rich AGB
 stars have been detected by observations: the three main types are
 amorphous carbon (AMC), silicon carbide (SiC), and magnesium sulphide
 (MgS). In our models the presence of MgS has been
   neglected. MgS has been first proposed as a candidate to explain the
   30$\umu m$ feature in evolved C-stars by \citet{Goebel1985},
   and this hypothesis has been strengthened by theoretical and
   observational analyses \citep[see][ for more details]{Zhukovska2008b}.
However, according to recent studies, to account for the feature in a typical C-rich evolved 
object one would require a much higher MgS mass
than available \citep{Zhang2009}. Also, MgS causes a mismatch between predicted and observed
spectral feature \citep{Messenger2013}. In addition,
the 30$\umu m$ feature is not ubiquitous: it is difficult to determine the ranges of stellar mass and mass-loss
   where the feature should be included \citep{Zhukovska2008b}, and
   therefore, in conclusion, we decide to ignore MgS.
   With respect to
   AMC and SiC we rely on the results by \citet{Suh2000}, who derived
   new opacities for the AMC that are consistent with the
   Kramers-Kronig dispersion relations and reproduce the observational
   data. The models improve upon previous studies
   \citep{Blanco1998,Groenewegen1998} and are characterized by two
   components, SiC and AMC. AMC and SiC influence the outgoing
 spectrum in different ways: whilst the effects of AMC propagate over
 the whole spectrum, those of SiC are limited to the 11 $\umu$m
 feature, as indicated by the observations.  \citet{LorenzMartins1994}
 and \citet{Groenewegen1995} suggest that the ratio SiC to AMC
 decreases at increasing optical depth of the dusty
 envelope. According to \citet{Suh1999}, for optically thin
 dust-shells ($\tau_{\textrm{10}} \leq \textrm{0.15}$, where
 $\tau_{\textrm{10}}$ is the optical depth at 10 $\umu$m) the strong
 $\textrm{11}$ $\umu$m feature requires about $\textrm{20} \%$ of SiC
 dust grains to fit the observational data; for dust-shells with
 intermediate optical thickness ($\textrm{0.15} \leq
 \tau_{\textrm{10}} \leq \textrm{0.8}$) about $\textrm{10} \%$ SiC
 dust grains are needed, whereas for shells with larger optical
 depths, where the 11 $\umu$m feature is either much weaker or missing
 at all, no SiC is required. The optical constants of $\alpha
 \textrm{SiC}$ by \citet{Pegourie1988} are adopted to calculate the
 opacity of SiC, and according to the above considerations we take two
 extreme compositions: the first one has 100$\%$ AMC only, whereas the
 second one has 80 $\%$ AMC and 20$\%$ SiC. The reference optical
 depth has been chosen at 11.33 $\umu$m for the 100$\%$ AMC mixture,
 and at 11.75 $\umu$m for the 80 $\%$ AMC and 20$\%$ SiC mixture
 \citep{Groenewegen2006}. \\
 \textit{(ii) M-stars}. In the circumstellar environment of M-stars a
 wide number of dust grains is formed, and a condensation sequence has
 been proposed by  \citet{Tielens1990}. At increasing  mass-loss the
 dust composition changes  from aluminium and magnesium
 oxides rich at low $\dot{\textrm{M}}$, to a mixture with both oxides and
 olivines, and finally to a composition dominated by the silicates,
 with amorphous silicates and crystalline silicates at high
 $\dot{\textrm{M}}$. This sequence seems to be able to reproduce the
 changes observed in the shape of the 10 $\umu$m feature. Even if this
 scheme is still a matter of debate \citep{vanLoon2006}, it is
   consistent  with the observations of different types of stars
   at different metallicities
 \citep{Dijkstra2005,Heras2005,Lebzelter2006,Blommaert2006}. We adopt
 the above sequence as a plausible scenario for the condensation of
 dust in oxygen-rich stars. Three possible compositions are included:
 (1) pure Al$_{2}$O$_{3}$ with optical properties taken from
 \citet{Begemann1997}; (2) mixed composition with 60$\%$
 Al$_{2}$O$_{3}$ and 40$\%$ silicates, with the optical properties
 taken from \citet{David1995}; (3) 100$\%$ silicates for high
 mass-loss rates, with two possible choices, i.e.  either a complete
 composition with optical properties from \citet{David1995} for
 comparison with \citet{Groenewegen2006}, or a more elaborate
 description based upon \citet{Suh1999,Suh2002}. The latter author
 adopted different silicates opacities at varying 10 $\umu$m feature,
 namely cold and warm silicates. The model then has been refined by
 \citet{Suh2002} taking into account crystalline silicates through the
 so-called crystallinity parameter $\alpha$, because in many AGB stars
 with high mass-loss rates, ISO hi-resolution observations reveal the
 presence of prominent bands of crystalline silicates, like enstatite
 ($\textrm{MgSiO}_{3}$) and forsterite
 ($\textrm{Mg}_{2}\textrm{SiO}_{4}$) \citep{Waters1996,Waters1999}.
 The adopted opacity functions for these latter are taken from
 \citet{Jaeger1998}.  Following \citet{Piovan2003}, we adopt here
 $\alpha = 0.1$ for stars with low mass-loss rates and moderately
 optically thick shells ($\tau_{10} <$ 15), whereas for oxygen-rich
 stars with high mass-loss rates and very thick shells ($\tau_{10} >$
 15) we prefer the value $\alpha = 0.2$. Finally, in all models the
 relative contents of enstatite $\left( \textrm{MgSiO}_{3}\right)$ and
 forsterite $\left( \textrm{Mg}_{2}\textrm{SiO}_{4}\right)$ are the
 same as in \citet{Suh2002}.
Finally, we take also into account the recent results by
\citet{McDonald2011}. They found that metallic iron seems to dominate
the dust production in metal-poor oxygen-rich stars. We therefore
adopt a 100$\%$ iron mixture to simulate the envelope of
metal-poor stars surrounded by a thin dust shell. The optical
properties of iron are taken from \citet{Ordal1988}. The following
reference wavelengths are adopted for the grid of $\tau$: 11.75
$\umu$m for both pure aluminum oxides and oxides plus silicates, and
10.20 $\umu$m for both pure silicates cases. All the above opacities are used as input for DUSTY. 
This radiative transfer code then applies the Mie theory
to calculate scattering and absorption efficiencies by a homogeneous spherical sphere. 
The grain size distribution is chosen between the options available in DUSTY
\citep{Ivezic1997}. In particular, we adopt single size grains with dimension a=0.1$ \umu m$ 
\citep[see][ for more details about this choice]{Piovan2003}. An analytical
dust density
profile, suitable for the modelling of AGB stars, is also selected from the available options.
This profile is appropriate in most
  cases and offers the advantage of a much reduced computational  time (see DUSTY
  manual at the url
  \textit{http://www.pa.uky.edu/$\sim$moshe/dusty/}). The
  envelope expansion is driven by radiation pressure on the dust
  grains.
\item[-] \textbf{\textsf{The temperature T at the inner boundary of
    the dust shell}}.
For this parameter we assume either 1000 K or 1500 K, depending on the
type of dust \citep{Piovan2003}.
\item[-] Finally, we comment on the luminosity of the central
  star. For C-stars we keep the luminosity specified by
  \citet{Aringer2009}.  Given that \citet{Fluks1994} does not specify
  the luminosity of the M-stars producing the empirical spectra, but
  gives only the specific intensity, we fix the luminosity of the
  M-stars at L=3000L$_{\odot}$.  The library of dusty stellar spectra
  is therefore calculated for a fixed luminosity of the underlying
  objects. This is not a problem, for the luminosity does not affect
  the solution of the radiative transfer \citep{Ivezic1997} and the
  shape of the outgoing SEDs. The resulting flux is scaled to the real
  luminosity of the AGB star we are considering.
\end{enumerate}

\begin{figure}
\begin{center}
 \subfigure
{
 \includegraphics[width=0.4\textwidth]{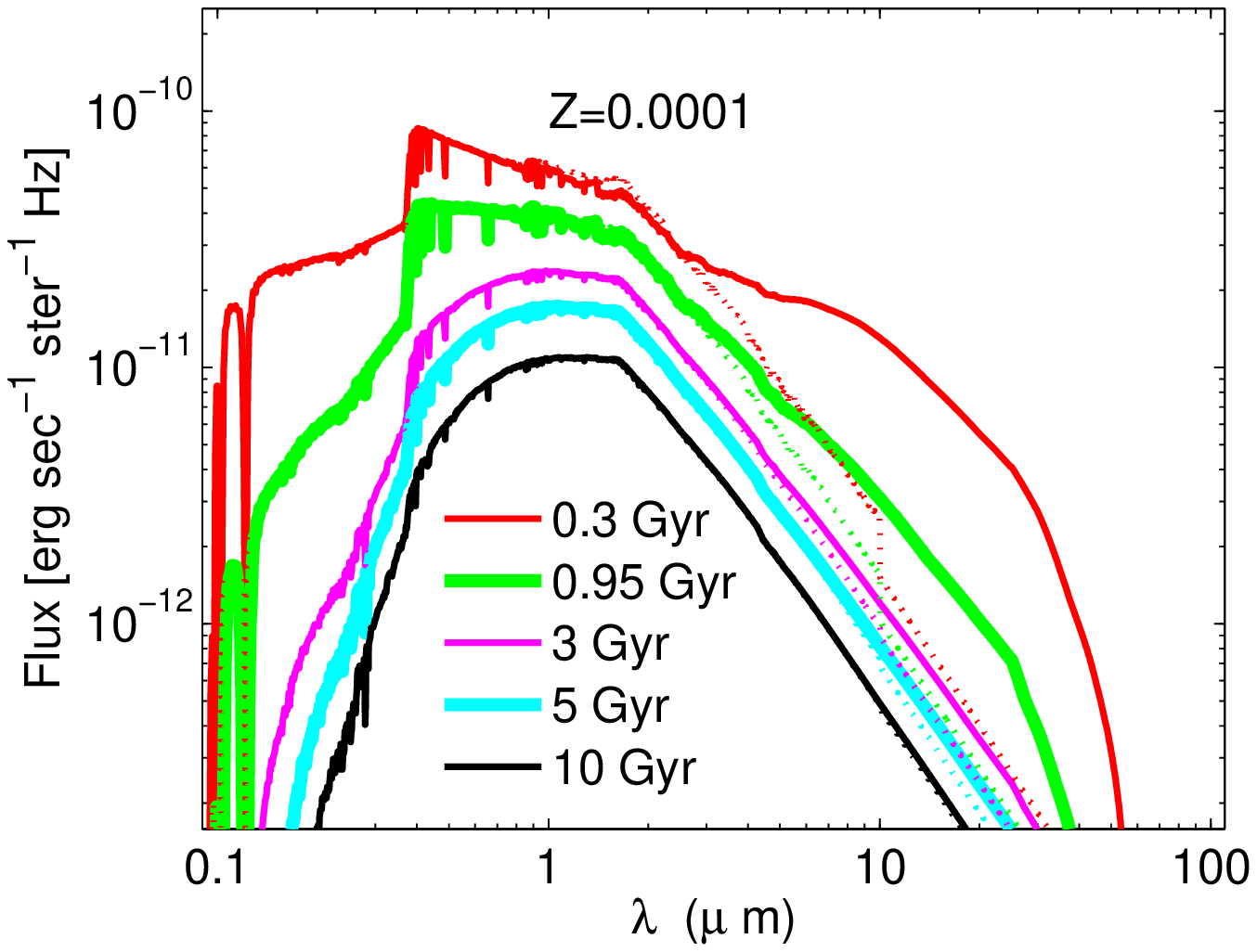} 
 }
\subfigure
{
 \includegraphics[width=0.4\textwidth]{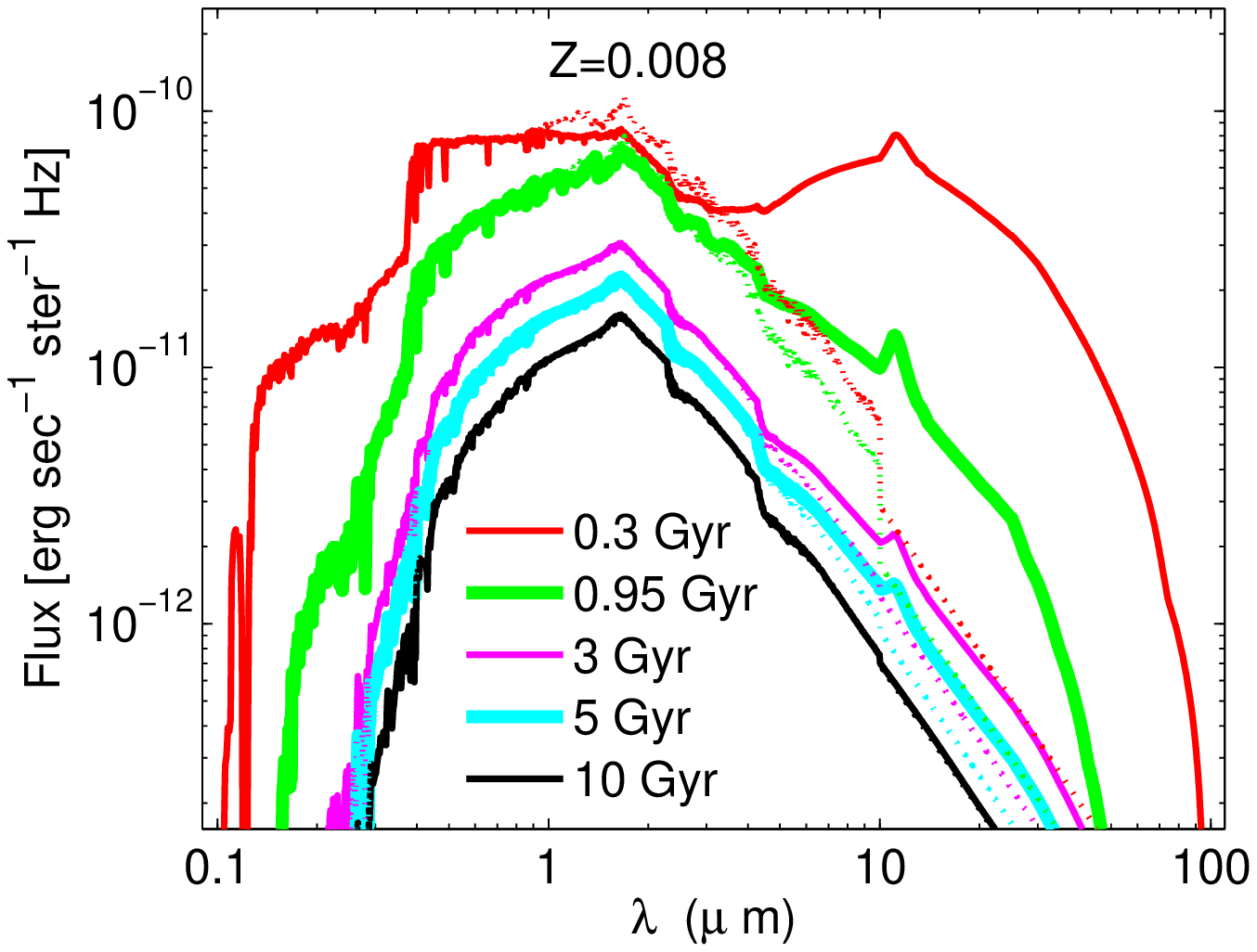}
 }
\subfigure
{
 \includegraphics[width=0.4\textwidth]{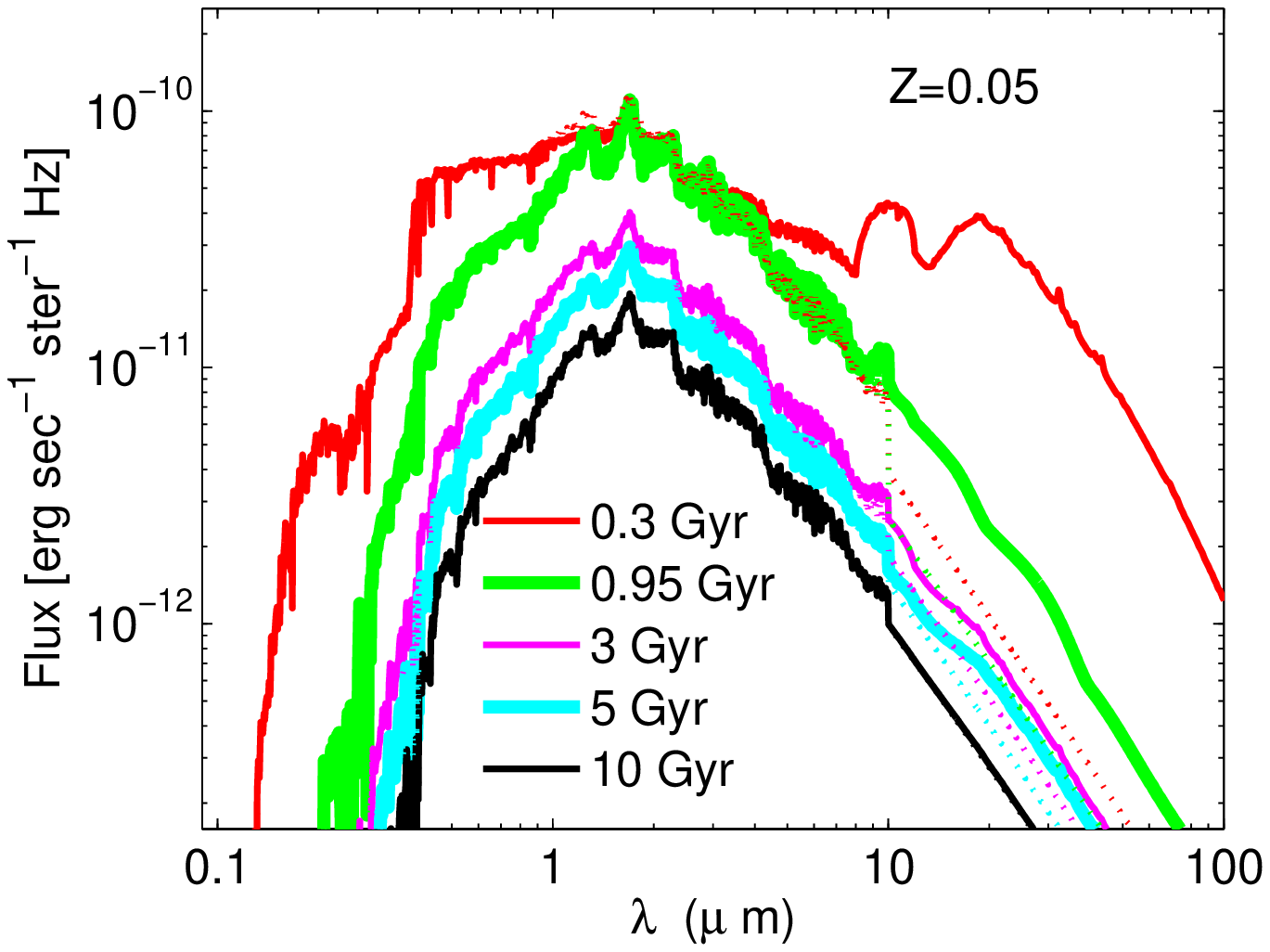}
 }
\caption{Detailed comparison of SEDs for old SSPs (dotted lines) and new SSPs (solid lines), for
the labelled ages and metallicities. Three  metallicities are
shown:  $Z$=0.0001 (\textbf{top panel}), $Z$=0.008 (\textbf{middle panel}) and  $Z$=0.05 (\textbf{bottom  panel}).}
\label{SSP_flux_some_123}
\end{center}
\end{figure}

Figure~\ref{MstarsTau_123} displays the SEDs of
O-rich AGB stars of our template library, for four values of the
optical depth. The input parameters are: T$_{\rm
  eff}$=2500 and L=3000L$_{\odot}$ of the central star, optical depths $\tau$=0.0224,
0.2081, 1.306 and 30.0, and dust composition according to the second
prescription for oxygen-rich stars (60$\%$ Al$_{2}$O$_{3}$ and 40$\%$
silicates). The lower-left panel shows the results for two different
gravities ($\log$ g=-1.02 and $\log$ g=0.5) of the input spectrum from
the \citet{Fluks1994} compilation. For the sake of clarity, the two
SEDs are artificially shifted by a small amount, otherwise the two
spectra would be coincident. Indeed, as expected and tested, there is no dependence on the gravity in the
\citet{Fluks1994} spectra. The stellar features in the UV-optical-near IR region disappear
with increasing optical depth, and an increasingly featureless SED appears: this is ultimately due
to the smooth optical properties of the selected composition. The
stellar light is shifted more and more toward longer wavelengths;
for the lowest optical depths
the input and output spectra are almost coincident, while the effect of dust is apparent
for the largest values of $\tau$.
In the lower panels we show: the fractional
contribution of the attenuated input radiation to the total flux (labelled as Att -- solid lines);
the fractional contribution of the scattered
radiation to the total flux (labelled as Ds -- dot-dashed lines), and finally the
fractional contribution of the dust emission to the total flux (labelled as De -- dotted lines).
As expected, at increasing optical depth $\tau$ (i) the fraction
 of not attenuated or scattered light escaping the dust-shell decreases, and (ii) the dust contribution becomes
significant at $\tau \sim$ 0.2 and dominant at $\tau \sim$ 1.\\
\indent Finally, Fig.~\ref{DustyStars} shows a sequence of
obscured spectra of AGB stars at increasing optical depth, both for
O-rich (top panel) and C-rich (bottom panel) stars. It is evident
  how the SEDs progressively shift towards longer wavelengths
  with increasing $\tau$. The spectra with 100$\%$ AMC dust composition represent a limiting case with no
SiC feature at 11.3 $\umu$m. It is worth noticing that for the oxygen-rich stars,
	when the optical depth is very high, the silicate feature at 9.7 $\umu$m appears in  absorption as indicated
	by the observational data \citep{Suh1999,Suh2002}.


\section{SEDs and colours of SSPs with dust enshrouded AGB Stars}\label{SSPs_dustyAGB}

With our new isochrones and library of dust-enshrouded
AGB stars we have calculate the SEDs of SSPs.
Figure~\ref{SSP_flux_123} displays the SEDs with (red lines) and
without (blue lines) dust-enshrouded  AGB stars for different ages and metallicities
$Z$=0.004 and $Z$=0.05, respectively.  Ages
  range from 0.1 to 2~Gyr and correspond to young and
intermediate-age SSPs. The effect of the dust-enshrouded AGB stars is
remarkable and cannot be neglected in the region from NIR to
FIR. Figure~\ref{SSP_flux1b} shows the
  SEDs for old ages from 6 to 10 Gyr, metallicities  $Z$=0.008 and
$Z$=0.05, respectively.
For old ages the effect of dust-enshrouded
AGB stars is small, mainly because of the short duration
  of the AGB phase (only a few thermal pulses) due to
  of the low total mass along the AGB. Only for
	the highest metallicity $Z$=0.05  the dust surrounding AGB stars has some effect on the SED.
Another example of the effect of the dust-shells around
AGB stars is shown by Fig.~\ref{SSP_flux_some_123}, that compares
the new and old SEDs for a few selected ages and
 for three metallicities, e.g. $Z$=0.0001, $Z$=0.008 and
$Z$=0.05. Similar results are found for all the remaining
metallicities.\\
\indent The old spectra without dust shells do not extend into the NIR and
FIR, but decline sharply at wavelengths longer than about 3-4 $\umu$m. The spectra
 of the new SSPs, instead, extend towards long wavelengths, and the amount of flux in the MIR and FIR is significant.
Differences start at about 1 $\umu$m; in the IR range up to 3-4 $\umu$m the flux
of dusty SSPs is lower than the old one, due to the fact that dusty envelopes shift
the emission of M- and C-stars towards longer wavelengths. It is worth noticing the evolution of
the features of silicon carbide at 11.3 $\umu$m, and amorphous silicate at 9.7 $\umu$m, at
 different metallicities. The amount of energy shifted to longer wavelengths is larger for the young ages, because of
more massive and luminous AGB. Considering the different metallicities, we note the following effects:

\begin{figure}
\begin{center}
{\includegraphics[width=0.41\textwidth]{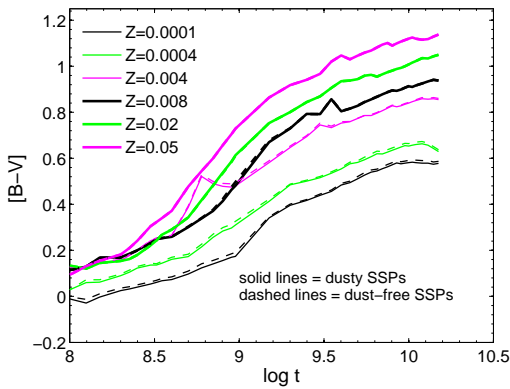}}
\caption{Integrated $[\textrm{B}-\textrm{V}]$ colour of SSPs as a function of age, in the range from 0.1 to 15 Gyr,
for the whole metallicity grid. Solid lines denote
the SSPs with \textit{dust-free AGB stars}, dashed lines colours including
\textit{dust enshrouded AGB} stars. The unit of time $t$ is yr.}\label{ssp_coloursBV}
\end{center}
\end{figure}

\begin{figure}
\begin{center}
{\includegraphics[width=0.41\textwidth]{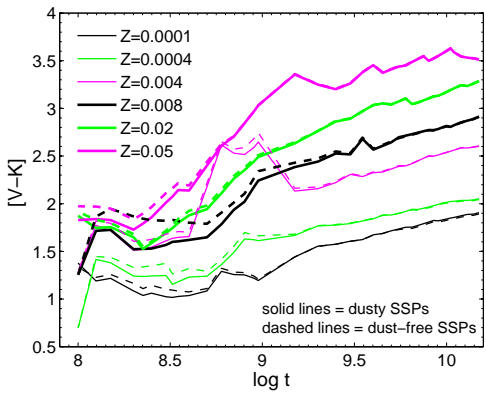}}
\caption{As in Fig.~\ref{ssp_coloursBV}, but for the integrated $[\textrm{V}-\textrm{K}]$ colour.}
\label{ssp_coloursVK}
\end{center}
\end{figure}

\begin{figure}
\begin{center}
{\includegraphics[width=0.41\textwidth]{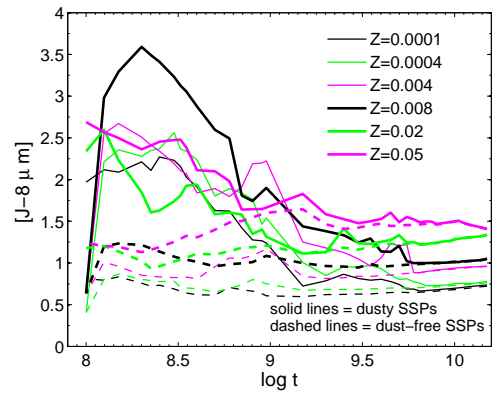}}
\caption{As in Fig. \ref{ssp_coloursBV}, but for [\textrm{J} - \textrm{8} $\umu$m]} 
\label{ssp_coloursJ8}
\end{center}
\end{figure}


\begin{figure*}
\begin{center}
{\hspace{-2mm}
 \includegraphics[width=0.33\textwidth]{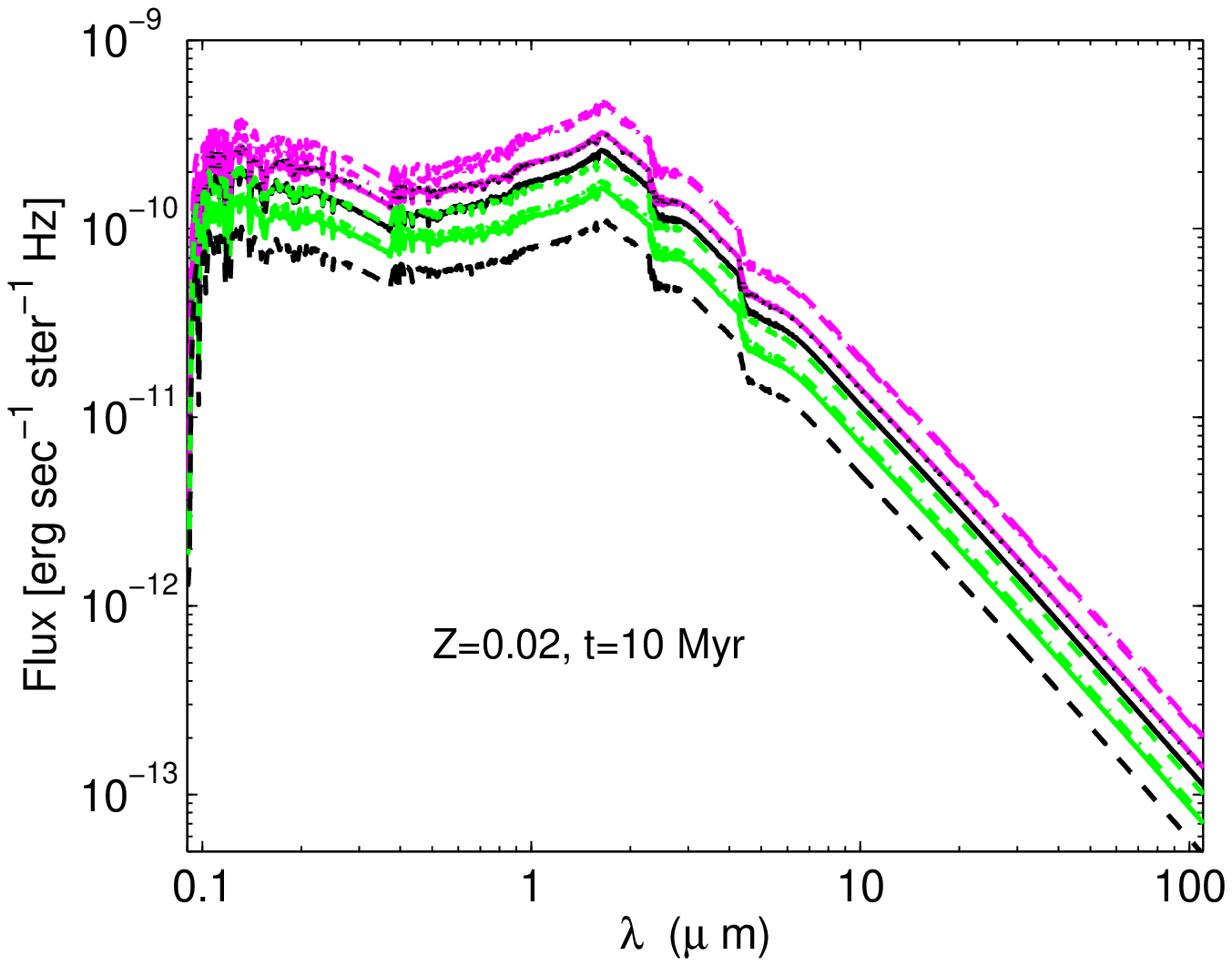}
 \hspace{-3mm}   
 \includegraphics[width=0.33\textwidth]{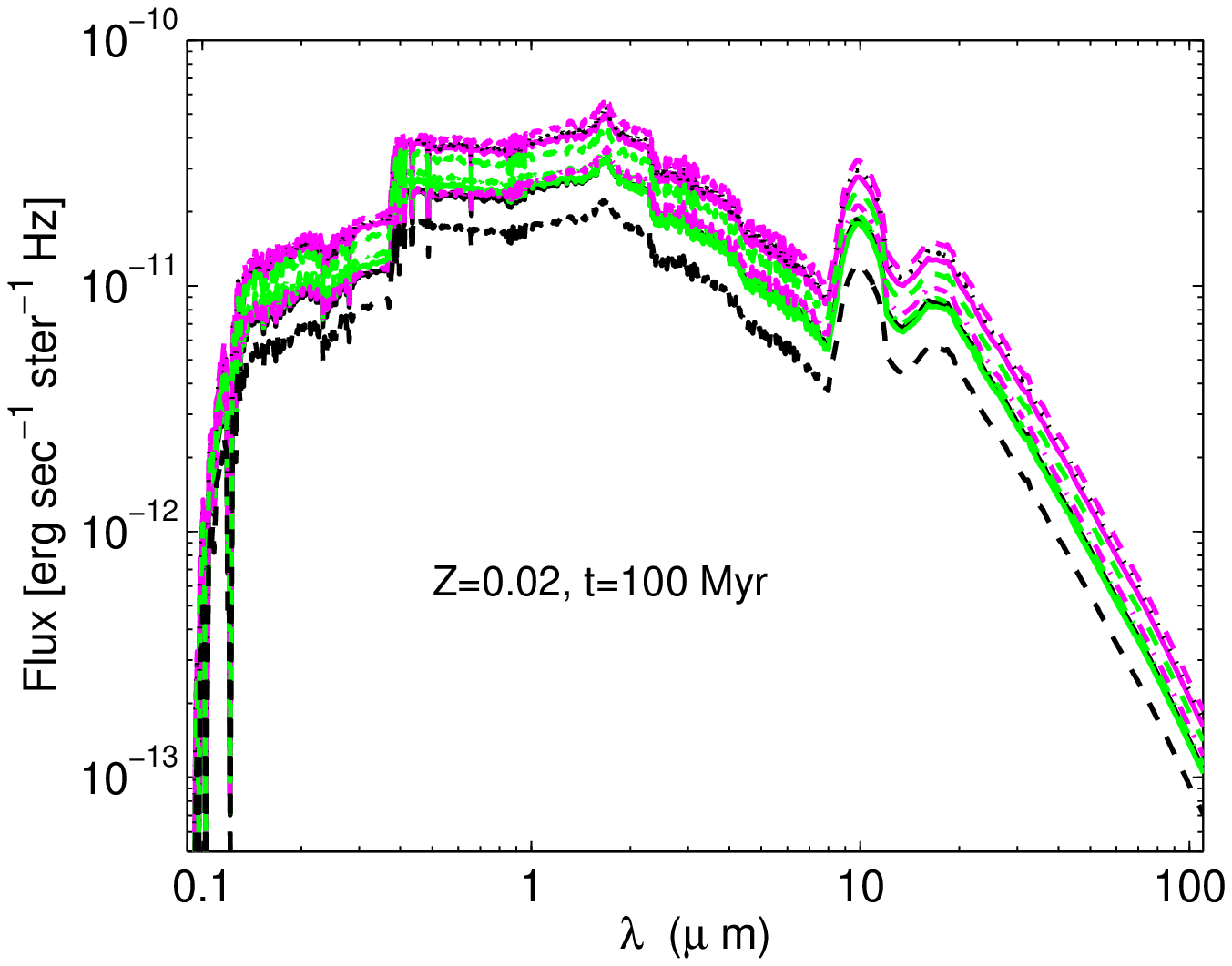}
 \hspace{-3mm} 
 \includegraphics[width=0.33\textwidth]{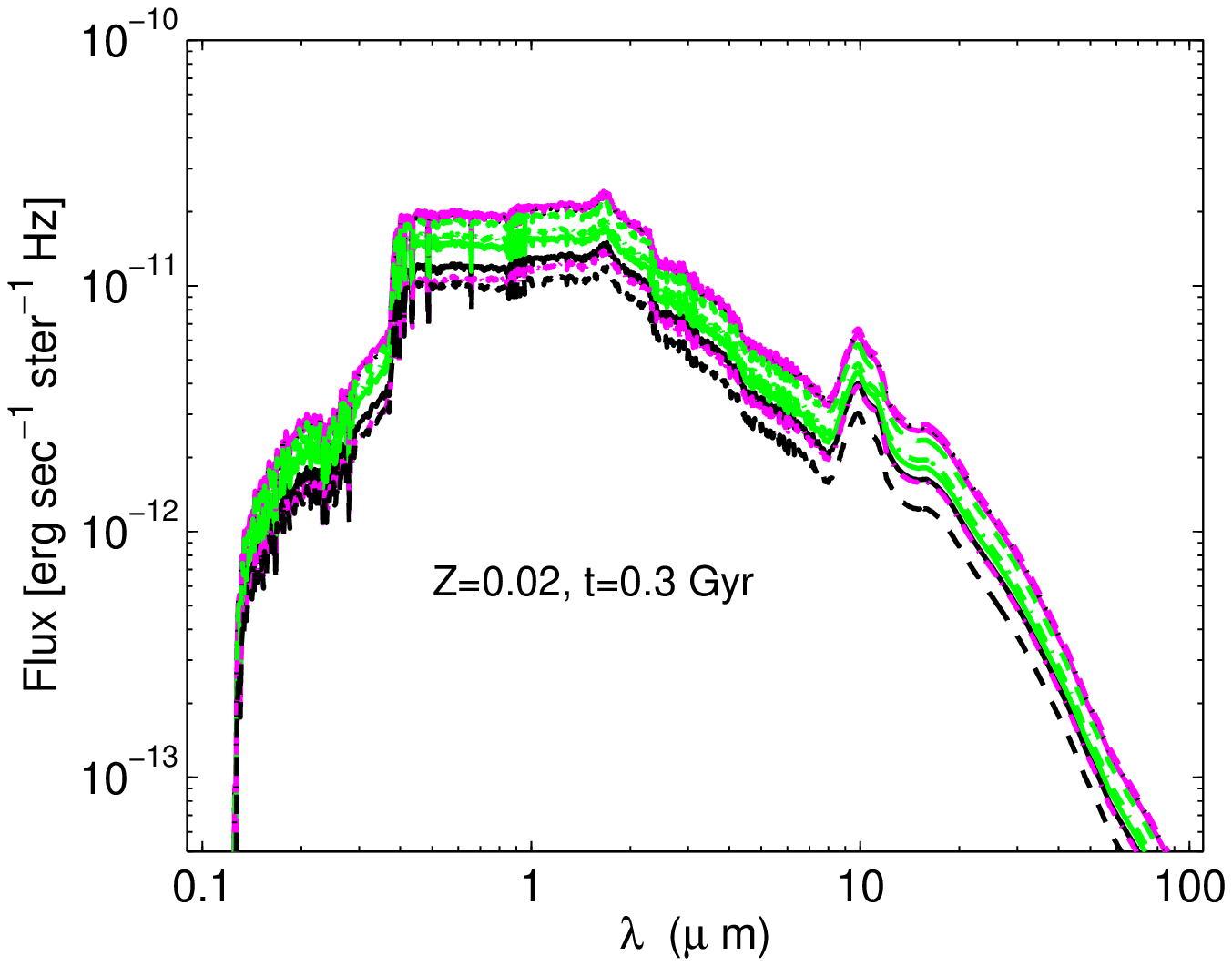}}
\caption{Comparison between the new dusty SEDs,
at varying age and IMF. The metallicity is 
$Z$=0.02. Three ages are considered: t=10 Myr (\textbf{left panels}), t=100 Myr 
(\textbf{middle panels}), and
t=300 Myr (\textbf{right panels}). Each line corresponds to a different IMF, 
according to the line-style and colour code of Fig.~\ref{imf_plot}.
Very similar curves are found for all other metallicities.}
\label{IMF_age}
\end{center}
\end{figure*}

\begin{description}
\item[-] $Z$=0.0004 and $Z$=0.0001: for stars of very low metallicity
  the TDU is particularly efficient during the AGB phase in enriching
  the surface with $^{12}$C and other products of He-burning. These
  stars display a C-rich surface for most of their evolution. The
  presence of dusty C-rich stars in young SSPs leads to featureless
  SEDs dominated by amorphous carbon. The small amounts of metals in
  the envelope inhibit high optical depths, and the amount of
  radiation re-emitted in the MIR/FIR region is smaller than in case
  of higher metallicities.
\item[-] $Z$=0.004 and $Z$=0.008: for most of the age range covered by
  the models, the spectrum does not show the features due to amorphous
  and crystalline silicates, because C-stars dominate (the 11.3
  $\umu$m feature of SiC is indeed prominent). This is quite
    different from the results by \citet{Piovan2003}, where the
    oxygen-rich phase at these metallicities played an important role,
    and a significant evolution of the MIR features was present. In
    \citet{Piovan2003}, C-stars dominated for young ages, whereas for
    intermediate ages (around 3~Gyr), the O-stars of low optical depth
    contributed to the integrated spectrum and the 9.7 $\umu$m feature
    could be seen in emission. Finally, for even older ages the high
    optical depth O-stars dominated, such that the spectrum became
    more articulated and the features due to crystalline silicates
    started to appear in the IR. All this does no longer occur (for
  metallicities in this range), simply because the evolution of the
  [C/O]-ratio in the \citet{Weiss2009} models of AGB stars leads to a
  different path.
Furthermore, it must be noted that, compared to
\citet{Piovan2003}, the amount of flux shifted towards
longer wavelengths is generally smaller. This is due to the lower optical
depths, now more realistically linked to the composition and
mass-loss of the underlying star. Finally, the present optical
depths agree well with those of  \citet{Groenewegen2006} for
similar mass-loss rates.
\item[-] $Z$=0.02: for the youngest ages, C-stars appear in a narrow luminosity range (Fig. \ref{agb_123}), and
  the stars at the AGB tip, with the highest mass-loss rate and the
  highest optical depth, are O-stars. The SSP spectrum is dominated by the 9.7 $\umu$m feature, that
	appears in emission and not in absorption, as expected in envelopes with rather small optical depths.
	No features of crystalline silicates show up in the spectrum, because according to the models by \citet{Suh2002} much
higher optical depths would be required.
\item[-] $Z$=0.05: AGB stars of super-solar metallicities  display
  only oxygen-rich surfaces (Fig. \ref{agb_123}) and do not reach the
  carbon-rich phase in the models by \citet{Weiss2009}. At 0.3 Gyr and for  young ages (see Fig.~\ref{SSP_flux_some_123}),
	the SSPs spectra display both amorphous silicates (at $\sim$ 10 $\umu$m)
and features due to cristalline silicates ($\sim$ 30 $\umu$m) because of the high optical depths.
\end{description}

\begin{table*}
\begin{center}
\caption{Ratios of the integrated fluxes predicted by several IMFs to the Salpeter IMF.
Two metallicities are considered: $Z$=0.004 and
$Z$=0.02. Three ages
are displayed: $t$=10 Myr, $t$=100 Myr, and $t$=300 Myr. All SSP fluxes are normalized to a total SSP mass of
  1~$M_\odot$. Given that for a given IMF, Z and age the ratios depend on the wavelength, we display for each case the minimum and maximum values.}
\label{ratios_imf}
\vspace{1mm}
\begin{tabular*}{159.3mm}{l l c c c c c c}
\hline
\noalign{\smallskip}
Age (Myr)                     &                    &  10     & 10     & 100      &  100     & 300     &  300    \\
\hline
\noalign{\smallskip}
Metallicity Z              &                    &  0.004  & 0.02   & 0.004    & 0.02     & 0.004   & 0.02      \\
\hline
\noalign{\smallskip}

Larson Solar Neighbourhood & IMF$_{\rm Lar-SN}$     &  0.87--0.99   & 0.90--1.05   & 1.33--1.41     & 1.29--1.42      & 1.46--1.55    & 1.44--1.54 \\
Larson  (Milky Way Disc)   & IMF$_{\rm Lar-MW}$     &  1.79--1.81   & 1.77--1.81   & 1.69--1.71     & 1.67--1.72      & 1.59--1.66    & 1.60--1.68 \\
Kennicutt                  & IMF$_{\rm Kenn}$       &  1.22--1.30   & 1.24--1.34   & 1.50--1.57     & 1.49--1.58      & 1.58--1.67    & 1.58--1.67 \\
Kroupa (original)          & IMF$_{\rm Kro-Ori}$    &  0.61--0.70   & 0.63--0.76   & 1.00--1.09     & 0.97--1.04      & 1.13--1.27    & 1.10--1.26 \\
Chabrier                   & IMF$_{\rm Cha}$        &  1.18--1.33   & 1.22--1.38   & 1.57--1.61     & 1.58--1.61      & 1.58--1.62    & 1.59--1.63 \\
Arimoto                    & IMF$_{\rm Ari}$        &  1.80--1.57   & 1.48--1.74   & 1.10--1.01     & 1.00--1.14      & 0.86--0.96     & 0.87--0.98 \\
Kroupa 2002-2007           & IMF$_{\rm Kro-27}$     &  0.66--0.76   & 0.68--0.82   & 1.08--1.18     & 1.05--1.19      & 1.22--1.37    & 1.19--1.35 \\
Scalo                      & IMF$_{\rm Sca}$        &  0.42--0.48   & 0.43--0.51   & 0.69--0.74     & 0.66--0.75      & 0.77--0.86    & 0.75--0.85 \\
\hline
\end{tabular*}
\end{center}
\end{table*}

\subsection{SSP colours}\label{temporal}

SEDs and colours of SSPs have a strong dependence on age, metallicity
and the presence of dust shells around the AGB stars.
This is clearly demonstrated by Figs. \ref{ssp_coloursBV}, \ref{ssp_coloursVK} and \ref{ssp_coloursJ8} for the $[\textrm{B}-\textrm{V}]$,
$[\textrm{V}-\textrm{K}]$ and [\textrm{J} - \textrm{8}$ \umu$m] colours.
The age range considered covers the onset of the AGB phase until the
very old ages when the contribution of AGB stars to the integrated flux is very
low. As well known, age and metallicity -- this latter enhanced by the
effects of AGB dust-shells -- drive the evolution of the SSP colours.\\
\indent The peak emission of the central AGB stars surrounded by dust shells is at
around 1-2 $\umu$m; the dust shifts the flux from \textit{J},
\textit{H} and \textit{K} bands to longer wavelengths.
This effect is stronger at shorter wavelengths (\textit{J}-band) than at the longer ones (like \textit{K}-band).
It is the combination of this effect together with the exact position of the emission peaks
of the central stars that determine the increase of the IR magnitudes.
Depending on the age and metallicity of the population, the
net effect is that sometimes the \textit{J} magnitude increases more than the \textit{K} magnitude,
while in other cases the opposite effect is true.
It is in the Spitzer 8 $\umu$m pass-band that we mostly see the radiation emitted by dust around AGB stars.
Even the coolest AGB star would provide a negligible contribution to that band
if the dust shells are ignored.
For the $[\textrm{V}-\textrm{K}]$ colour, dust mainly makes the $\textit{K}$ magnitude fainter, and affects only slightly
the $\textit{V}$ flux, producing overall bluer colours.
The effect of dust is small in the UV/optical pass-bands (see Fig. \ref{ssp_coloursBV}), and the  $[\textrm{B}-\textrm{V}]$
colours are practically unaffected by the presence of the dust
surrounding AGB stars.

\subsection{SEDs and colours of SSPs  for variations in the IMF}
\label{SSP_col_imf}

  Recalling the definition of the monochromatic flux emitted by an
  SSP of age $t$ and metallicity $Z$, the choice of an IMF implies
  that along the corresponding isochrone, between $M_l$ and $M_{u}(t)$
  (the most massive living stars at that age), the
  relative number of stars per mass intervals $dM$ is defined. The
  mass range spanned by all evolutionary stages beyond
  the main sequence turn-off decreases from a few solar masses to a
  few hundredths of a solar mass as the age increases from very young
  (a few Myr) to very old (a few Gyr).  This means that but for very
  young SSPs, changing the IMF has little impact on the integrated
  magnitudes and colours of SSPs.  Main sequence stars that are one or two
  magnitudes fainter than the turn-off contribute significantly to the
  SSP mass, but little to magnitudes and colours.

\begin{figure}
\begin{center}
\subfigure
{\includegraphics[width=0.30\textwidth]{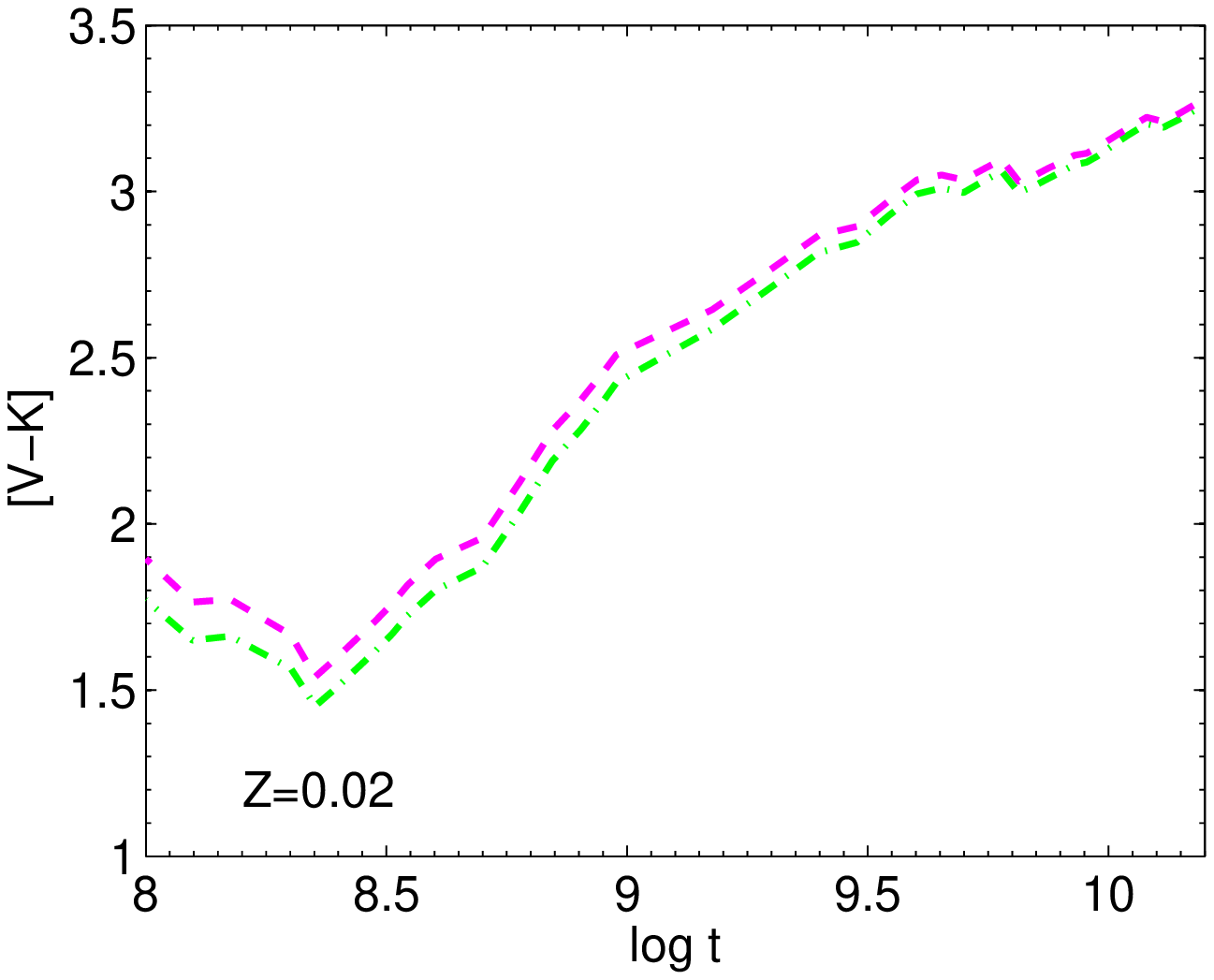}}
{\includegraphics[width=0.30\textwidth]{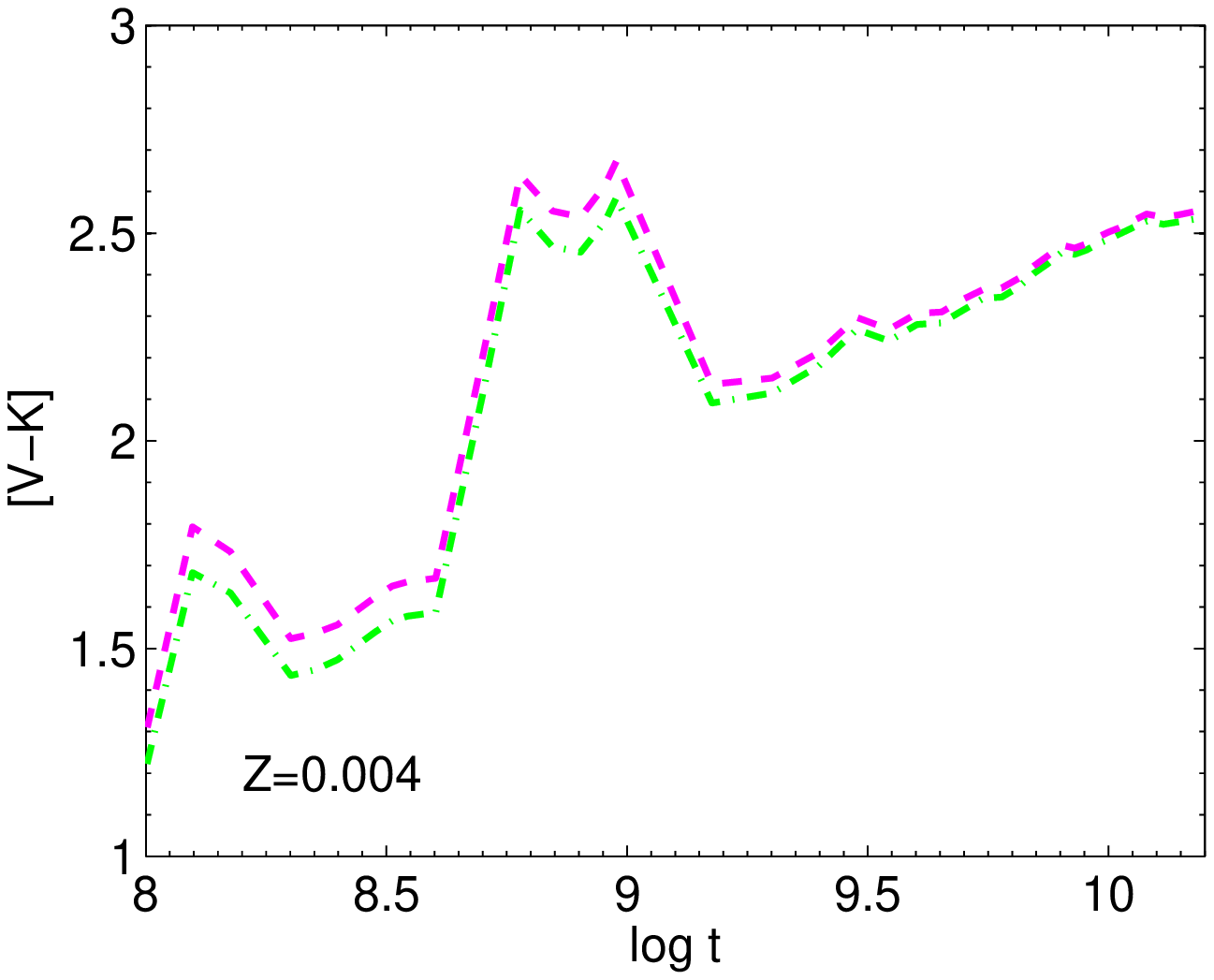}}
\subfigure
{\includegraphics[width=0.30\textwidth]{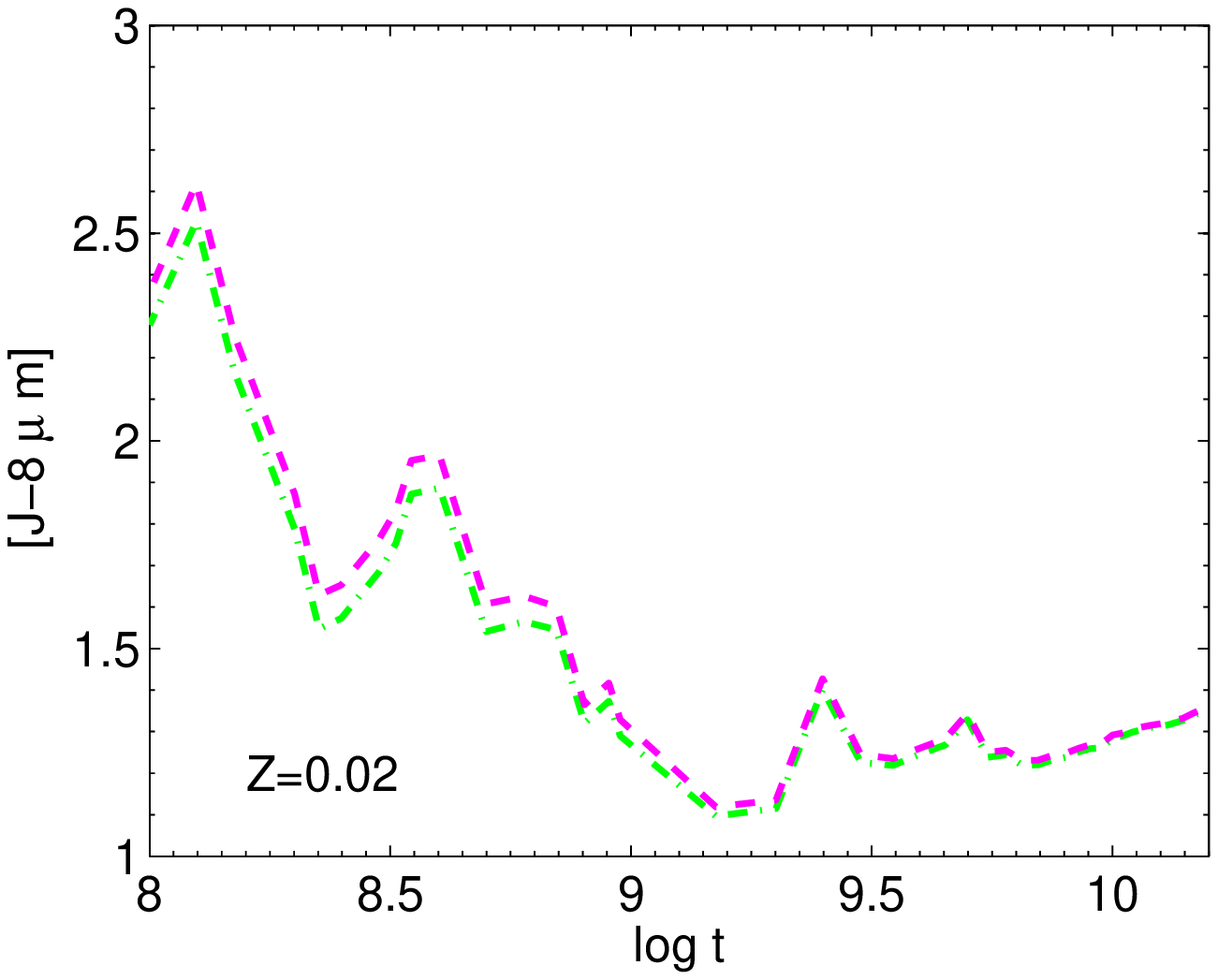}}
{\includegraphics[width=0.30\textwidth]{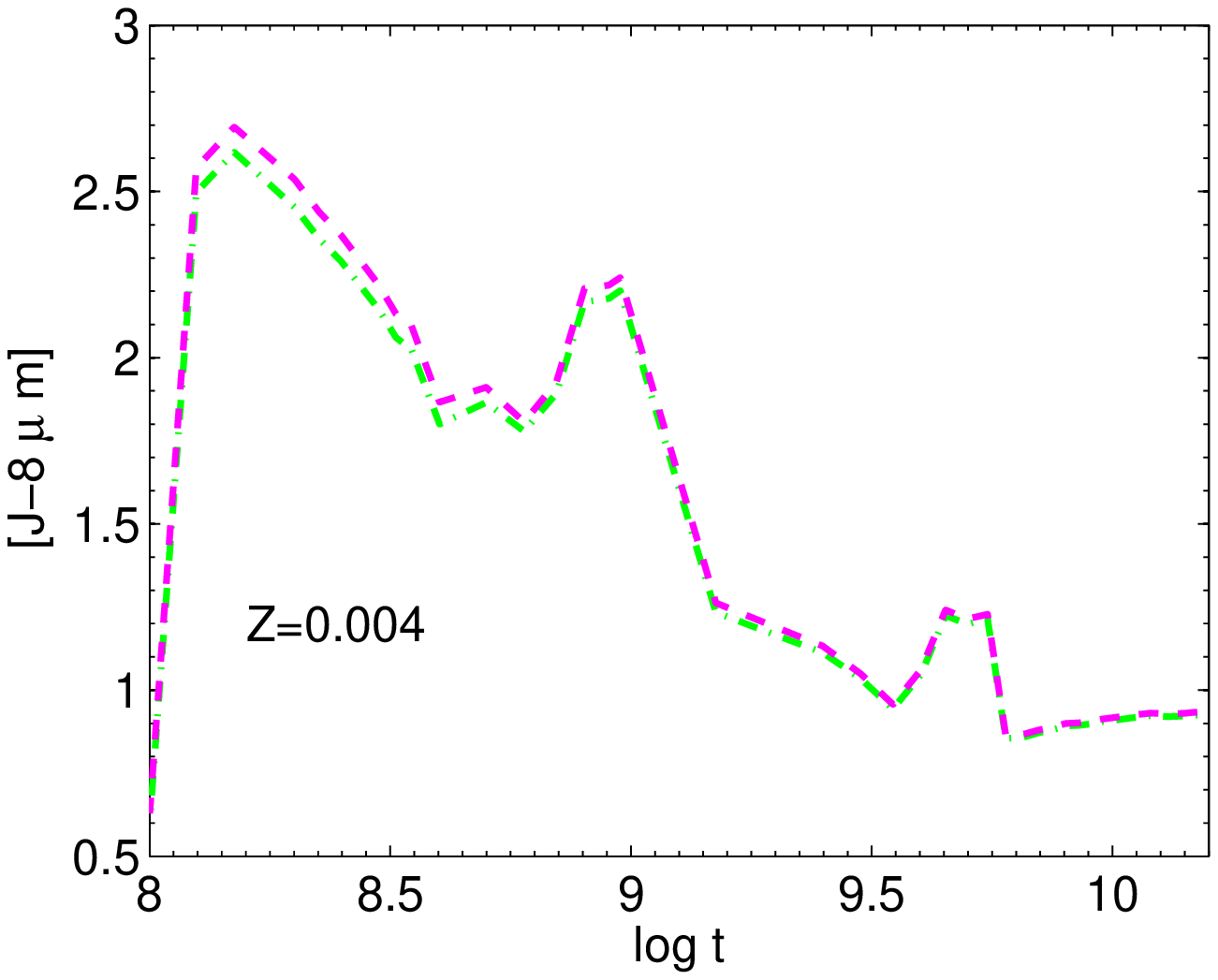}}
\caption{Theoretical $[\textrm{V}-\textrm{K}]$ and [$\textrm{J}$ -
    \textrm{8} $\umu$m] colours as a function of age (from 0.1 to
  15 Gyr) for the labelled metallicities, and the two IMFs that cause the
most extreme variations: results for the other IMFs discussed in this paper
lie between the lines displayed.
  Colour and line-style for the results with different IMF are as in the
  Fig.~\ref{imf_plot}. Time $t$ is again in yr.}
\label{VK_J8_IMF}
\end{center}
\end{figure}

 In relation to this, we recall that the various IMFs
  do not have the same $M_l$ and do not predict the same
  percentages of stars and hence stellar mass in different mass
  intervals. Looking at the entries of Table \ref{tab_imf}, IMF$_{\rm
    Kro-27}$, IMF$_{\rm Lar-MW}$, IMF$_{\rm Lar-SN}$,
  IMF$_{\rm Cha}$, and IMF$_{\rm Ari}$ are all defined down
  to a lower mass limit of 0.01 $M_{\odot}$, whereas the remaining
  ones have a lower mass limit of 0.1$M_{\odot}$
  \citep{Piovan2011II}. Some IMFs, like IMF$_{\rm Kro-Ori}$, and
  IMF$_{\rm Sca}$, predict a number fraction of massive stars --
  including SNe --
  much smaller than the number fraction of AGB stars. Others,
  like IMF$_{\rm Lar-MW}$, IMF$_{\rm Kenn}$, and IMF$_{\rm
    Cha}$ behave in a different way and, even if again they predict a
  fractional mass of AGB stars higher than that of SNe, the relative
  contribution of SNe is much more significant. Only for the peculiar
  IMF$_{\rm Ari}$ the trend is reversed and SNe outnumber AGB stars.
The different number ratios of stars going through the AGB to stars that become
  SNe have two effects. First, we expect that for the IMFs with
  a greater AGB mass fraction the effect of AGB circumstellar dust
 on magnitudes and colours is more significant than in the
  other cases. Second, the IMF choice affects the amount of  dust
  injected into the ISM. Before the main process of dust formation happens
  in the ISM, the partition between AGB and SNe drives the
  amounts of star-dust injected into the ISM
  \citep{Zhukovska2008,Piovan2011II}. In each generation of stars, the
dust production by either SNe or TP-AGB stars therefore depend on
  the IMF. \\
\indent Small differences in the IMFs shown in Fig.~\ref{imf_plot}, especially for
$M> 10 M_{\odot}$, have a significant effect on young SSPs. To illustrate
this point we calculate the SEDs of SSPs with different chemical
composition and age, for all the IMFs listed in Table~\ref{tab_imf}
and for the metallicities $Z$=0.004 and $Z$=0.02. The SEDs are
presented in Fig. \ref{IMF_age} for three selected ages, namely 10,
100, and 300 Myr, and are limited to a metallicity of $Z$=0.02
(similar results are found for the other metallicities). First of all,
the SEDs run nearly parallel for the full wavelength range of
interest. We therefore expect the ratio of the flux $F_\lambda
(\lambda, IMF_1)/F_\lambda(\lambda, IMF_2)$ between the SEDs of any
two IMFs to remain similar over most of the spectrum. This is shown
by the entries of Table~\ref{ratios_imf} that lists the fluxes
predicted by our IMFs for the three selected ages
presented in Fig. \ref{IMF_age}, two metallicities and a total mass M$_{SSP}$=1$M_\odot$, normalized to the values predicted by
the Salpeter IMF. For each case, the minimum and maximum flux ratios obtained across the wavelength range
from 0.1 to 100 $\umu$m, are displayed.\\
\indent Looking at the entries of
Table~\ref{ratios_imf},  we see that IMF$_{\rm Ari}$ and
IMF$_{\rm Lar-MW}$ predict the largest SSP fluxes for the youngest age, 
followed by IMF$_{\rm Kenn}$ (IMF$_{\rm Cha}$ is practically coincident with IMF$_{\rm Kenn}$),
in agreement to what is suggested by Fig.~\ref{imf_plot}. IMF$_{\rm Ari}$ and IMF$_{\rm Lar-MW}$ 
indeed present the greatest fractional mass
contribution of high-mass stars.
At SSPs ages $t$=100 Myr and 300 Myr,  IMF$_{\rm Kenn}$, IMF$_{\rm Lar-MW}$ and IMF$_{\rm Cha}$ 
have the highest flux ratio (as expected from the entries of Table~\ref{tab_imf}).
These IMFs predict the greatest fraction of stars with mass between $1\leq M_\odot < 6$.  
The same trend is visible for $Z$=0.004. For the ages considered,  IMF$_{\rm Sca}$
predicts the lowest SSP flux.  It has a very small
fraction of massive stars,
favouring low-mass stars in comparison to the other IMFs: in fact,
68\% of the mass is contained in stars with $M <1 M_{\odot}$ (see
Table~\ref{tab_imf}).  This causes, as discussed, a lower integrated
flux (see the panels of Figs.~\ref{IMF_age}.\\
\indent We now turn to
  colours like $[\textrm{V}-\textrm{K}]$ and [\textrm{J} - \textrm{8
      $\umu$m}] that are affected by AGB stars. Figure \ref{VK_J8_IMF}
  shows that the effect of the IMF is stronger for the youngest ages
  where an AGB is present, up to
0.2~mag in $[\textrm{V}-\textrm{K}]$. For the oldest ages, changing the
  IMF does not affect significantly the integrated colours
  of the SSPs. For both metallicities ($Z$=0.02 and $Z$=0.004 -- the
  remaining ones behave in the same way), when AGB stars appear
  at approximately $\log t$ =8, the IMF$_{\rm Lar-MW}$, predicts
  the reddest $[\textrm{V}-\textrm{K}]$ colour, whereas the
  IMF$_{Kro-27}$ predict the bluest one. This is due to
  the combined effect on the $\textrm{K}$ band of the varying fractional mass
  of AGB stars and the circumstellar dust. The former effect
  dominates and makes the colour redder; it is increasing
  when going from IMF$_{Kro-27}$ to IMF$_{\rm Lar-MW}$.
  The effect of circumstellar dust effect is less important and
  opposite. As we have seen, while the V magnitude is not affected
  by circumstellar dust, the K magnitude is influenced, producing
  bluer colours (see
  Fig. \ref{ssp_coloursVK}).  In case of the [\textrm{J} - \textrm{8}
    $\umu$m] colour, different IMFs do not change appreciably its
  evolution and the effect is even smaller than in
  $[\textrm{V}-\textrm{K}]$.\\
  Finally, the effect of the IMF
  on the SED of SSPs does not depend appreciably on the chosen
  metallicity.  This is shown by the entries in
  Table~\ref{ratios_imf}: the ratios are almost identical for both
  metallicities at all ages considered.

\section{Comparing theoretical results with observations}\label{StarClusters}

The key quantities to determine from the analysis of a stellar
system are age and metallicity distribution
of the parent stars, to reconstruct the star formation and chemical
enrichment history of the whole population. A widely used method to
determine age and metallicity of unresolved stellar systems is
  to compare their
observed colours with the predictions of EPS models (see
\citet{Bressan1994}, \citet{Tantalo1998}, \citet{Bruzual2003},
\citet{Buzzoni2005}, \citet{Piovan2006a,Piovan2006b},
\citet{Galliano2008a}, and \citet{Popescu2011} just to mention a few).
If instead the stellar populations can be resolved into single
  stars, the star-by-star EPS technique enables one to build
  synthetic CMDs
  \citep{Chiosi1986,Bertelli1994,Aparicio2004,Chung2013} to derive
  the star formation history of the population under scrutiny.\\
\indent To this aim, we wish to validate the new SSPs and study the
effect of the new AGB models and
the libraries of dusty AGB spectra, by means of comparisons
with observed magnitudes and colours of resolved and unresolved stellar populations.
The ideal laboratory for this are both the massive star clusters and the
rich CMDs of field stars of the Magellanic Clouds, that host
significant populations of intermediate-age,
populations not easily
accessible in the Milky Way \citep{Pessev2008}.

Due to the large number of studies on the subject,
no attempt is made here to summarize previous work, and we
will limit ourselves to mention only the sources of the data used in our
analysis.
Particularly interesting are data in the NIR region of the spectrum,
because they allow to, at least partially, break  the
age-metallicity degeneracy of the integrated colours,
especially for stellar populations older
than $\sim$ 300-400 Myr
\citep{Goudfrooij2001,Puzia2002,Hempel2004,Pessev2006,Pessev2008}.
In the following, we compare our theoretical models with:
\begin{description}
\item[-] CMDs of field stars in the Large and Small
  Magellanic Clouds (LMC and SMC respectively);

\item[-] broad-band colours of star clusters in the LMC and SMC,
using SEDs with and without \textit{AGB dust shells}.
\end{description}

\subsection{Field Stars of the Magellanic Clouds}

We compare here selected CMDs of Magellanic Clouds' fields, with our isochrones with and without AGB dust shells.
The best available data come from extensive near and mid-infrared surveys like
SAGE (\textit{Surveying the Agents of Galaxy Evolution})
\citep{Blum2006,Bolatto2007}, that matches IRAC (or MIPS) data with 2MASS
photometry. The survey has been described by \citet{Meixner2006}.

\textbf{\textsf{CMDs of Field Stars}}. We start comparing the CMDs in NIR and MIR photometry,
with both our new and old isochrones,
for ages in the interval between $t$=0.09~Gyr and 2~Gyr. We consider a
metallicity $Z$=0.008 for the LMC, and $Z$=0.004 for
the SMC.

\begin{figure*}
\begin{center}
{\includegraphics[width=0.41\textwidth]{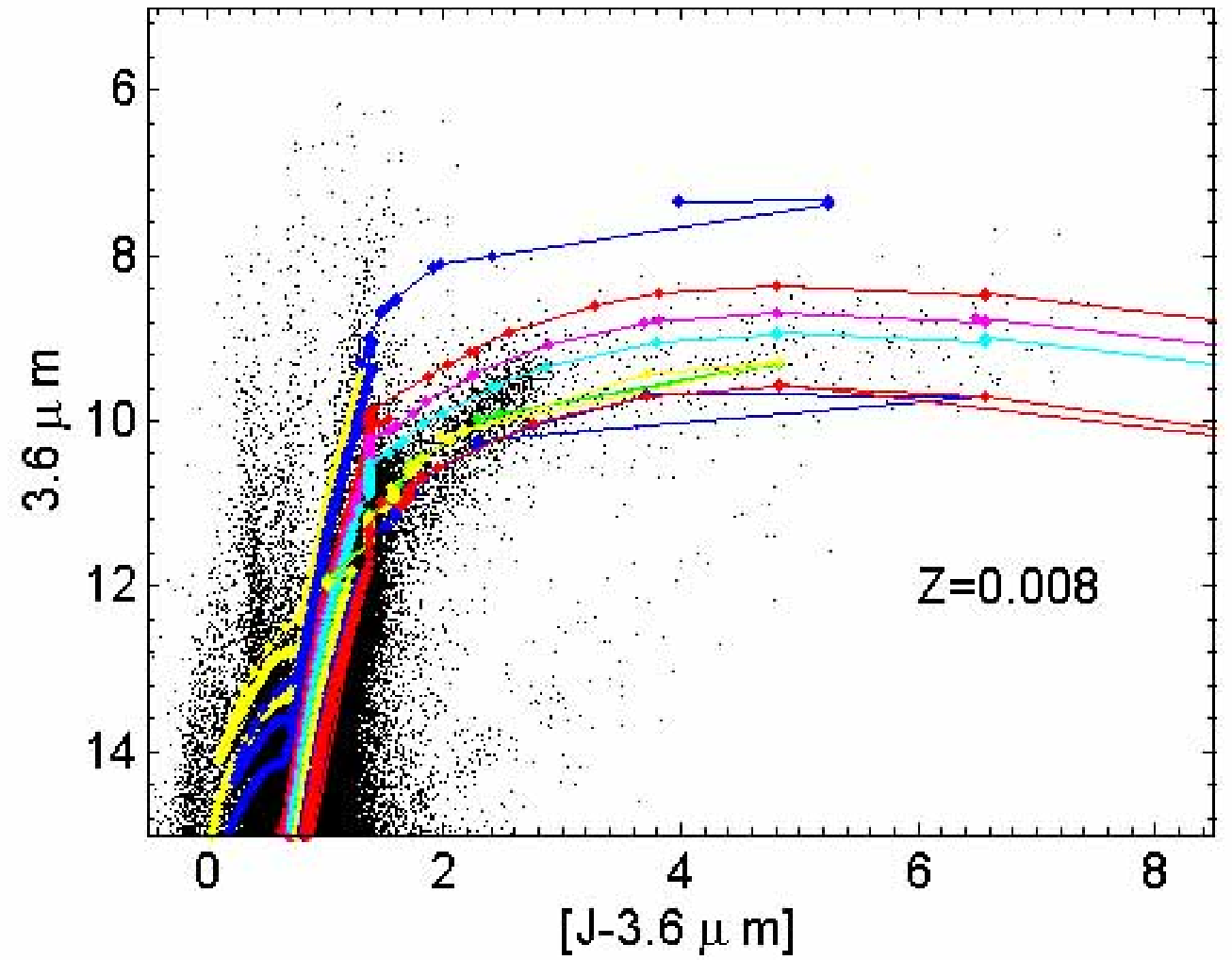}  
 \includegraphics[width=0.41\textwidth]{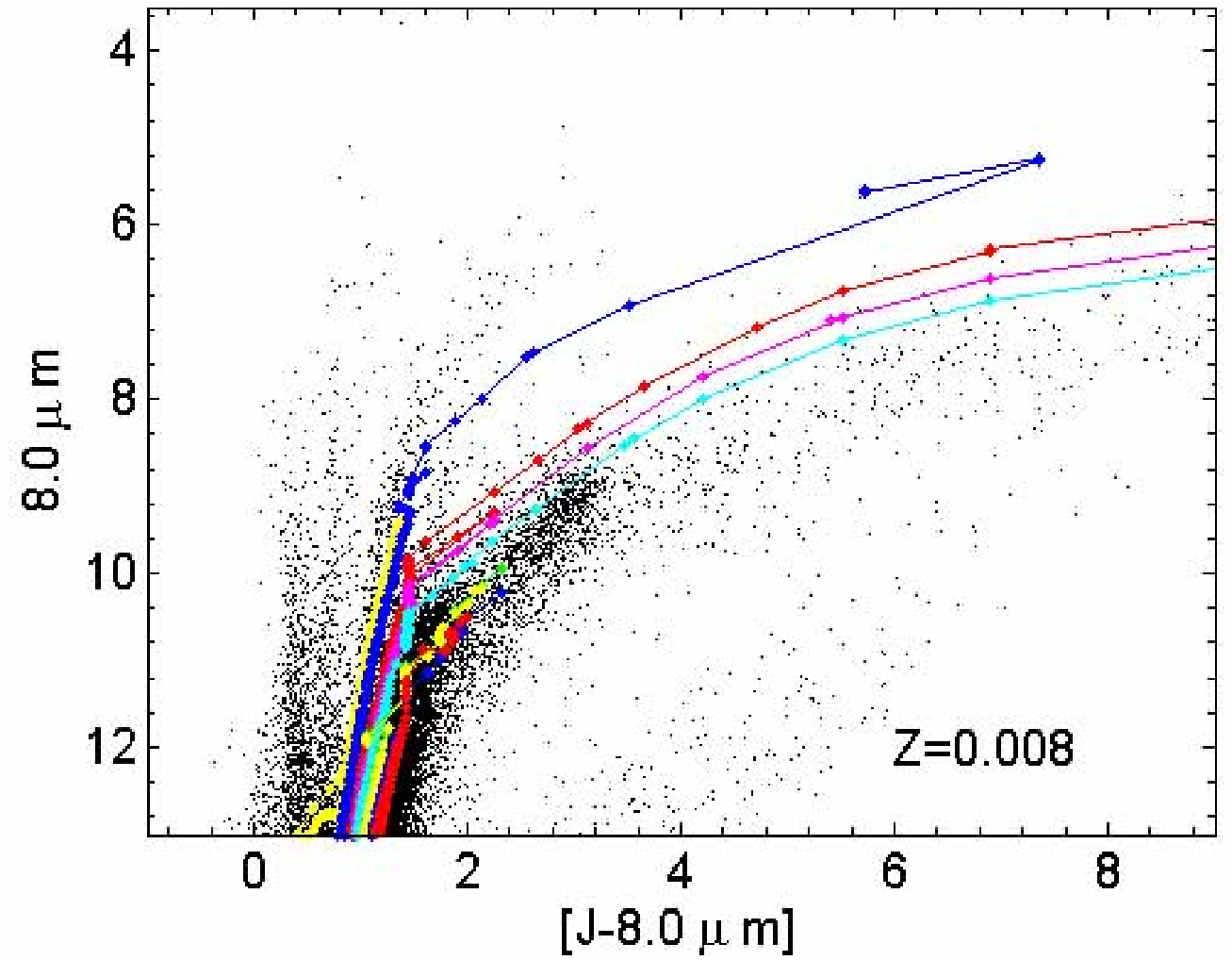}}
{\includegraphics[width=0.41\textwidth]{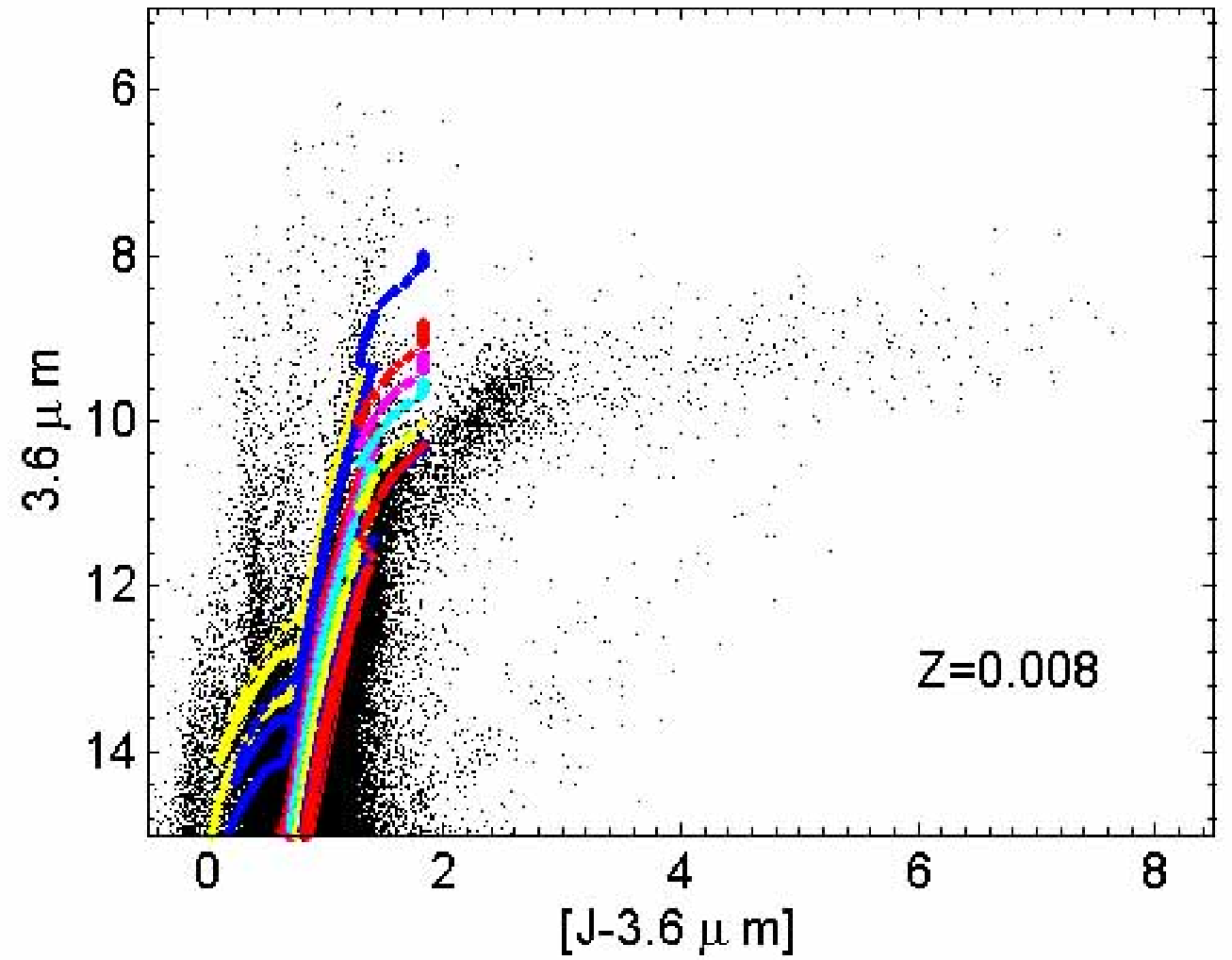} 
 \includegraphics[width=0.41\textwidth]{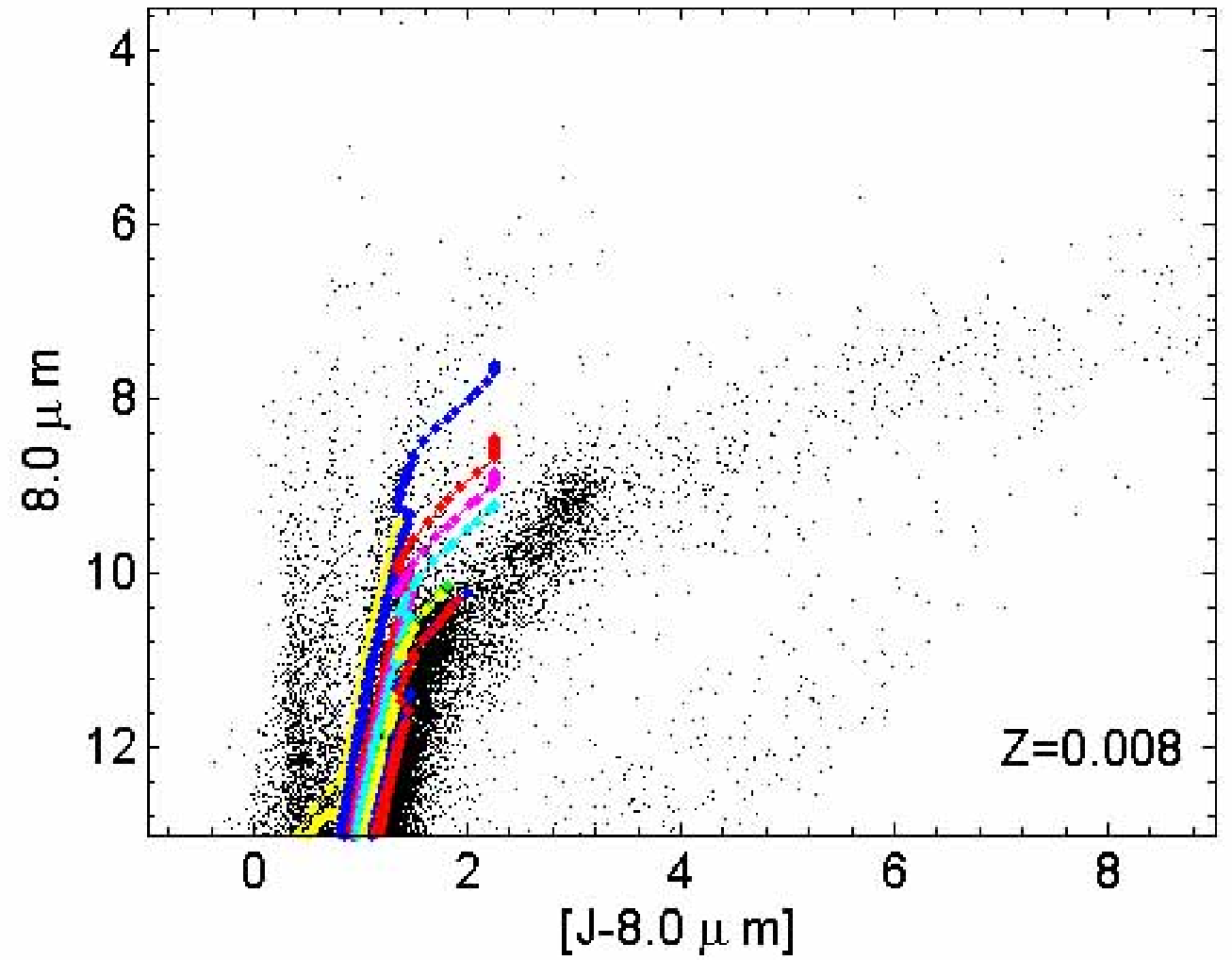}}
\caption{Field stars of the LMC in the 2MASS+IRAC surveys:
The [3.6 $\umu$m] vs [$\textrm{J}$ - \textrm{3.6} $\umu$m] (\textbf{left panels}) and [8 $\umu$m] vs.
[$\textrm{J}$ - \textrm{8} $\umu$m]  (\textbf{right panels}) CMDs are compared to
isochrones with metallicity $Z=0.008$,
representative of the LMC composition.
The isochrones are displayed in different colours
according to their age ($\log t$=7.95, 8.10, 8.48, 8.60, 8.70,
  8.90, 8.95, 9.18, 9.30, with $t$ in yr. The oldest isochrones are in
  the bottom right part of each CMD).
The top panels show isochrones including AGB dust shells.
The bottom panels show models without circumstellar dust shell.} 
\label{LMC_j36_j8}
\end{center}
\end{figure*}

\begin{figure*}
\begin{center}
{\includegraphics[width=0.41\textwidth]{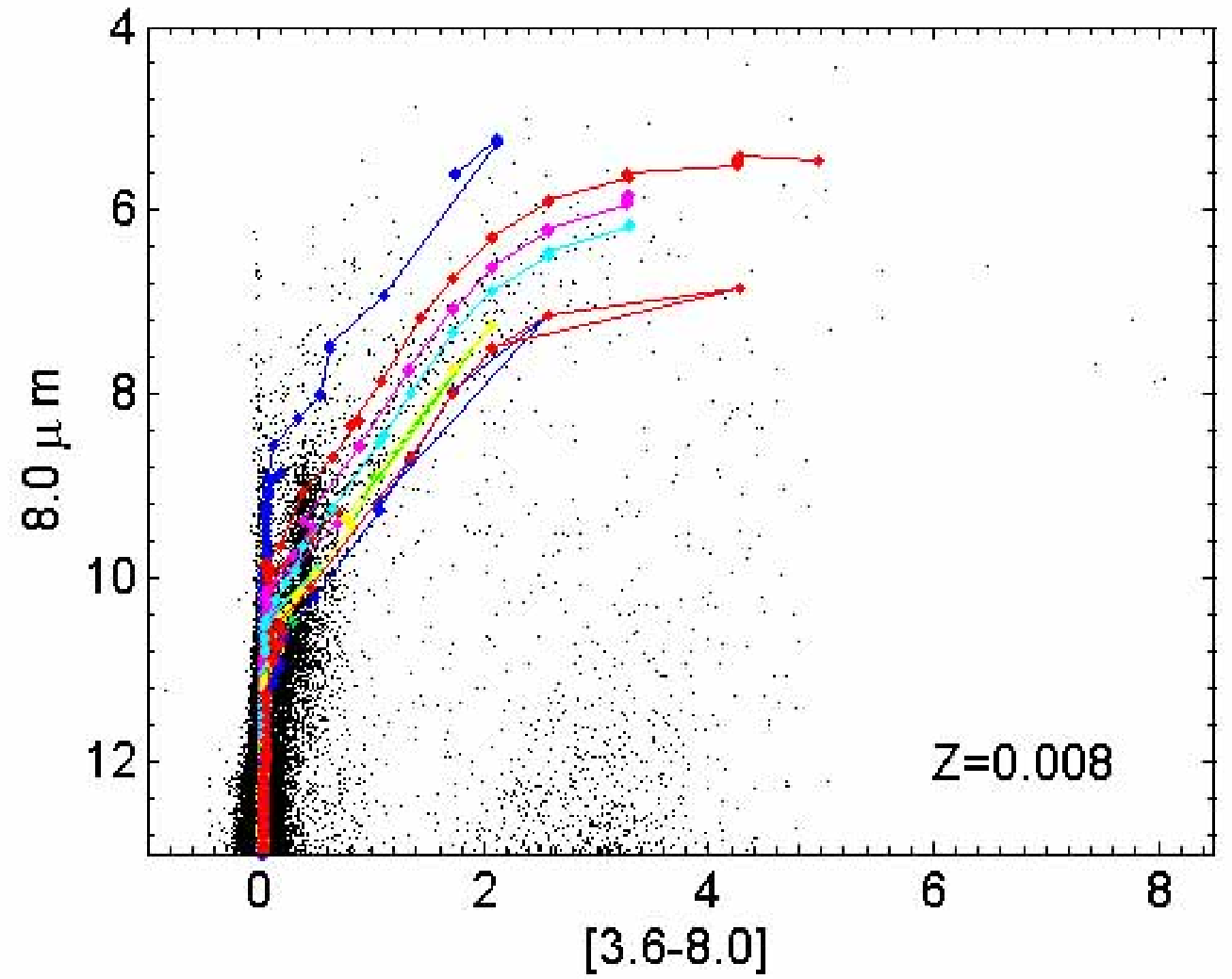}   
 \includegraphics[width=0.41\textwidth]{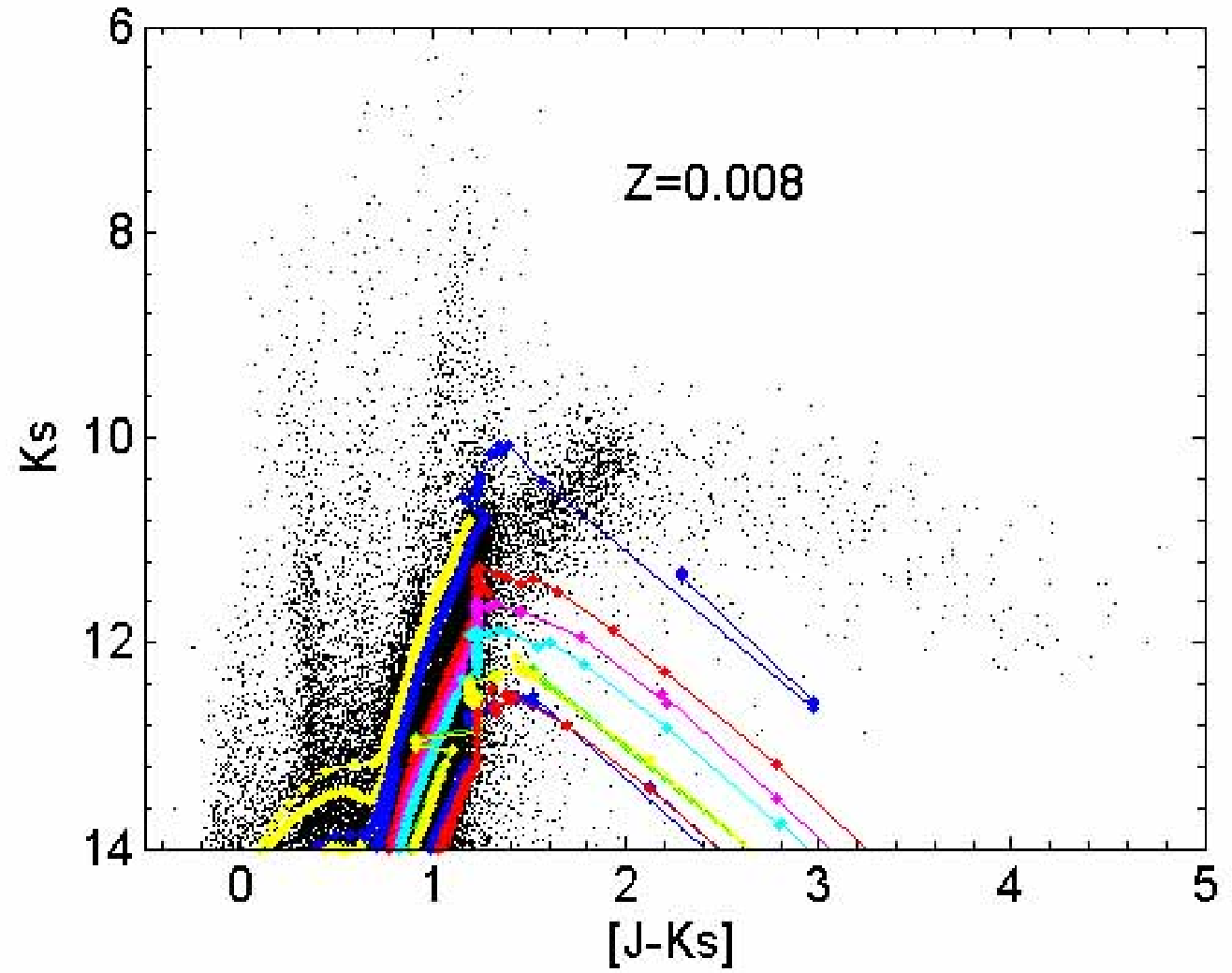} }
{\includegraphics[width=0.41\textwidth]{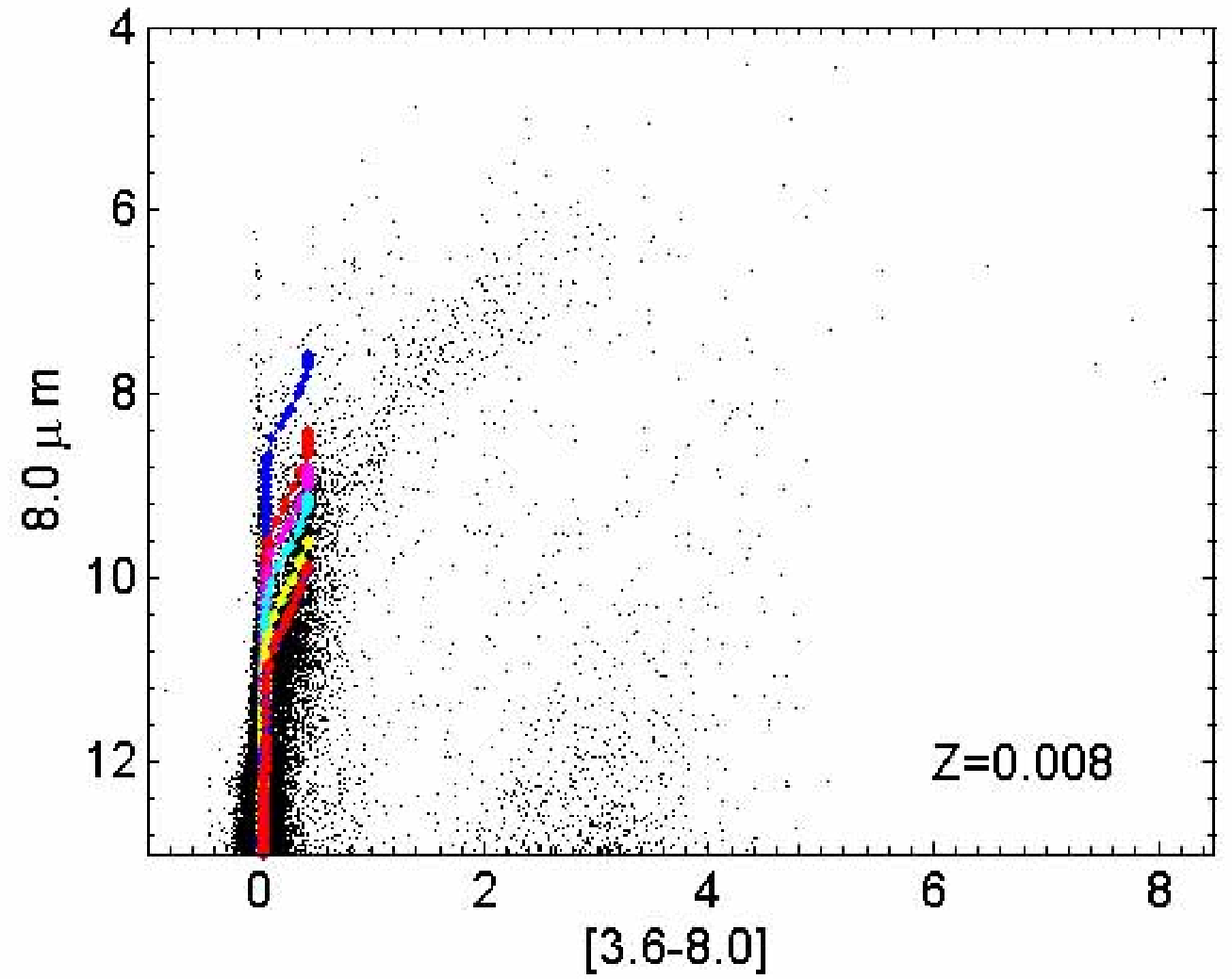} 
 \includegraphics[width=0.41\textwidth]{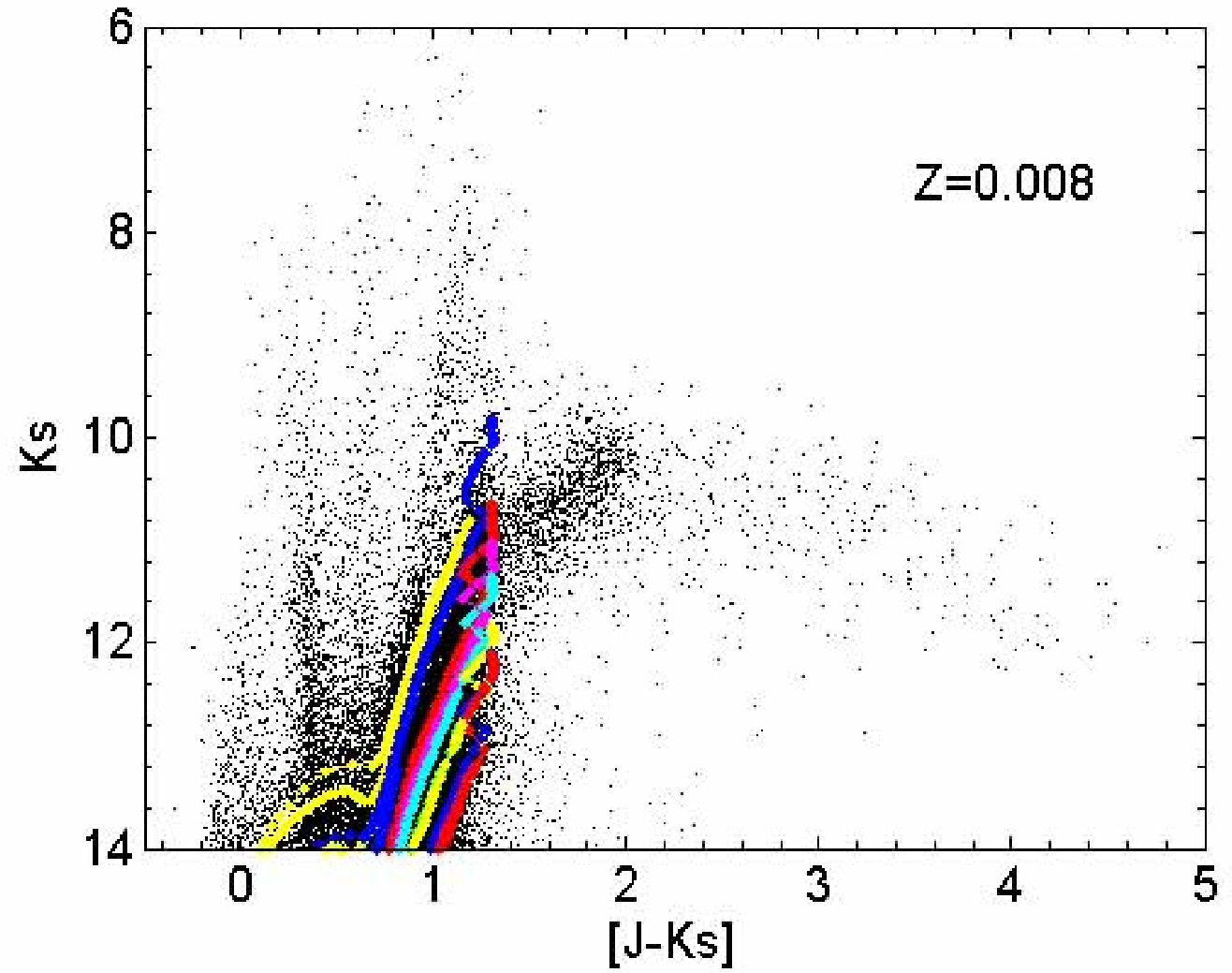}}
\caption{As Fig.~\ref{LMC_j36_j8}, but for IRAC
    [8 $\umu$m] vs [3.6 - 8] CMDs (\textbf{left panels}) and $\textrm{K}_{s}$
  vs.  [\textrm{J} - $\textrm{K}_{s}$]  (\textbf{right panels}).}
\label{LMC_368_2MASS}
\end{center}
\end{figure*}

\textbf{\textsf{Large Magellanic Cloud}}. For the LMC  we adopt a standard
distance modulus of 18.5~mag \citep{Pessev2008}.  Following  \cite{Marigo2008}, the data have
been limited to a circular area of $\pi$ square degrees, centred on the LMC bar
($\alpha_{2000}=5^{h} 23^{m}.5$, $\delta=-69^{\circ}45^{'}$), to include thousands of bright
AGB stars and exclude foreground stars likely belonging to the Galactic Disk  \citep{Marigo2008}. 
The observed CMDs are shown
in the various panels of Figs.~\ref{LMC_j36_j8}
and  \ref{LMC_368_2MASS}. A nearly vertical blue plume is always present, 
probably due to foreground stars of the Galaxy.
There is also some
contamination in the bottom right part of the CMDs, likely caused by  background galaxies. 
Moreover, in Fig.~\ref{LMC_368_2MASS},
the objects with roughly [3.6 - 8] $\gtrsim$ 2 (the vertical finger in
the bottom right part of the CMD) are not detected by 2MASS; they are
 probably background
galaxies and young pre-main sequence stars. The top panels refer to
the new isochrones, whereas the bottom panels are for the old
ones. The  isochrone ages are given
at the top of each panel. The most
remarkable feature is the much wider extension of the new AGB phase, 
in better agreement with the observations.

\begin{figure*}
\begin{center}
{\includegraphics[width=0.41\textwidth]{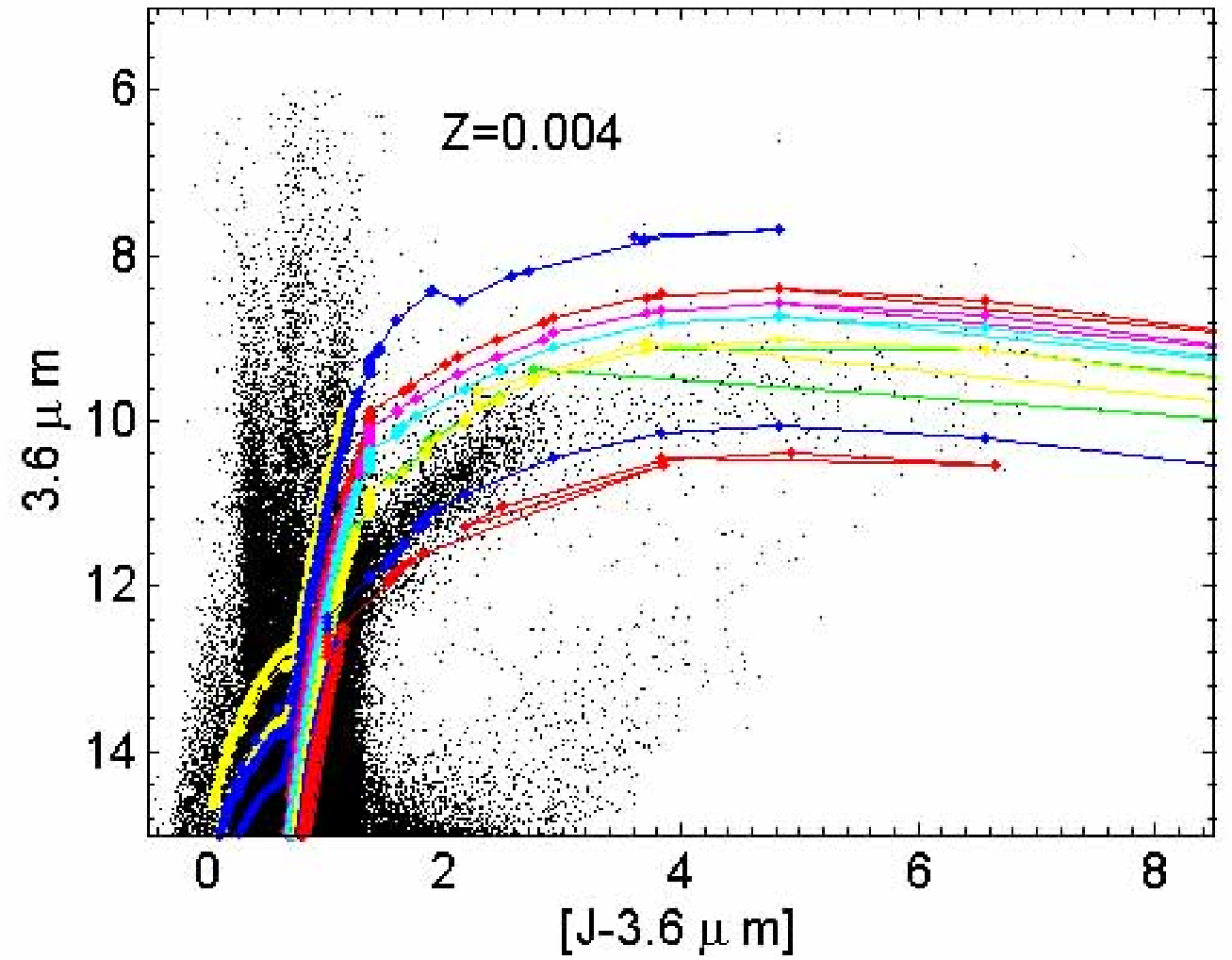}  
 \includegraphics[width=0.41\textwidth]{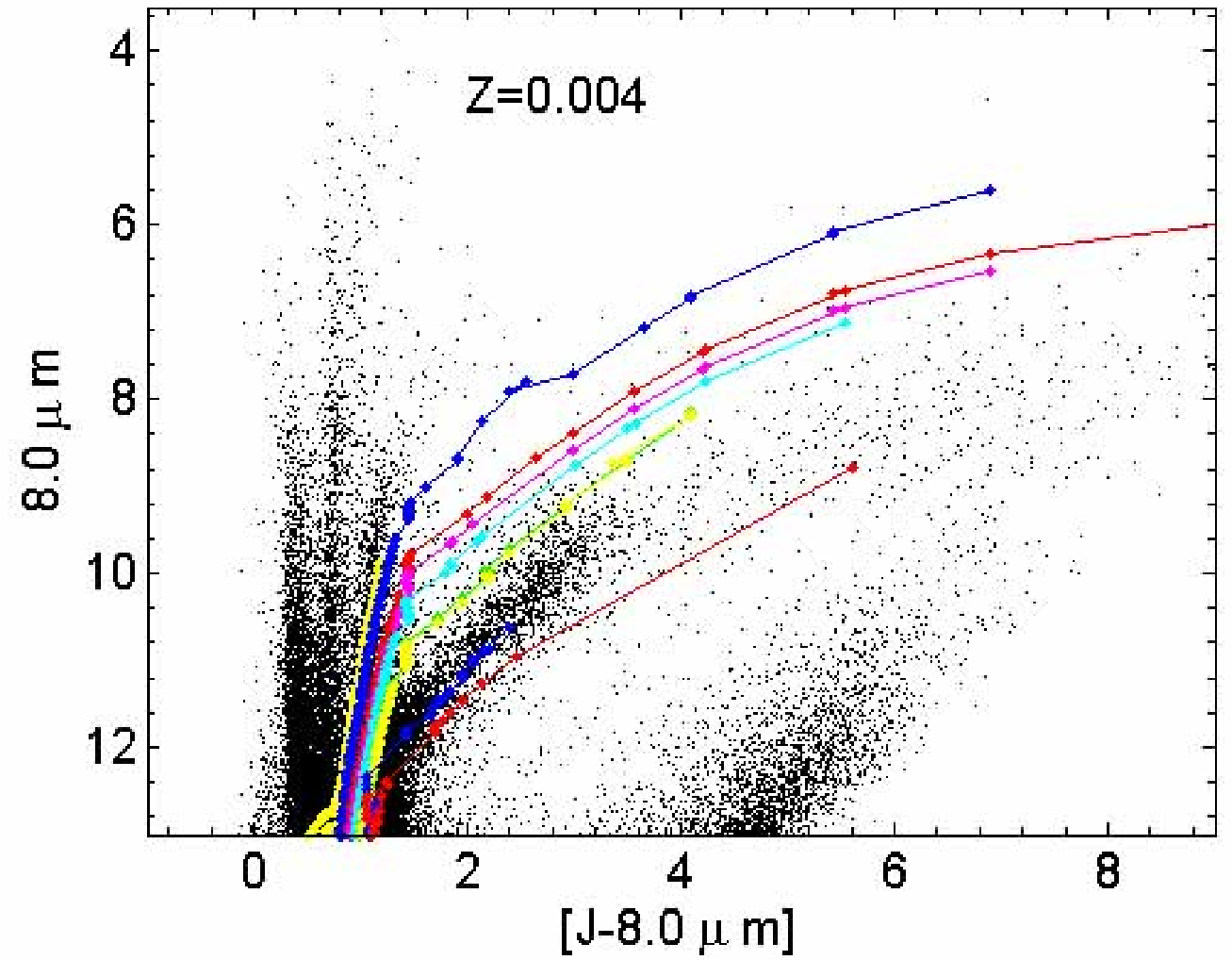}}
{\includegraphics[width=0.41\textwidth]{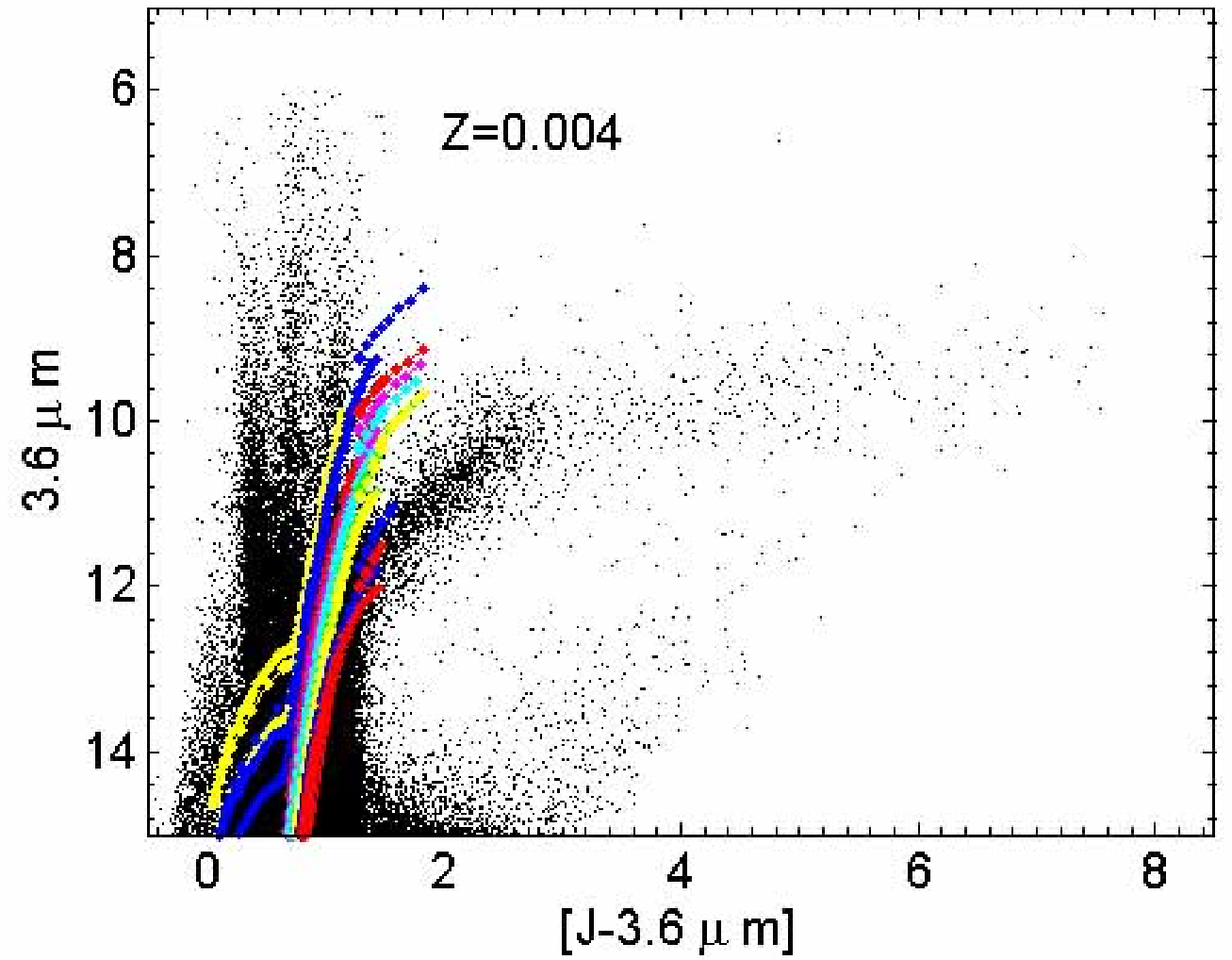} 
 \includegraphics[width=0.41\textwidth]{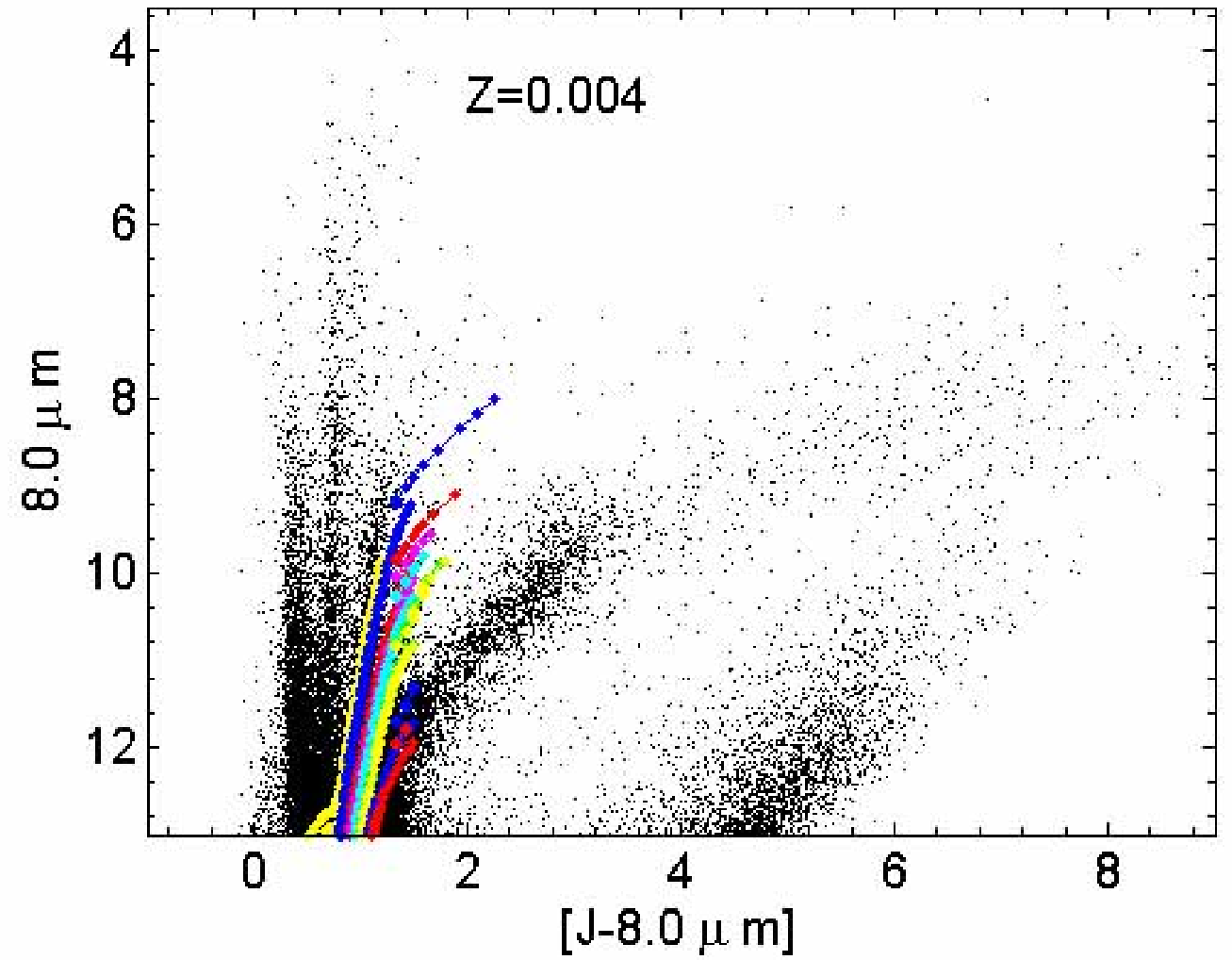}}
\caption{
 As Fig.~\ref{LMC_j36_j8}, but for the
  SMC, using isochrones with $Z=0.004$.}

\label{SMC_j36_j8}
\end{center}
\end{figure*}

\begin{figure*}
\begin{center}
{\includegraphics[width=0.41\textwidth]{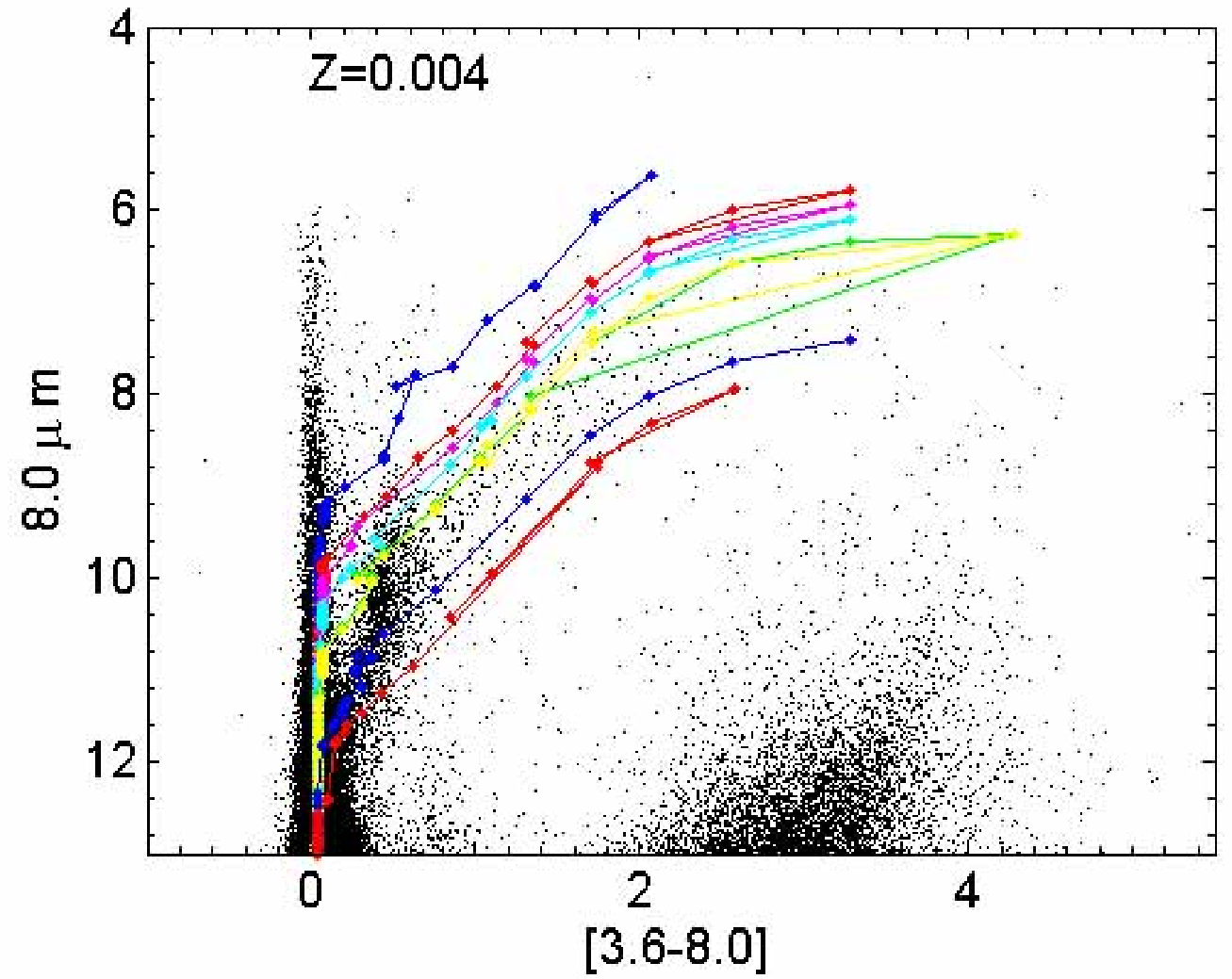}   
 \includegraphics[width=0.41\textwidth]{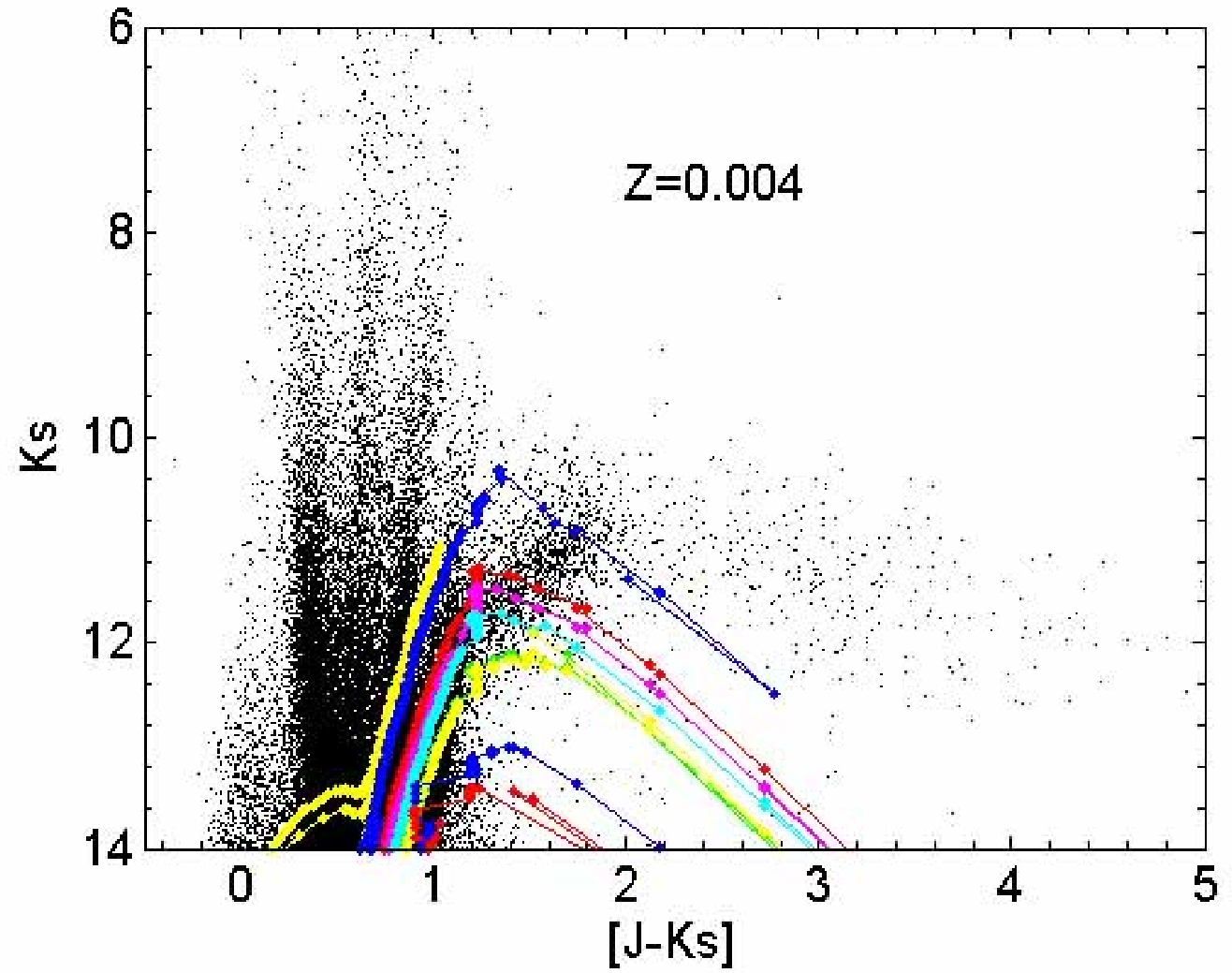} }
{\includegraphics[width=0.41\textwidth]{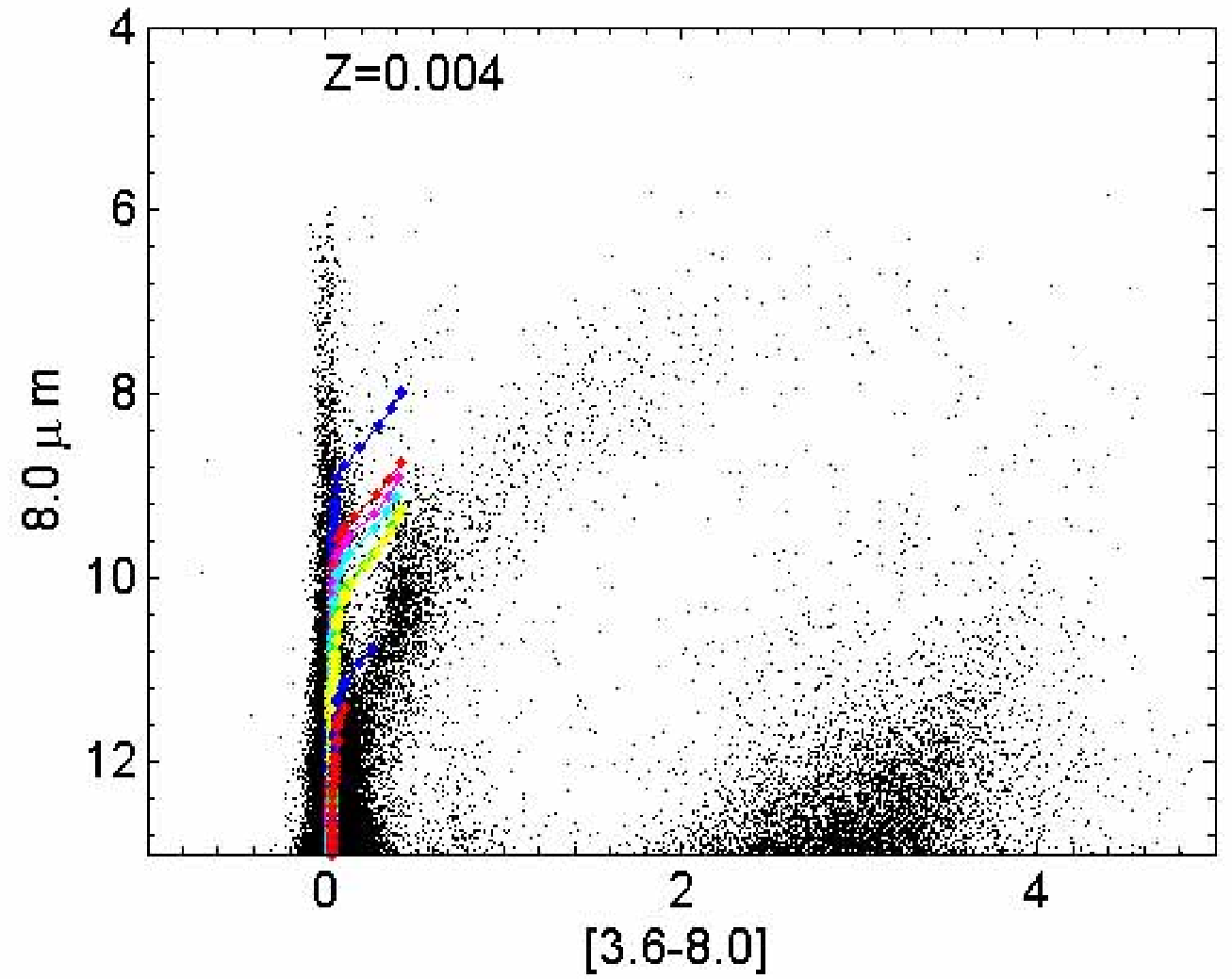} 
 \includegraphics[width=0.41\textwidth]{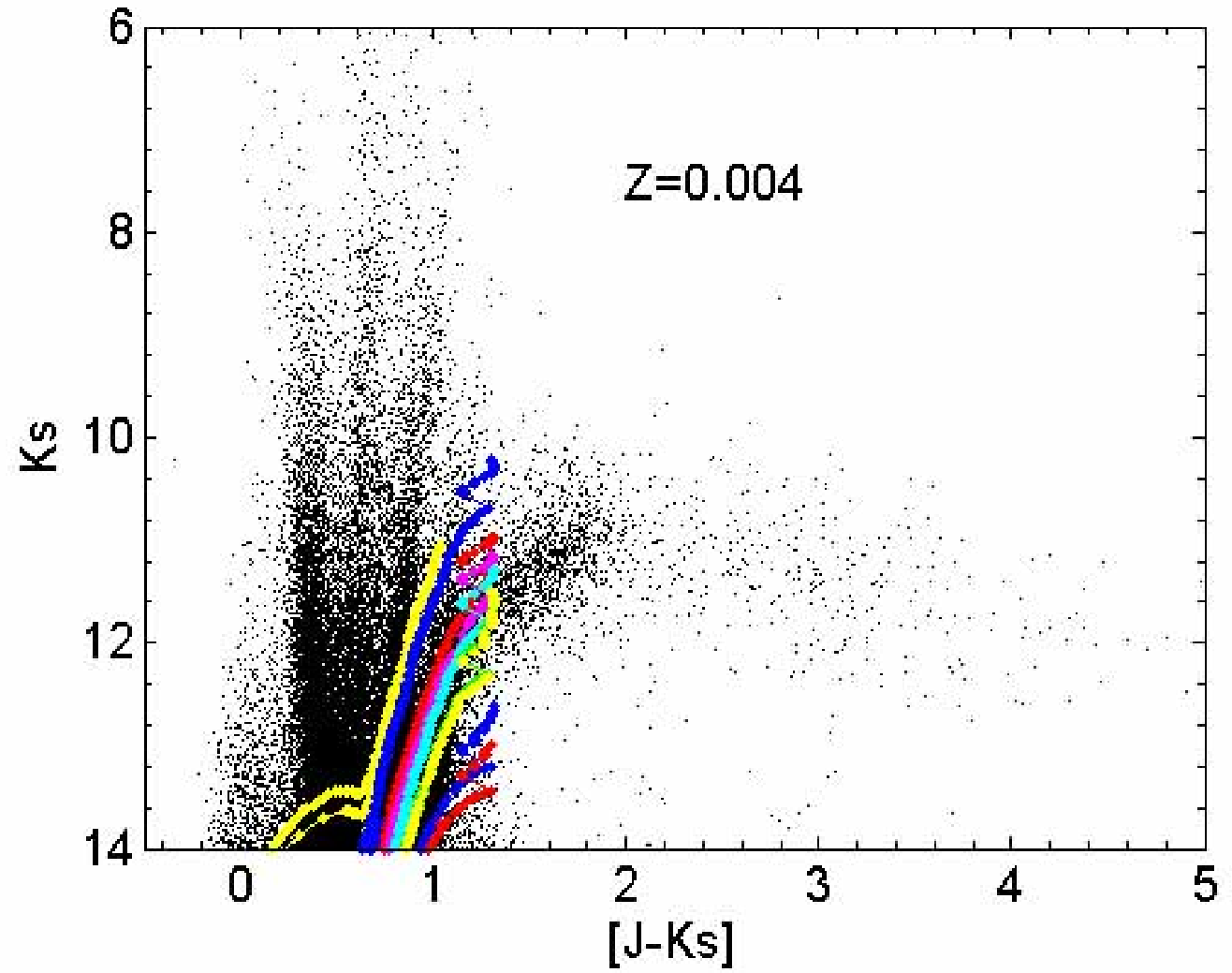}}
\caption{As Fig.~\ref{LMC_368_2MASS}, but for the SMC.}
\label{SMC_368_2MASS}
\end{center}
\end{figure*}

\textbf{\textsf{Small Magellanic Cloud}}.
For the SMC we adopted a distance modulus of 19.05~mag \citep{Pessev2008}, and the CMDs are shown in
Figs.~\ref{SMC_j36_j8} and \ref{SMC_368_2MASS}.
The contamination by objects not belonging to the SMC
is  visible also in this case, due to the same type of stars as for the LMC.
Finally, like in the LMC, objects
with roughly [3.6 - 8] $\gtrsim$ 2 in Fig.~\ref{SMC_368_2MASS}, are not detected by 2MASS.

\textbf{\textsf{Discussion of field star CMDs}}.
All CMDs of the LMC and SMC field stars, show the clear effect  of the dusty circumstellar shells
around AGB stars on NIR and MIR colours. Without dust,
the theoretical colours are \textit{not able} to reproduce the observed AGB populations.
In fact, without dust the AGB phase does not extend towards very red colours, as observed.\\
\indent In general, isochrones of the appropriate age reproduce the observed AGB sequences,
even though in some cases we do not have an exact match.
The reason is the too coarse grid of ages we are using. Nevertheless, the
fit of the observed CMDs with synthetic CMDs generated with
population synthesis techniques is beyond the aims of this study. For our purposes,
it is enough to test whether the new isochrones (and SSPs)
can cover the observed colour range of AGB stars. By inspecting the various CMDs we
draw the following conclusions:

\begin{description}
\item[-] at very young ages ($\log t$=7.95; the
  yellow line on the upper-left corner of the CMDs, well visible in
  all CMDs except in Figs.~\ref{LMC_368_2MASS} and
  \ref{SMC_368_2MASS}), AGB stars are not present.
The most prominent NIR emitters are the red supergiant stars.

\item[-] in the age range $8.10\leq \log t \leq 8.90$ the TP-AGB phase is well developed.
This is the age range in which AGB models
best reproduce the red tail of carbon-rich stars.
For these
metallicities the
intermediate-mass stars develop an extended carbon-rich envelope
  during the AGB phase.  The duration of the carbon-rich phase is determined by the interplay
between dredge-up and mass-loss.
As already pointed out, the efficiency of the TDU increases with decreasing metallicity; it is
very efficient for the compositions we are considering.

\item[-] for ages in the range $9.18 \leq \log t \leq 9.30$ the TP-AGB phase gets shorter:
the carbon-rich envelope is less important, even if its contribution to the red tail is still relevant.

\end{description}

\noindent
The agreement between our new models and observations is satisfactory for most
of the CMDs, and is a significant step forward towards a more realistic description of AGB stars.
Similar results have been obtained by \citet{Marigo2008} using
synthetic TP-AGB models and a library of dusty spectra of
AGB stars. We may also observe that our
isochrones, thanks to the new libraries of spectra for M- and C-stars,
have a more regular behaviour in the CMDs compared to those of \citet{Marigo2008}. This is likely due to
the different optical depths adopted for the models (in particular different recipes for the mass-loss rate).

Only in the 2MASS [$\textrm{K}_{s}$]
vs.  [\textrm{J} - $\textrm{K}_{s}$] CMDs the models did did not match the data
satisfactorily.
Our AGB isochrones bend towards fainter $\textit{K}_{s}$ magnitudes and redder [\textrm{J} - $\textrm{K}_{s}$]
colours more than the observations, that would suggest a nearly constant $\textit{K}_{s}$. It is hard to trace back the reasons for the
discrepancy, considering that in all other
NIR/MIR bands the agreement is very good. It could be due to some
unrealistic absorption effect in the
$\textit{K}_{s}$-band. Work is in progress to clarify this issue.

\subsection{Integrated Colours of Star Clusters and SSPs} \label{comp_ssp}

\begin{figure}
\begin{center}
\subfigure
{\includegraphics[width=0.41\textwidth]{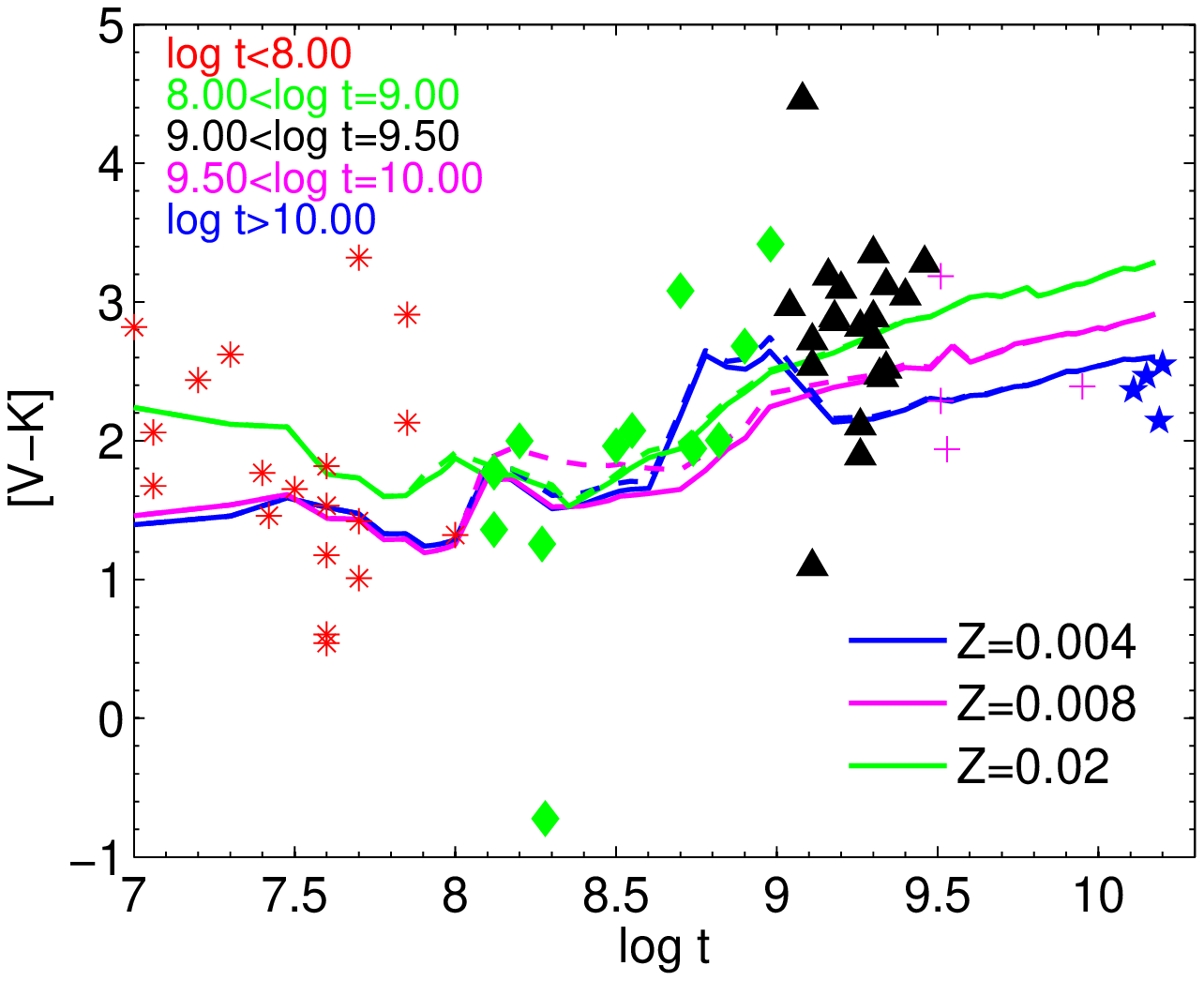}}
\subfigure
{\includegraphics[width=0.41\textwidth]{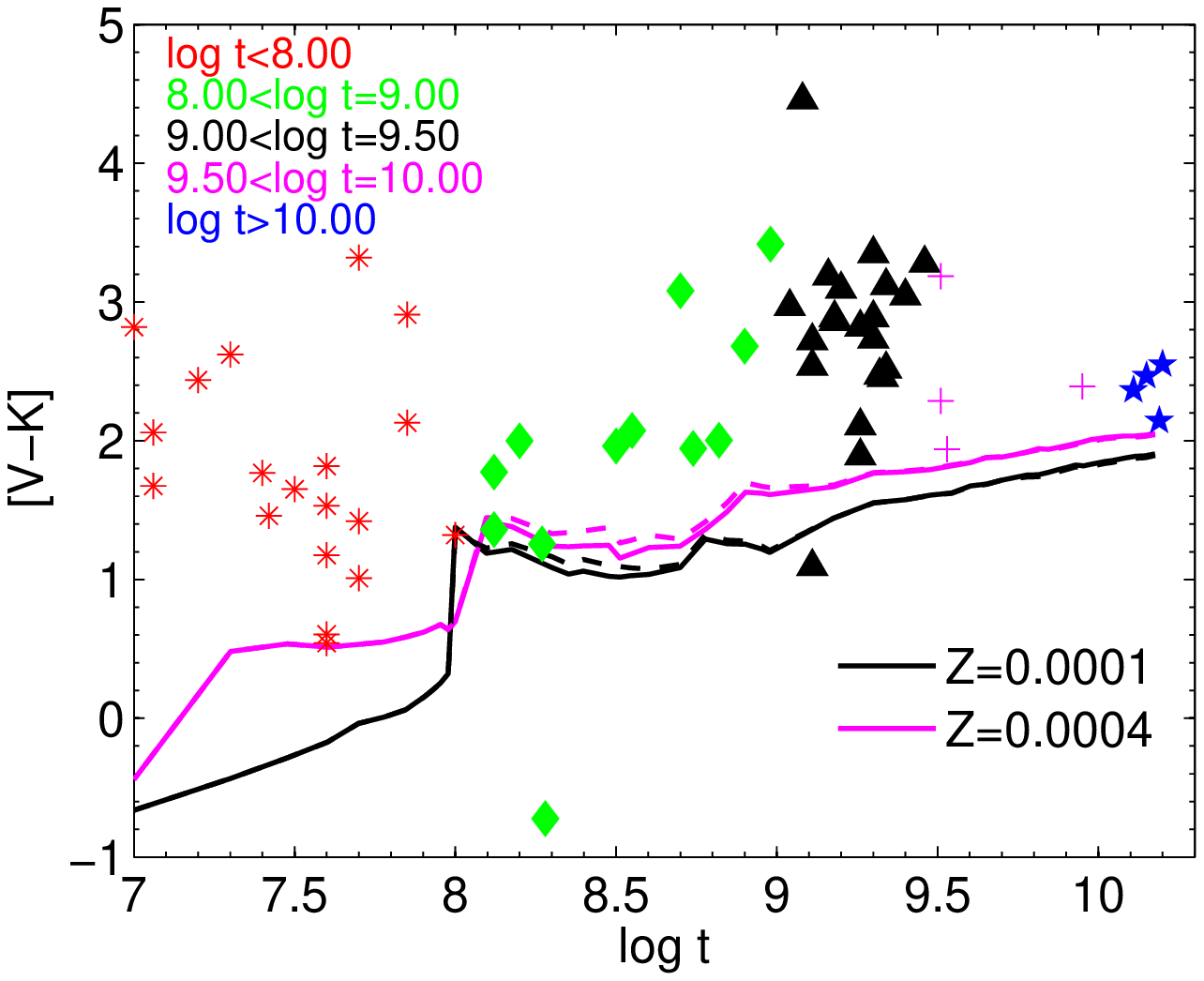}}
\caption{\textbf{Upper panel}: theoretical $[\textrm{V}-\textrm{K}_{s}]$ colours of our SSPs,
with (solid lines) and without (dashed lines)
the contribution of circumstellar dust shell
around AGB stars, as a function of age (from 0.01 to 15 Gyr, with t in yr).
We considered metallicities $Z$=0.004 (blue
lines), $Z$=0.008 (magenta lines), $Z$=0.02 (green lines). The SSPs
are superimposed on the integrated colours of LMC star
clusters, taken from the compilation by \protect \citet{Pessev2006, Pessev2008}; the data are plotted in
different colours according to the cluster age, as labelled. \textbf{Lower panel}: as in the
upper panel, but we superimposed  SSPs of metallicity $Z$=0.0001 (black lines) and $Z$=0.0004 (magenta lines).} \label{ssp_VK}
\end{center}
\end{figure}

\begin{figure}
\begin{center}
\subfigure
{\includegraphics[width=0.41\textwidth]{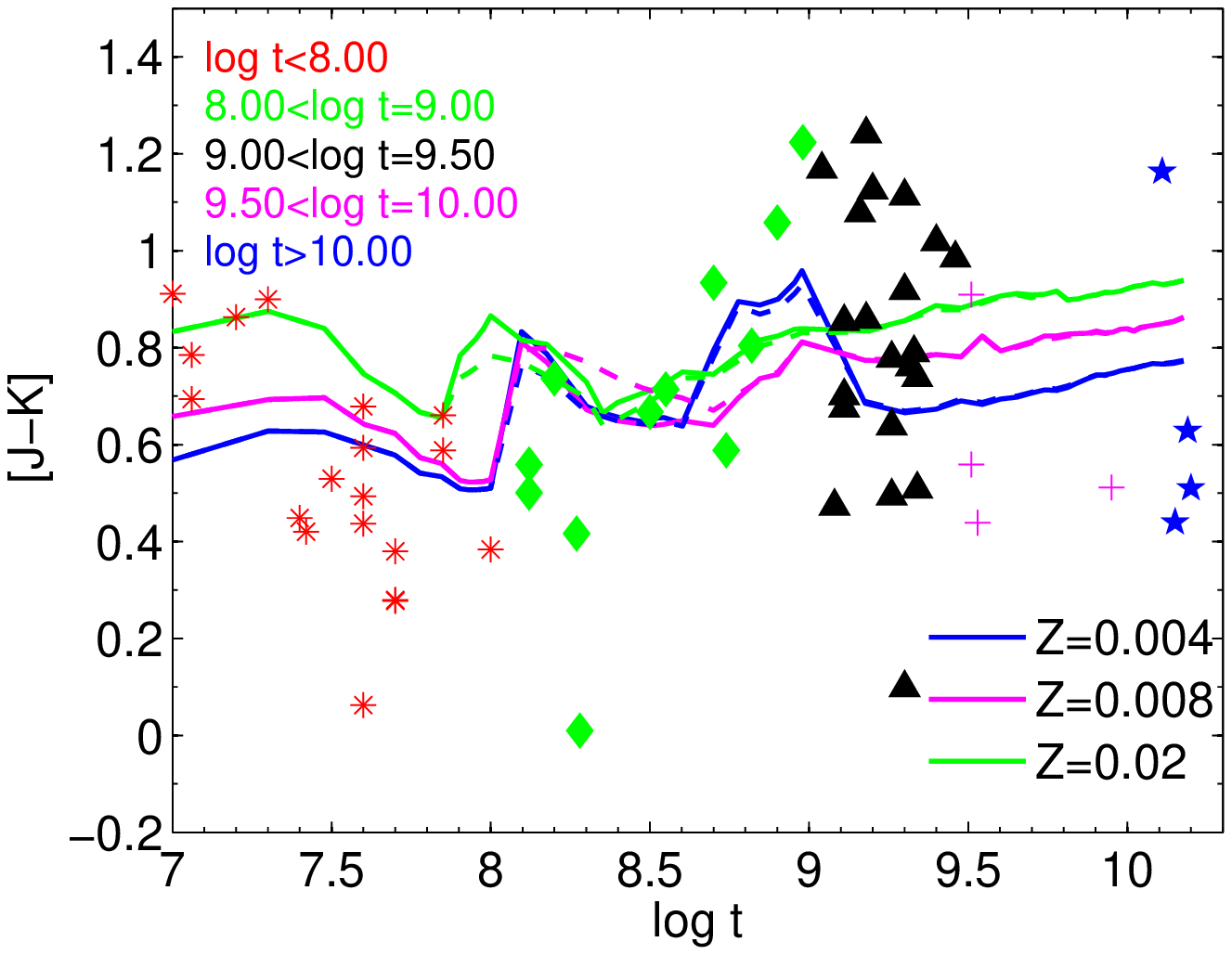}}
\subfigure
{\includegraphics[width=0.41\textwidth]{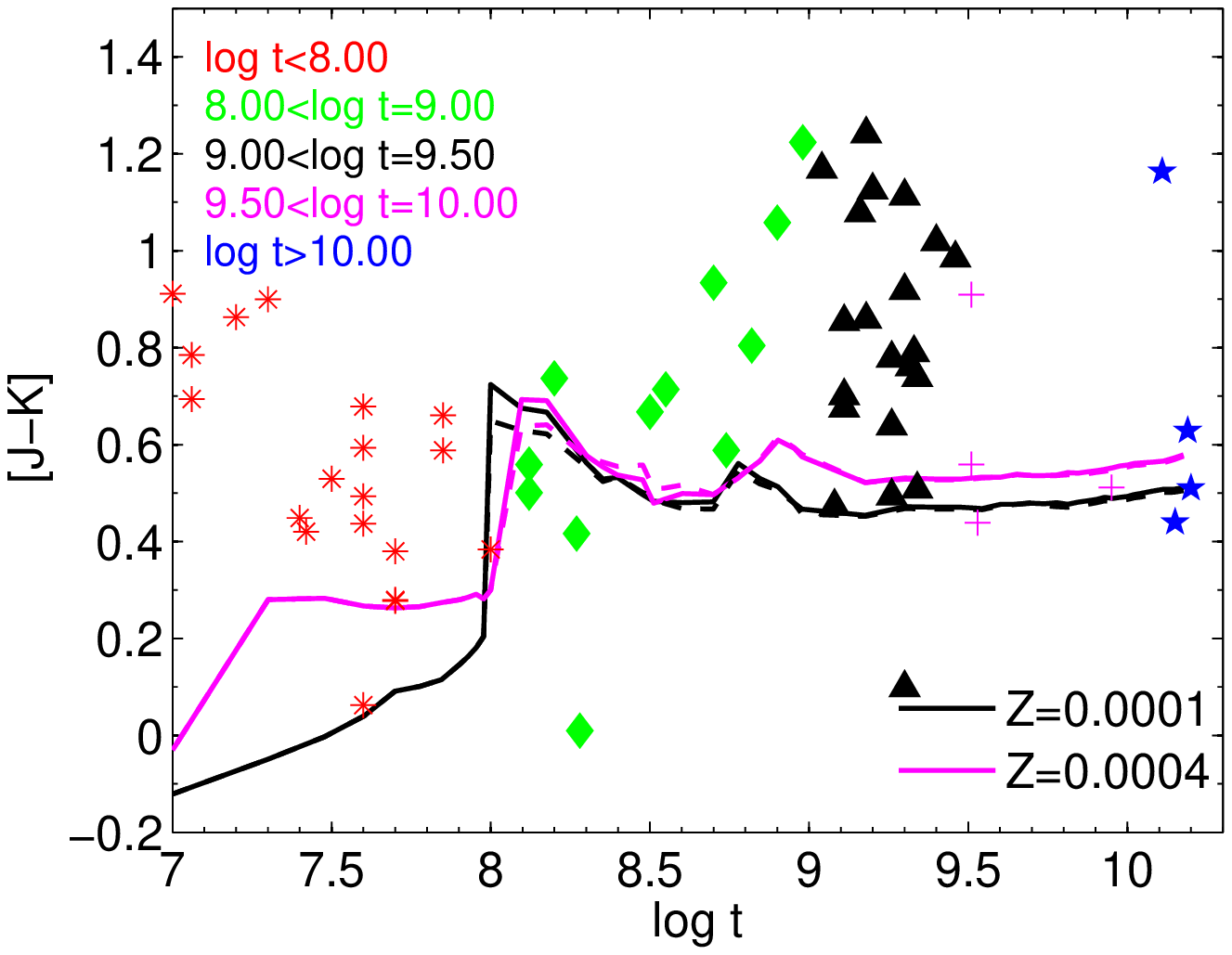}}
\caption{As Fig.~\ref{ssp_VK}, but for the
$[\textrm{J}-\textrm{K}]$ colour.} \label{ssp_JK}
\end{center}
\end{figure}

We compare now our models with integrated photometry of star clusters.
Figures~\ref{ssp_VK} and \ref{ssp_JK}  show the time evolution of
theoretical $[\textrm{V}-\textrm{K}_{s}]$ and
$[\textrm{J}-\textrm{K}_{s}]$ colours for SSPs with and without the
  contribution of dusty shells. 
They are superimposed on optical and near-infrared colours of LMC star clusters,
taken from the database by \citet{Pessev2006,Pessev2008}.

Recently \citet[][and references therein]{Pessev2006} presented
integrated NIR magnitudes (\textit{$\textrm{J}$},
\textit{$\textrm{H}$}, \textit{$\textrm{K}_{s}$}) and colours for a
large sample of star clusters in the LMC and SMC, using 2MASS
data. These clusters have a good estimate of both age and metallicity, and
can be used as a calibration set for SSPs models.
Many clusters have ages in the age range between 0.3 and 3 Gyr, and
consequently their integrated properties are heavily affected by AGB stars, that are extremely
luminous in the NIR \citep{Pessev2006}. \citet{Pessev2008}
combined these data with new photometry for nine additional objects: the whole resulting set
forms the largest existing database of
integrated NIR magnitudes and colours of
LMC/SMC star clusters. Moreover, the 2MASS data have also been merged with optical photometry from the works by
by \citet{Bica1996} and \citet{vandenBergh1981}.
This data set provides a very good coverage of the age-metallicity parameter space of LMC/SMC star clusters.

As in similar studies by
\citet{Elson1985a,Elson1985b,Elson1988}, \citet{Chiosi1986,Chiosi1988},
\citet{Girardi1995}, and \citet{Pessev2008}, we compare  theoretical
predictions and observed cluster colours as a
function of age, to test the models, to single out possible ages
where discrepancies do appear, and to
study the evolutionary history of the photometric properties of a
star cluster. As pointed out long ago by \citet{Chiosi1986,Chiosi1988}
and more recently by \citet{Pessev2008}, the comparison
  between theory and observations yields good results only  when
	stochastic effects on the integrated colours are taken into account.
The issue is the following: the calculation of integrated properties of SSPs
assumes that all stellar evolutionary phases are well sampled. Therefore, a
model will match the observations only if the observed integrated magnitudes/colours sample a number of stars large enough.
As a consequence, magnitudes/colours of clusters of similar ages and metallicities
hosting AGB stars may display a large spread due to the effect of stochastic fluctuations 
in the number of bright, short-lived AGB stars, caused by the small, finite number of
objects populating a real cluster. In other words, the integrated magnitudes/colours of
star clusters with a small, stochastically fluctuating number of AGB
stars, may take a large range of values that can be very different from the value calculated 
for a fully sampled AGB phase, with the number of stars
appropriately scaled according to the AGB lifetime and the cluster total mass, 
in compliance with the Fuel consumption Theorem
\citep{RenziniBuzzoni1983}.\\
\indent Although the star clusters of the
Magellanic Clouds are richer in stars than Galactic open clusters, stochastic effects can still be sizeable. 
Attempts to compare theory and observations by means of mean values of the integrated colours of clusters grouped 
into age bins \citep{Noeletal2013}, may led to misleading results.
Furthermore, the luminosity and effective temperature of stars during
the TP-AGB phase vary significantly,  affecting the colours by deviating from the mean, 
smoothed isochrone discussed in Sect.~\ref{iso_state_art}. This issue will be treated in detail in a 
forthcoming paper of this series (Salaris et al.\ 2013, in preparation).\\
\indent The temporal variation of the integrated $[\textrm{V}-\textrm{K}_{s}]$ colours of LMC clusters is 
presented in Fig.~\ref{ssp_VK}. For the sake of clarity, the upper panel
contains theoretical colour-age relations for $Z$=0.004, $Z$=0.008 and $Z$=0.02, while the lower
panel shows the same but for SSPs with $Z$=0.0001 and $Z$=0.0004.
Solid lines denote the case with dusty AGB stars, while
dashed lines display dust-free results. Figure~\ref{ssp_JK} shows the same comparisons but
for the $[\textrm{J}-\textrm{K}_{s}]$ colour.
The data have been selected (and plotted with different
  colours) according to their metallicity and age, the latter estimated
  by \citet{Pessev2006,Pessev2008} from the cluster CMDs (see this
  paper for further details).  Four age groups are considered. All
these clusters have [Fe/H] $>-1.71$, therefore they can be classified
as \textit{metal-rich}.  The data agree reasonably well with
  the theoretical age-colour relationships, in particular when
  appropriate metallicities for the clusters of the LMC are taken into
  account. This is displayed in the top panels of Figs.~\ref{ssp_VK}
  and \ref{ssp_JK}, for both the colours with and without dust around
  AGB stars, whereby the difference between the two cases starts when
  the first AGB stars appear. Finally, we call attention to two
'bumps' toward
redder colours displayed by the theoretical colour-age relationships:
the first one occurs at an age of about 0.1 Gyr and is caused by the onset
of the  AGB phase; the colour becomes
redder because of the increased flux in the near-IR pass-bands. The
second 'bump' occurs at an age between 1 and 2~Gyr, and it is likely due to
the development of the RGB phase; this is more evident when a higher
metallicity is taken into account. See \citet{Chiosi1988},
\citet{Girardi1995}, and \citet{Bressan1994} for a detailed discussion
of the subject.

\begin{figure}
\begin{center}
{\includegraphics[width=0.41\textwidth]{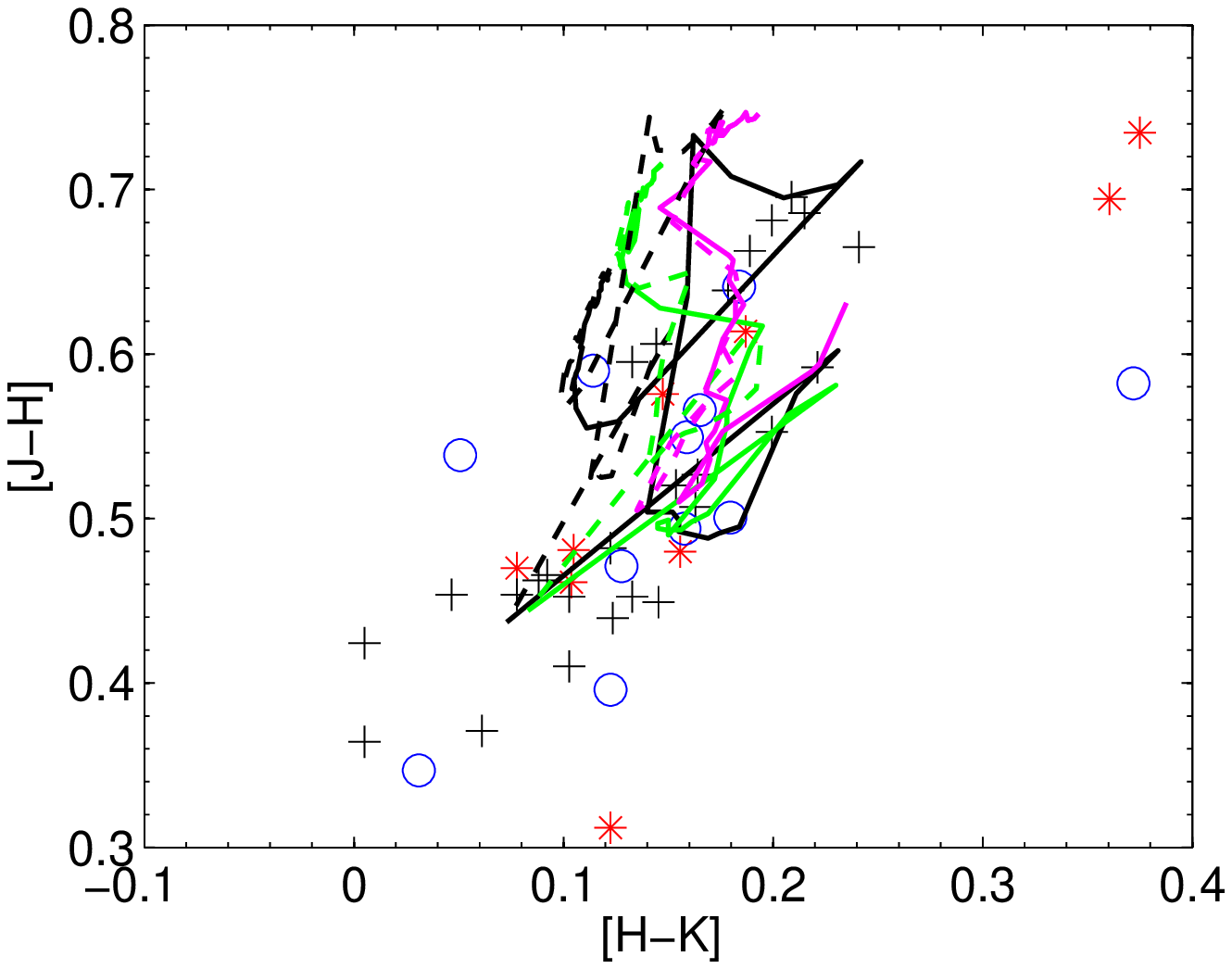}}
\caption{The \textit{two colour diagram}
$[\textrm{H}-\textrm{K}]$ vs. $[\textrm{J}-\textrm{H}]$ for young star
  clusters in the Magellanic Clouds. Open circles are LMC clusters
  selected by \protect \citet{Mouhcine2002c} from the catalogue of
  \protect \citet{Persson1983}, while red stars denote
  the SMC counterpart. Cross-shaped points are LMC clusters whose IR colours
  have been collected by \protect \citet{Pretto2002} using 2MASS
  data. All data have been reddening corrected. The lines
  show the colour range spanned by the new SSPs with (solid lines)
  and without (dashed lines) the contribution of circumstellar dust
  shells. Results for different values of the metallicity are shown:
  $Z$=0.02 (magenta), $Z$=0.004 (black) and $Z$=0.008 (green). The
  range of SSP ages goes from 0.1~Gyr to 15~Gyr.}  \label{colour16}
\end{center}
\end{figure}

\begin{figure}
\begin{center}
{\includegraphics[width=0.41\textwidth]{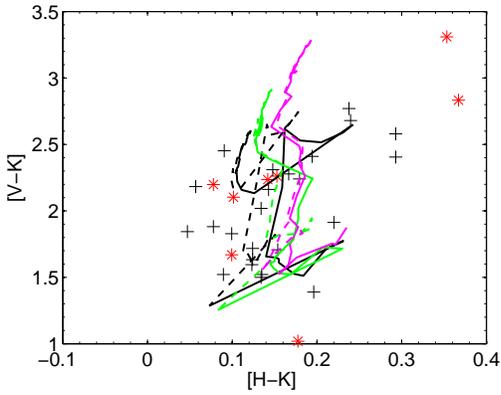}}
\caption{The same as in Fig. \ref{colour16}, but for $[\textrm{H}-\textrm{K}]$ vs. $[\textrm{V}-\textrm{K}]$ colours.
The meaning of the symbols is the same as in Fig. \ref{colour16}.}
\label{colour17}
\end{center}
\end{figure}

\begin{figure}
\begin{center}
{\includegraphics[width=0.41\textwidth]{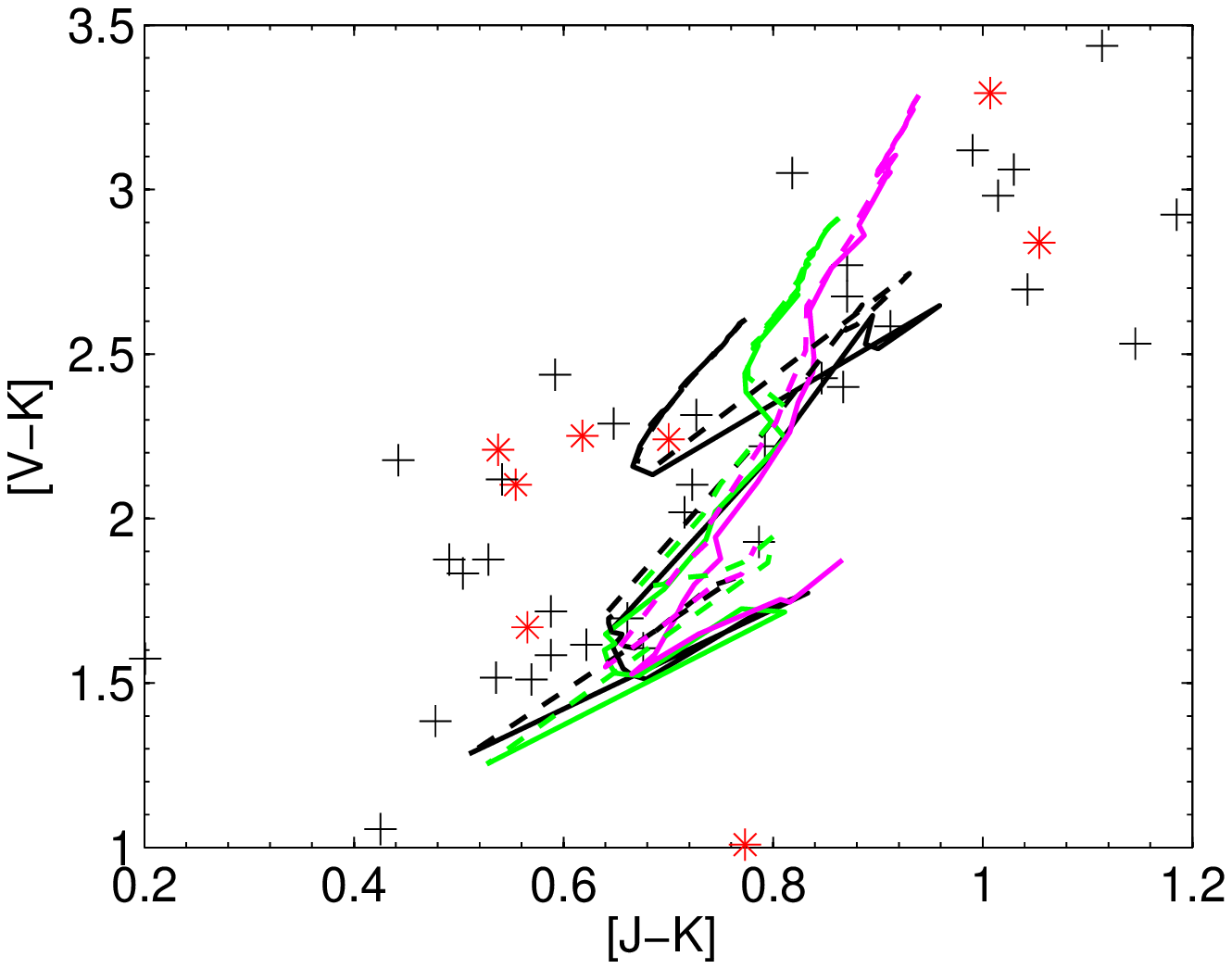}}
\caption{The same as in Fig. \ref{colour16}, but for $[\textrm{J}-\textrm{K}]$ vs. $[\textrm{V}-\textrm{K}]$ colours.
The meaning of the symbol is the same as in Fig. \ref{colour16}.}
\label{colour18}
\end{center}
\end{figure}

\subsection{The colour-colour diagrams}
Two-colour diagrams are powerful diagnostics of several photometric
properties of star clusters \citep[see][for a detailed discussion of
  the issue]{Girardi1995}.  Following \citet{Piovan2003},
we selected a small sample of clusters whose AGB population provides a
strong contribution to the integrated light. These photometric data come
from the \textit{2MASS Second Incremental Data Release} and the
\textit{Image Atlas}, that contains 1.9 millions of images in the
infrared bands $J$, $H$, and $K_{s}$. \citet{Pretto2002} calculated the
integrated magnitudes and made them available to us. Moreover, we considered the IR 
colours for the LMC and SMC clusters by \citet{Persson1983} used by \citet{Mouhcine2002c}.
As for their ages, we rely on the compilation by
\citet{Pietrzynski2000}, who presented age determinations for about 600
star clusters belonging to the central part of the LMC. They
are younger than 1.2~Gyr, thus their AGB stars provide a major
contribution to the integrated light.
Figures~\ref{colour16}, \ref{colour17}, and \ref{colour18} display the
two-colour diagrams
 $[\textrm{H}-\textrm{K}]$ vs.  $[\textrm{J}-\textrm{H}]$,
 $[\textrm{H}-\textrm{K}]$ vs.  $[\textrm{V}-\textrm{K}]$, and
 $[\textrm{J}-\textrm{K}]$ vs.  $[\textrm{V}-\textrm{K}]$, respectively. Each diagram shows the
data by \citet{Persson1983} for the LMC (open circles) and SMC (black stars), and the data 
by \citet{Pretto2002} (cross-shaped points), all
corrected for the reddening. The upper panel of each figure shows the new SSPs with
(solid lines) and without (dashed lines)  dust shells around  AGB stars,
whereas the lower panels display the old SSPs by \citet{Bertelli1994}.
For both types of SSPs, different metallicities are considered, namely
$Z$=0.02 (magenta lines), $Z$=0.004 (black lines), and $Z$=0.008
(green lines); the age ranges goes from 100 Myr (when AGB stars start
contributing  to the integrated light of the stellar population) to 13
Gyr, when only low-mass AGB stars are present.

The new dusty SSPs cover the observed colour ranges better than both the old SSPs by \citet{Bertelli1994} 
\citep[see][for more details]{Piovan2003} and
the new ones without dust shells around AGB stars.
The dust-free SSPs span a narrower range in
$[\textrm{H}-\textrm{K}]$, while the new dusty SSPs extend over a larger
$[\textrm{H}-\textrm{K}]$ interval and to bluer
$[\textrm{J}-\textrm{H}]$ and $[\textrm{V}-\textrm{K}]$ colours.  We
recall that the  $[\textrm{V}-\textrm{K}]$
colour is particularly suited to study the AGB phase of stellar
populations. In fact the AGB phase greatly contributes to the NIR
light of SSPs between $\sim$100 Myr and $\sim$1 Gyr, and causes abrupt changes in the NIR luminosity,
while producing only small changes in the optical.
Our new SSPs extend towards bluer colours, approximately like the SSPs by \citet{Mouhcine2002c},
as shown in \citet{Piovan2003}. Differences between our SSPs and the models by \citet{Mouhcine2002c}
are due to their adoption of empirical spectra for O- and C-stars, that are limited to the
objects they observed, and do not cover the full range of relevant parameters (age, metallicity, [C/O]-ratio).
In our work we use instead theoretical spectra of M-stars, that cover all relevant parameters' range,
but of course cannot match perfectly the empirical spectra. More work is
required to improve the theoretical spectra of cool stars to be
included in population synthesis studies.

\begin{figure}
\begin{center}
\subfigure
{\includegraphics[width=0.41\textwidth]{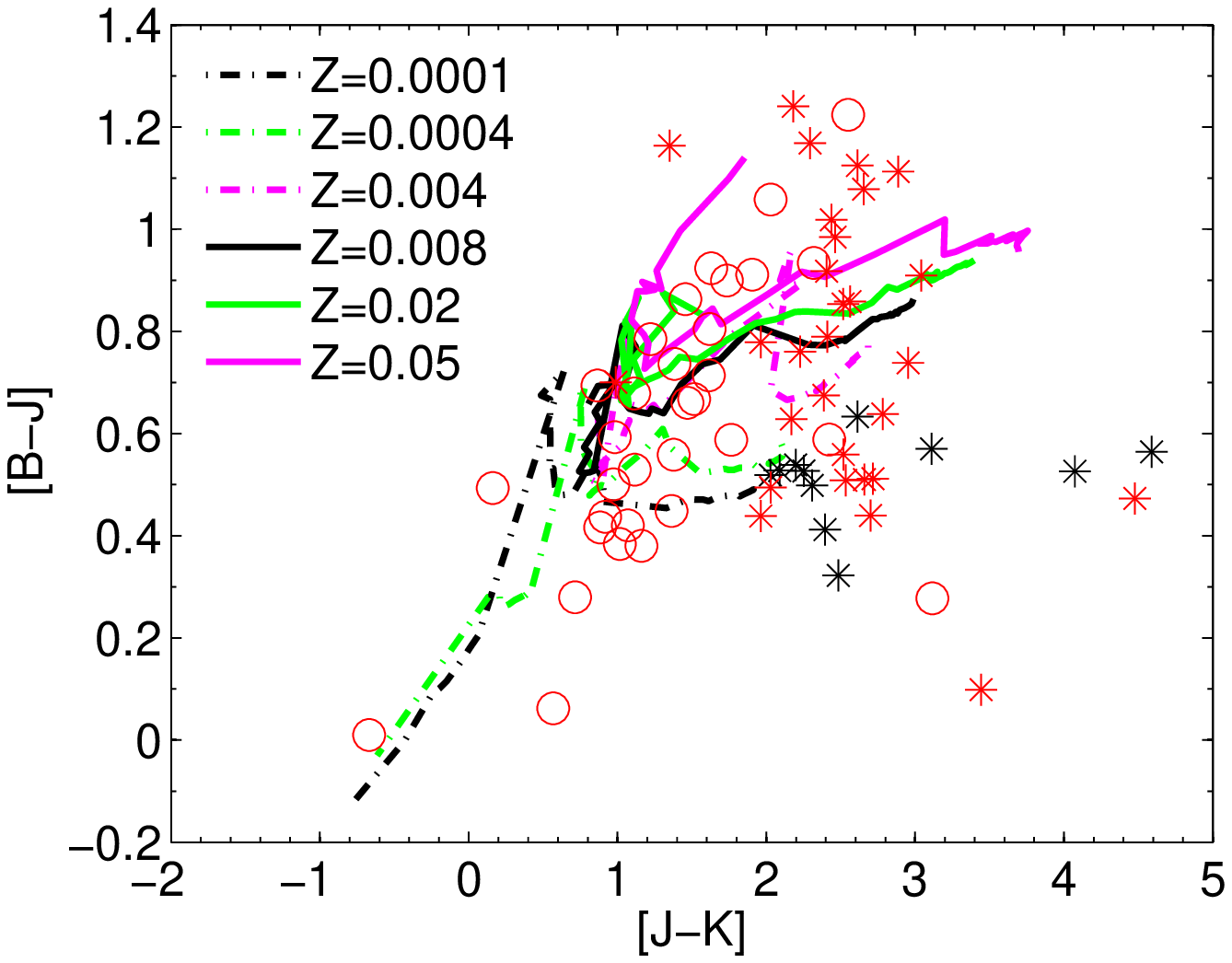}}
\subfigure
{\includegraphics[width=0.41\textwidth]{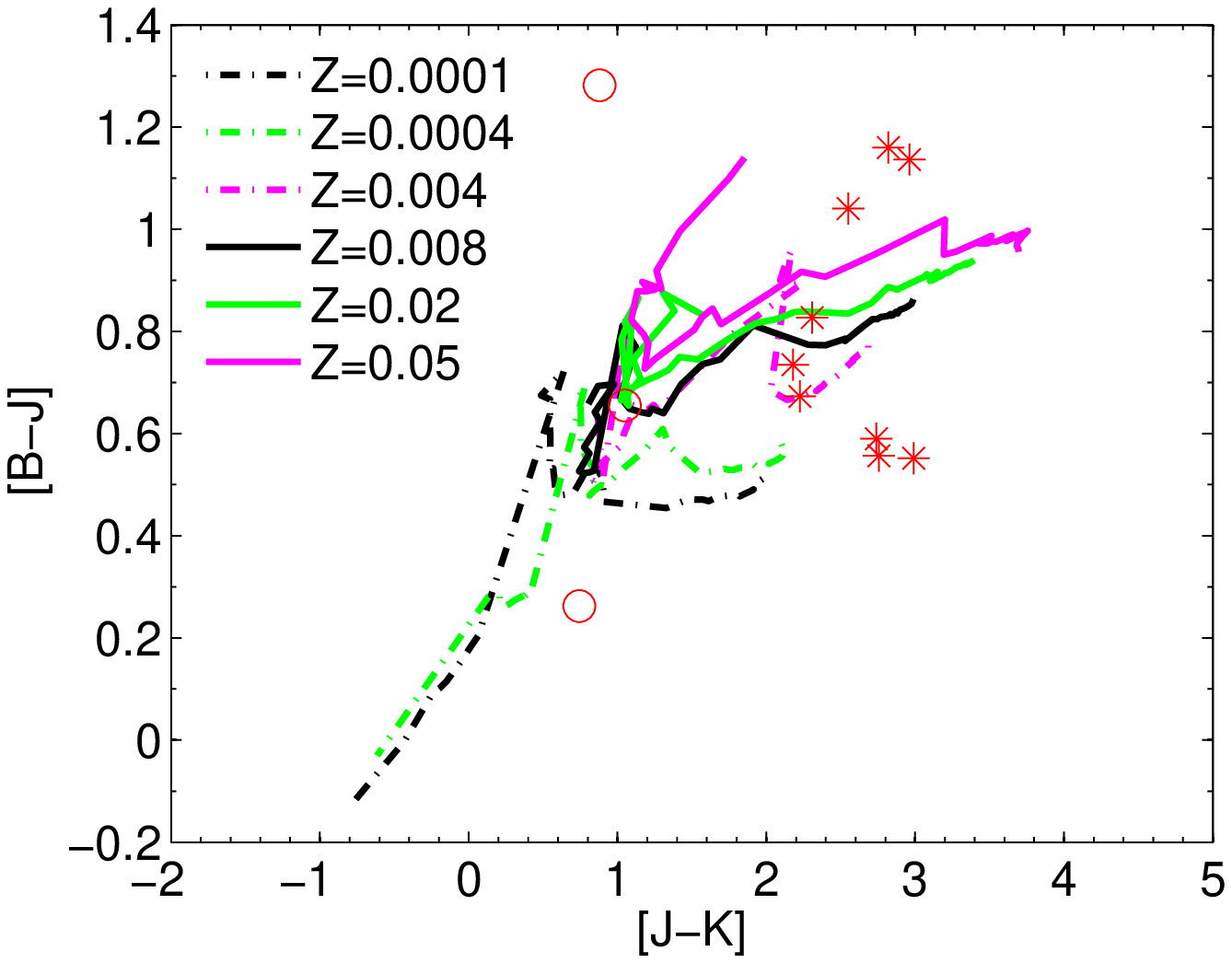}}
\caption{ $[\textrm{B}-\textrm{J}]$ vs.  $[\textrm{J}-\textrm{K}]$
  diagrams for a sample of LMC (\textbf{upper panel})
  and SMC (\textbf{lower panel}) star clusters. Three combinations of age and metallicity are
  considered: open red circles denote young ($t< 0.95$ Gyr)
  and metal-rich ([Fe/H] $>-1.71$) clusters, red stars denote
  old and metal-rich ones, and black stars display old and
  metal-poor objects. Superimposed on the data we show the new SSPs
  for different values of the metallicity, as labelled.} \label{PessevBJ}
\end{center}
\end{figure}

\indent It is interesting to compare colours that help to break the
 age-metallicity degeneracy.  Following \citet{Pessev2008},
we consider the $[\textrm{B}-\textrm{J}]$ vs.  $[\textrm{J}-\textrm{K}]$ diagram.
\citet{Puzia2007} have shown that the colour $[\textrm{B}-\textrm{J}]$
is well suited for age estimates,  while
$[\textrm{J}-\textrm{K}]$ is more sensitive to metallicity (except for
the small age interval around the onset of the  AGB phase, where
$[\textrm{J}-\textrm{K}]$ can show a weak age dependence).
We again use the database by \citet{Pessev2006,Pessev2008}, considering the whole
metallicity grid and ages from 10~Myr to 13~Gyr from our new dusty SSPs.
Data and theoretical predictions are compared in Fig.~\ref{PessevBJ}.
The age increases from  left to right along the theoretical sequences.
We consider here LMC and SMC clusters of any age, split
into two groups: young  ($t< 0.95$ Gyr) and old ones (age above this limit).
Furthermore, we subdivide the clusters into \textit{metal-rich}
([Fe/H] $>-1.71$), and \textit{metal-poor} ones ([Fe/H] above this limit).

The open red circles denote young and metal-rich clusters,
red stars display the old and metal-rich ones, and finally
the black star denote the old and metal-poor
objects. The upper panel shows the LMC clusters
and the bottom panel the SMC ones. In case of the the LMC we
have clusters with all combinations of age and metallicity (only young metal poor clusters
are lacking), while the SMC the database
does not include any metal-poor object.
There is a clear separation between the different age and
metallicity groups: our theoretical colours
overlap with the data, particularly in the region of the youngest
metal-rich clusters. Instead, they fail to match the old
clusters of the LMC between  0.3 $\lesssim$
$[\textrm{B}-\textrm{J}]$ $\lesssim$ 0.6 and 2 $\lesssim$
$[\textrm{J}-\textrm{K}]$ $\lesssim$ 5, that are mostly old
metal-poor (even if some metal-rich clusters appear in this
region). We encountered some problems also in the region  between  1
$\lesssim$  $[\textrm{B}-\textrm{J}]$ $\lesssim$ 1.3 and 2 $\lesssim$
$[\textrm{J}-\textrm{K}]$ $\lesssim$ 3, populated by the oldest
metal-rich clusters. Of course this partial disagreement with the
observations suggests that some of the ingredients of the models should
be improved, like the spectra for the cool stars in our stellar
  spectra library. It is however important to notice, as already pointed
out, that stochastic fluctuations of the number of giant stars will always
cause a spread of cluster colours at any age and metallicity. This
effect is obviously amplified in case  the SMC clusters,
considering that we are analyzing \textit{only} a very small number of
objects. Stochastic fluctuations could  explain the disagreement
between data and theory in the above colour intervals as well.

\begin{figure}
\begin{center}
\subfigure
{\includegraphics[width=0.41\textwidth]{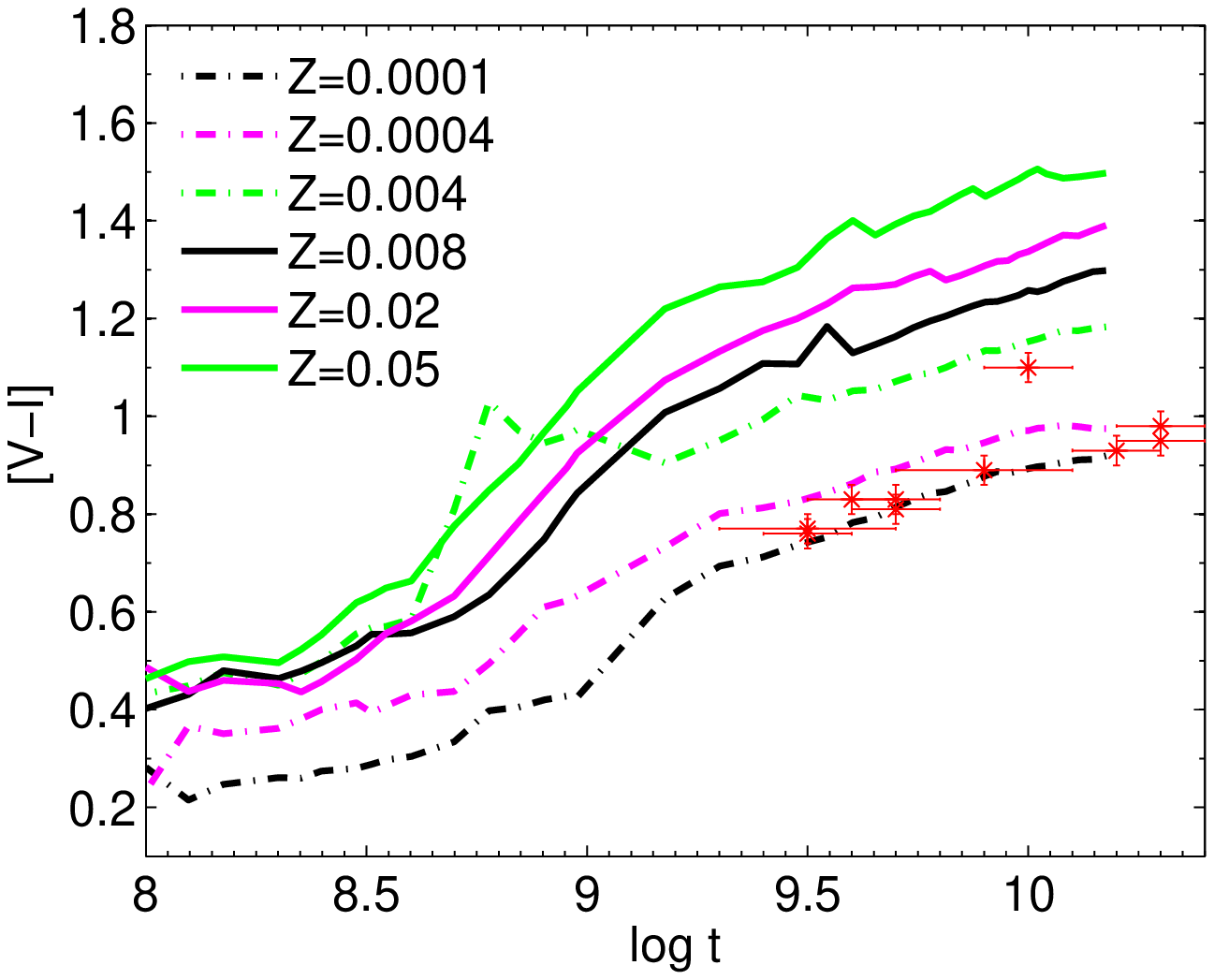}}
\subfigure
{\includegraphics[width=0.41\textwidth]{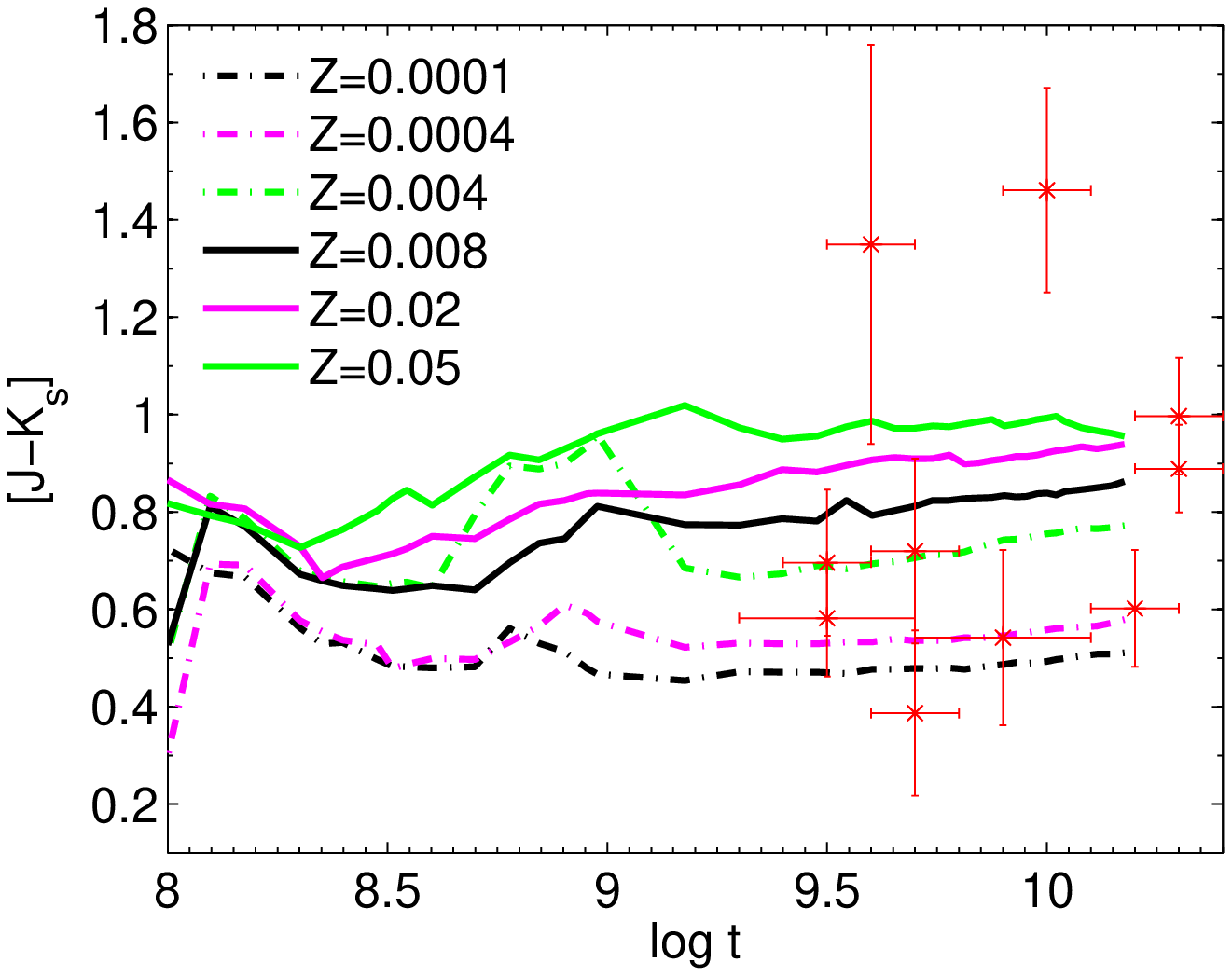}}
\caption{Theoretical evolution of the  $[\textrm{V}-\textrm{I}]$
  (\textbf{upper panel}) and  $[\textrm{J}-\textrm{K}]$ (\textbf{lower panel}) colours as a function of age.
SSPs for different metallicity are shown, as labelled. The time $t$ is in yr. 
The data and their uncertainties are from \protect\citet{Ma2012}.}
\label{Jumaage}
\end{center}
\end{figure}

\subsection{Clusters of  M31}

Before concluding, we examine a sample of star clusters in M31, another Local Group galaxy, with a
distance modulus of 24.47 mag \citep{Holland1998,Stanek1998,McConnachie2005}.
 M31 is the ideal Local Group galaxy to study globular clusters, for it contains more globulars than all
 other Local Group galaxies combined \citep{Battistini1987,Racine1991,Harris1991,FusiPecci1993}. The globular cluster system of M31
  has  been extensively studied over many decades  \citep[see][for a recent review of the subject]{Ma2012}.
In the following we focus on clusters located in the outskirts of this galaxy, because they are useful probes of the substructures
expected to form in the outer region of the host galaxy \citep{Ma2012}, as predicted by the popular $\Lambda$CDM model for galaxy
formation. This model predicts that these substructures continue to grow in time, from
the accretion and the disruption of companions satellite.

\indent We make use of the $J$, $H$, and $K_s$ magnitudes from 2MASS
imaging of 10 globular clusters in the outermost regions of M31,
analyzed by \citet{Ma2012}. These data have been combined with
$\textit{V}$ and $\textit{I}$ photometry
  \citep{Galleti2004,Galleti2006,Galleti2007,Huxor2008}.  The sample
of M31 halo globular clusters is taken from
\citet{Mackey2006,Mackey2007}, who evaluated the metallicities
\textbf{($-2.14 <\mathrm{[Fe/H]}<-0.70$)}, distance modulus and reddening from
their CMDs, using deep images obtained with
ACS/\textit{HST}. The cluster ages have been determined by
  \citet{Ma2012}, by comparing the integrated photometry with the
  SSP models by \citet{Bruzual2003},
  based upon the \citet{Fagotto1994a,Fagotto1994b,Fagotto1994c}
  evolutionary tracks and a Salpeter IMF from 0.1M$_{\odot}$ to
  100M$_{\odot}$. We adopt these age determinations, the same
  $[\textrm{V}-\textrm{I}]$ and $[\textrm{J}-\textrm{K}]$ colours and
  associated uncertainties, and compare these data to our SSPs with
  circumstellar dust around AGB stars.
The upper panel of  Fig. \ref{Jumaage} displays the theoretical evolution of $[\textrm{V}-\textrm{I}]$
as a function of age, whereas  the bottom  panel
does the same but for the  $[\textrm{J}-\textrm{K}]$ colour. The $[\textrm{V}-\textrm{I}]$
colour appears best suited to match  the age and metallicity of these clusters.
Moreover, the uncertainties associated with the observational data are lower for this colour compared to $[\textrm{J}-\textrm{K}]$.
As a last remark, we note that for some clusters \citet{Ma2012} assigned a very old age, not only beyond the
upper limit of our SSPs (this is not a problem because they can be easily
extended), but more importantly well beyond the limit set by WMAP-5 data  on the age of the Universe, e.g.
13.772 $\pm$ 0.059 Gyr \citep{Dunkley2009}.

\section{Summary and Conclusions}\label{disc_conc}

Using state-of-the-art AGB models of low and intermediate-mass stars,
and accounting for the effect on the model SED of dust shells surrounding the AGB central star, 
we have  revised  the Padua library of isochrones  \citep{Bertelli1994} to cover a wide range  of chemical compositions (using a helium to heavy elements enrichment ratio $\Delta Y/ \Delta Z=2.5$) and ages.

The TP-AGB phase is taken from the detailed full evolutionary calculations by \citet{Weiss2009}. 
Two spectral libraries, containing about 600 dust enshrouded SEDs of AGB stars each, one for oxygen-rich M-stars and one for carbon-rich C-stars, have been calculated. Each library
considers different values of the necessary input parameters (like the optical depth $\tau$, 
dust composition, and temperature on the inner boundary of the dust shell). This is one of the few 
theoretical studies of stellar SEDs with dusty AGB stars, suitable for use in EPS models. 
The AGB star SEDs have been
implemented into larger libraries of theoretical stellar spectra, to cover all relevant regions of the 
HRD where stars with a large range of initial stellar mass and chemical composition are found.

Starting from these isochrones and SEDs, we have calculated the integrated
SED, magnitudes, and colours of SSPs with several different metallicities (6 values), 
ages (more than fifty values from to $\sim$3 Myr to 15 Gyr),
and nine choices of the IMF.

The new isochrones and SSPs have been compared with CMDs of stellar
populations in the LMC and SMC, with particular emphasis on the match to the observed AGB
sequences. We also compared our theoretical predictions with
integrated colours of star clusters in the same galaxies, using
data from the SAGE catalogues \citep{Blum2006,Bolatto2007},
which match IRAC (or MIPS) data with 2MASS photometry. Finally, we have
examined the integrated colours of star clusters located in the
outskirts of M31.

The agreement between theory and observations is generally good, even
if some discrepancies still occur.
In particular, the new SSPs reproduce the very extended red tails of AGB stars in CMDs.

The whole libraries of spectra, isochrones and SSPs are
made freely available to the scientific community, and can be obtained
from the authors upon request.

\section{Acknowledgments}
We would like to thank the anonymous referee whose comments improved the
quality of the manuscript.
We are deeply grateful to A. Buzzoni for fruitful discussions. L. P. Cassar\`{a}, L.
Piovan and M. Salaris acknowledge the Max-Planck-Institut f\"ur Astrophysik (Garching - Germany) for the warm and
friendly hospitality and the
computational support during the visits when part of this study has been carried out. L. P. Cassar\`{a}
is also grateful to P. Marigo for useful discussions, and to D. Maccagni and M. Polletta for their support. 
This study makes use of data products from 2MASS,
which is a joint project of the University of Massachusetts and the Infrared Processing and Analysis
Centre/California Institute of Technology, founded by the National Aeronautics and Space Administration
and the National Science Foundation.

\begin{small}
\bibliographystyle{mn2e}                    
\bibliography{mnemonic,biblio}              
\end{small}
\end{document}